%% file: cosmo2.tex
\documentclass[12pt]{article}

\usepackage{epsfig,multicol,multirow}
\usepackage{amsmath,subfigure,latexsym,amssymb}

\def\lsi{\raise0.3ex\hbox{$<$\kern-0.75em\raise-1.1ex\hbox{$\sim$}}}
\def\gsi{\raise0.3ex\hbox{$>$\kern-0.75em\raise-1.1ex\hbox{$\sim$}}}
\def\backder{\raise1.4ex\hbox{$\leftarrow$\kern-0.75em\raise-1.4ex\hbox{$\partial$}}}
\def\forder{\raise1.4ex\hbox{$\rightarrow$\kern-0.75em\raise-1.4ex\hbox{$\partial$}}}

\newcommand{\excleq}{\mathop{\stackrel{!}{=}}}

\newcommand{\backderi}{\mathop{\backder}}
\newcommand{\forderi}{\mathop{\forder}}

\newcommand{\lsim}{\mathop{\lsi}}
\newcommand{\gsim}{\mathop{\gsi}}

\newcommand{\be}{\begin{equation}}
\newcommand{\ee}{\end{equation}}
\newcommand{\nn}{\nonumber}
\newcommand{\bea}{\begin{eqnarray}}
\newcommand{\eea}{\end{eqnarray}}

\newcommand{\uno}{1 \!\! 1}
\newcommand{\Z}{Z \!\!\! Z}
\newcommand{\R}{{\kern+.25em\sf{R}\kern-.78em\sf{I} \kern+.78em\kern-.25em}}
\newcommand{\RR}{{\kern+.25em\sf{R}\kern-.6em\sf{I} \kern+.6em\kern-.25em}}
\newcommand{\N}{{\kern+.25em\sf{N}\kern-.78em\sf{I} \kern+.78em\kern-.25em}}
\newcommand{\C}{{\kern+.25em\sf{C}\kern-.45em\sf{I} \kern+.45em\kern-.25em}}

\makeatletter
\@addtoreset{equation}{section}
\makeatother

\begin{document}

\input{titel2}

\newpage

\input{preface2}

\tableofcontents

\newpage

\section{Cosmic rays}

\input{cosmicrays2}

\section{Lorentz Invariance and its possible violation}

\input{LIV2}

\section{Cosmic $\gamma$-rays}

\input{gammaray2}

\vspace*{-1.5mm}

\section{Conclusions}

\input{conclu2}

\appendix

\section{New development since Nov.\ 2007: \\ The AGN Hypothesis}

\input{AGNhyp2}

\section{Table of short-hand notations}

\input{short2}

\input{acknow2}

\input{refscosmo2}
\end{document}

%% file: titel2.tex


\vspace*{0mm}

\begin{center}

{\Large\bf Cosmic Rays and the Search for} \\
\vspace*{5mm}
{\Large\bf 
a Lorentz Invariance Violation}

\vspace*{1cm}

Wolfgang Bietenholz\footnote{Present address: Institut
f\"{u}r Theoretische Physik, Universit\"{a}t Regensburg, \\ \indent
~\, D-93040 Regensburg, Germany. ~ E-Mail: bietenho@ifh.de} \\

\vspace*{3mm}

John von Neumann Institut (NIC) \\

\vspace*{1mm}
Deutsches Elektronen-Synchrotron (DESY)

\vspace*{1mm}

Platanenallee 6, D-15738 Zeuthen, Germany \\

%

\vspace*{3mm}

{\tt DESY-08-072} \\

\end{center}

\vspace*{3mm}

{\small This is an introductory review about the on-going search 
for a signal of Lorentz Invariance Violation (LIV) in cosmic rays. 
We first summarise basic aspects of cosmic rays, 
focusing on rays of ultra high energy (UHECRs).
We discuss the Greisen-Zatsepin-Kuz'min (GZK) 
energy cutoff for cosmic protons, which
is predicted due to photopion production
in the Cosmic Microwave Background (CMB). This 
is a process of modest energy in the proton rest frame. 
It can be investigated to a high precision in the laboratory, 
{\em if} Lorentz transformations apply even at 
factors $\gamma \sim O(10^{11})$. 
For heavier nuclei the energy attenuation is even faster due to
photo-disintegration, again {\em if} this process is Lorentz invariant.
Hence the viability of Lorentz symmetry up to tremendous 
$\gamma$-factors --- far beyond accelerator tests --- is a central issue.

Next we comment on conceptual aspects of
Lorentz Invariance and the possibility of its
spontaneous breaking. 
This could lead to slightly particle dependent
``Maximal Attainable Velocities''.
We discuss their effect in decays, \v Cerenkov radiation, 
the GZK cutoff and neutrino oscillation in cosmic rays.

We also review the search for LIV in cosmic $\gamma$-rays.
For multi TeV $\gamma$-rays we possibly encounter another puzzle 
related to the transparency of the CMB, similar to the GZK cutoff,
due to electron/positron creation and subsequent inverse Compton
scattering. The photons emitted in a 
Gamma Ray Burst occur at lower energies, but their very long
path provides access to information not far from
the Planck scale. We discuss conceivable non-linear photon 
dispersions based on non-commutative geometry or effective approaches.

No LIV has been observed so far.
However, even extremely tiny LIV effects
could change the predictions for
cosmic ray physics drastically.

An Appendix is devoted to the recent hypothesis
by the Pierre Auger Collaboration, 
which identifies nearby Active Galactic Nuclei --- or objects
next to them --- as probable UHECR sources. \vspace*{3mm} }

%% file: preface2.tex
\noindent
{\bf Preface :} \ This overview is designed for non-experts
who are interested in a cursory look at this exciting
and lively field of research. We intend to sketch 
phenomenological highlights and theoretical concepts
in an entertaining and self-contained form (as far as possible),
avoiding technical details, and
without claiming to be complete or rigorous.

%% file: cosmicrays2.tex
\subsection{Discovery and basic properties of cosmic rays}

Cosmic rays consist of particles and nuclei of cosmic origin
with high energy. Part of the literature restricts this term
to electrically charged objects, but in this review we adapt
the broader definition, which also embraces cosmic neutrinos
and photons.

The {\em discovery} of cosmic rays dates back to the
beginning of the $20^{\rm th}$ century.
At that time electroscopes and electrometers were 
developed to a point which enabled 
the reliable measurement of ionising $\gamma$-radiation. 
Around 1909 the natural radioactivity on the surface 
of the Earth was already well-explored, 
and scientists turned their attention to the radiation
above ground. If its origin was solely radioactivity in the Earth, 
a rapid reduction would have been predicted with increasing hight.

In 1910 first observations on balloons led to contradictory
results: K.\ Bergwitz did report such a rapid decrease \cite{Bergwitz},
whereas A.\ Gockel could not confirm it \cite{Gockel}.
Still in 1910 Th.\ Wulf --- who had constructed an improved
electrometer --- performed measurements on 
the Eiffel tower. 
He found some reduction of the $\gamma$-radiation compared to the ground,
but this effect was clearly 
weaker than he had expected
\cite{Wulf}. Although he admitted that the
iron masses of the tower could affect the outcome, he concluded
that either the absorption in the air is weaker than expected,
or that there are significant sources above ground.

In 1912 V.F.\ Hess evaluated his measurements during seven
balloon journeys \cite{Hess}: he found only a very mild reduction of
the radiation (hardly 10 \%) for heights around 1000 m.
Rising further he observed a slight increase, so that the radiation
intensity around 2000 was very similar that that on ground.
As he went even higher (up to 5350 m) he found a {\em significant 
increase,} which was confirmed in independent balloon-borne experiments
by W.\ Kolh\"{o}rster in 1913 up to 6300 m \cite{Kol}.

In Ref.\ \cite{Hess2} V.F.\ Hess presented a detailed analysis of
his measurements in heights of $(1000 \dots 2000)~{\rm m}$, which 
were most reliable. These results agreed essentially with those by
A.\ Gockel, who had also anticipated the observation of an
intensified radiation when he rose even higher \cite{Gockel}.
Hess ruled out the hypothesis of sources mainly
inside the Earth. He added that sources in the air would 
require {\it e.g.}\ a RaC density in high sectors of the atmosphere,
which exceeds the density measured near ground by about
a factor of $\approx 20$. He concluded that a large fraction
of the ``penetrating radiation'' does apparently {\em originate
outside the Earth and its atmosphere.} In a specific test during
an eclipse in 1912, and in the comparison of data
that he obtained at day and at night, he did not observe
relevant changes. From this he inferred further that the sun 
is unlikely to be a significant source of this radiation 
\cite{Hess},\footnote{Radiation from the sun occurs at
low energy, up to $O(10^{8})~{\rm eV}$.
It won't be addressed here, and we adapt a notion
of cosmic rays starting at higher energy.}
which we now denote as {\em cosmic rays} (the earlier German term
was ``H\"{o}henstrahlung'').

Another seminal observation was made by P.\ Auger in the Alps (1937)
\cite{Auger1}. Several Geiger counters, which were
separated by tens or hundreds of meters, detected an event at
practically the same time.
Such a correlation, in excess of accidental coincidence, was also reported
in the same year by a collaboration in Berlin \cite{KolMatWeb}.
Later B.\ Rossi commented that he had made similar observations
already in 1934. P.\ Auger gave the
correct interpretation of this phenomenon, namely the occurrence of 
{\em extended air showers} caused by cosmic rays: a high energy primary
particle of cosmic origin triggers a large cascade in the terrestrial
atmosphere \cite{Auger1}. Based on the comparison of
air shower measurements at sea level and on Jungfraujoch 
(in Switzerland, 3500 m above sea level), and some simple 
assumptions about the shower propagation, P.\ Auger {\it et al.}\ 
conjectured that primary particles occur at energies 
of at least $10^{15} ~{\rm eV}$ \cite{Auger2}.

\begin{figure}
\begin{center}
\includegraphics[angle=0,width=1.\linewidth]{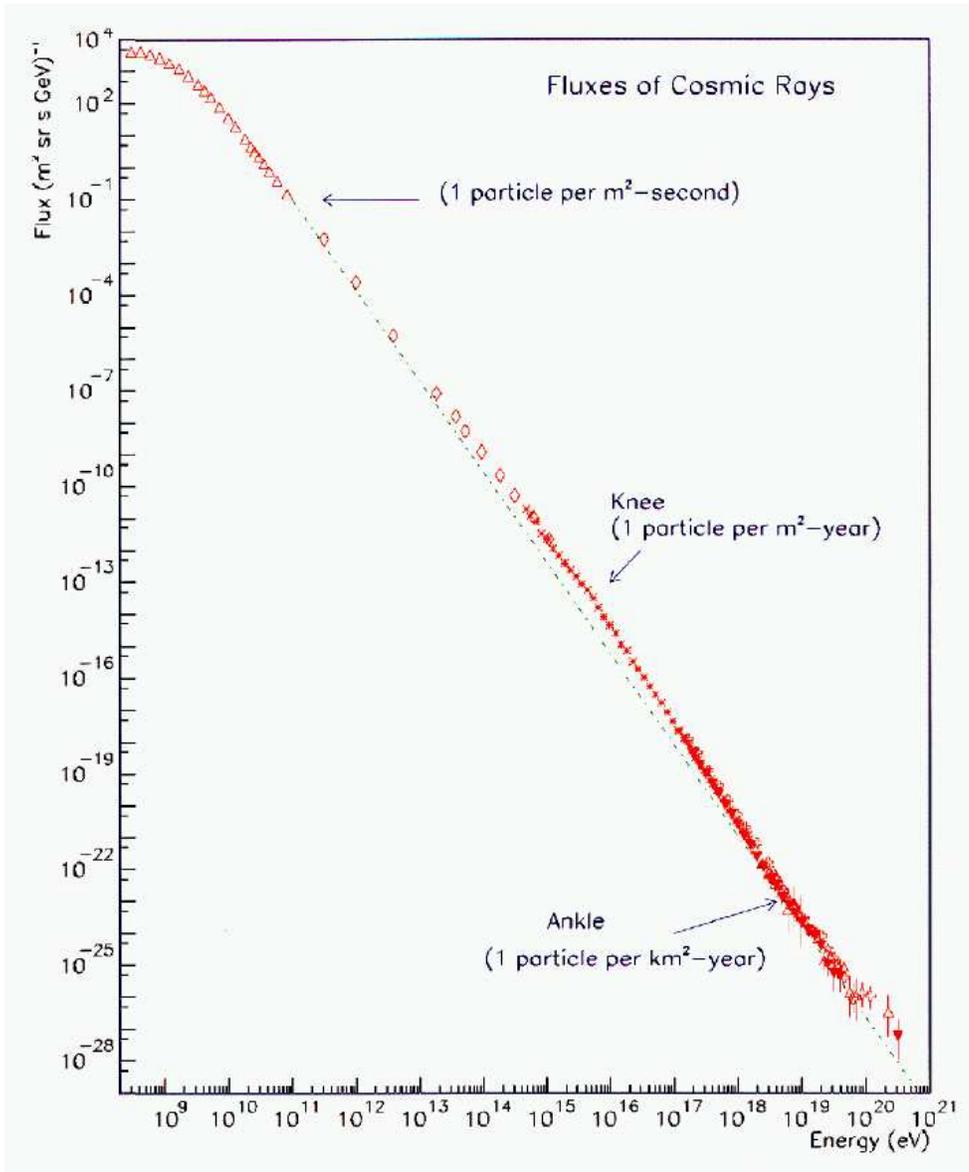} 
\end{center}
\vspace*{-5mm}
\caption{\it The flux intensity of cosmic rays as a function
of the energy (plot adapted from the HiRes Collaboration \cite{hires}). 
Over a broad energy interval it falls off essentially like
${\rm flux} \propto E^{-3}$ (or $ \propto E^{-2.7}$, according
to recent high precision measurements). 
We are most interested in the UHECRs, with energies
$E \gsim 10^{18.5}~{\rm eV}$. 
Around $E = 6 \cdot 10^{19}~{\rm eV}$ an abrupt flux reduction is 
predicted. This is the GZK cutoff, to be discussed in Subsection 1.3
and beyond.}
\label{fluxfig}
\end{figure}

Today we know that the primary particles carry
energies of about $E \approx (10^{9} \dots  
10^{20})~{\rm eV}$. The emerging cascades typically 
involve (in their maxima)
$O(1)$ secondary particle for each GeV of the primary energy,
so an air shower can include up to $O(10^{11})$ particles.
The energy dependence of the flux is illustrated in Figure \ref{fluxfig}. 
Its intensity falls off with a power law $\propto E^{-3}$
(new fits shift the exponent to $-2.7$), up to minor deviations 
(``knee''\footnote{In addition, a small ``second knee'' has been observed
around $10^{17.5}~{\rm eV}$. It is not visible in Figure \ref{fluxfig},
but we will comment on it in Subsection 1.3.\label{knee2}}, ``ankle'').
We will come back extensively to the behaviour in the highest
energy sector.

The {\em composition} of the rays depends on the energy. 
Here we are particularly interested in {\em ultra high energy cosmic rays
(UHECRs),} with energies $E \gsim 10^{18.5}~ {\rm eV}$, which are
mostly of extragalactic origin.
There one traditionally assumed about $90 \, \%$ of the primary 
particles to be protons, about $9 \, \%$ helium nuclei ($\alpha$), 
plus small contributions of heavier nuclei. 
At lower energies these fractions change.
In that regime, the magnetic field of $ \sim 3 ~ \mu {\rm G}$ 
in our galaxy confines charged particles as diffuse radiation, 
so that the galactic component dominates \cite{GMT}. 
In the range $(10^{15} \dots 10^{17})~{\rm eV}$ the rays
mainly consist of heavier nuclei
\cite{KASCADE}, and for $(10^{12} \dots 10^{15})~{\rm eV}$ 
there are $\sim 50 \, \%$ protons, $\sim 25 \, \%$ $\alpha$
and $\sim 13 \, \%$  C, N and O nuclei, according to Ref.\ \cite{hires}.
As we consider still lower energies, also leptons and 
photons contribute significantly, see Sections 2 and 3.
We quote here typical numbers from the literature, but many aspects 
of the energy dependent composition are controversial \cite{GMT}.

The arrival directions of charged rays are essentially {\em isotropic}.
This is explained by their deviation in interstellar 
magnetic fields: such fields occur at magnitudes of 
$\mu{\rm G}$ inside the galaxies, and of nG extragalacticly, 
which is --- over a large distance --- sufficient for 
a sizable deflection. As a consequence the origin of charged rays 
from far distances can hardly be located.\footnote{Recent 
developments could change this picture for the UHECRs,
see Appendix A. Of course, the approximately straight
path length increases with the energy, so that UHECR directions
may be conclusive if the source is nearby. As an example, a proton
of $E = 10^{20}~{\rm eV}$ has in our galaxy 
a Larmor radius,
$r_{\rm L} {\rm [kpc]} \approx E {\rm [ EeV]} / B [ \mu {\rm G}]$,
which exceeds the galaxy radius $\approx 15 ~ {\rm kpc}$ 
($1~{\rm EeV} = 10^{18}~{\rm eV}$, cf.\ Appendix B).} 

The {\em origin} of cosmic rays is still mysterious.
The first proposal was the {\em Fermi mechanism} (``second order''),
which is based on particle collisions in an interstellar magnetic 
cloud \cite{Fermi}. Collisions 
which are (almost) head-on are statistically favoured and lead to
acceleration. A later version of the Fermi mechanism (``first order'')
refers to shock waves in the remnant after a supernova \cite{Fermi2}.
However, these mechanisms can explain the cosmic rays at best
in part; they cannot provide sufficient energies for UHECRs.
In particular the first order mechanism 
could only attain about $E= O(10^{14})~{\rm eV}$, 
and it predicts a flux $\propto E^{-2}$ ---
but the observed flux is close to a behaviour $\propto E^{-3}$, as
Figure \ref{fluxfig} shows.
A variety of scenarios has been suggested later,
for comprehensive reviews we refer to Refs.\ \cite{Rev2002}.
They can roughly be divided into two classes:
\begin{itemize}

\item {\underline{Bottom-up scenarios}} : Certain celestial objects are 
equipped with some mechanism to accelerate particles to these tremendous
energies (which can exceed the energies reached in
terrestrial accelerators by at least 7 orders of magnitude). 
The question what these objects could be is puzzling.
Pulsars\footnote{Pulsars are rotating magnetised 
neutron stars.} \cite{pulsars} and quasars\footnote{A quasar, or 
quasi-stellar radio source, is an extremely bright centre of a 
young galaxy.\label{quasar}} \cite{FarBier} 
are among the candidates that were considered, and recently 
Active Galactic Nuclei (AGN)\footnote{An Active Galactic Nucleus 
is the environment of a super-massive black hole in the centre of a 
galaxy (its mass is estimated around $(10^{6} \dots 10^{10}) M_{\odot}\,$,
where $M_{\odot} \simeq 2 \cdot 10^{30}~{\rm kg}$ is the solar mass).
It absorbs large quantities of matter and emits highly energetic
particles, in particular photons, electrons and positrons
(we repeat that the acceleration mechanism is not known). 
Only a small subset of the galaxies have active nuclei. 
Our galaxy does not belong to them; the nearest AGN are located in
Centaurus A and Virgo; this will be of importance in 
Appendix A.\label{AGNf}}
attracted attention as possible UHECR sources, see Appendix A. 
However, a convincing explanation for the acceleration mechanism 
is outstanding (see Ref.\ \cite{UHECRsources} for a recent discussion, 
and  Ref.\ \cite{DarRu} for a modern theory).

\item {\underline{Top-down scenarios}} : 
The UHECRs originate from the decay or annihilation
of super-heavy particles spread over the Universe. 
These particles were generated in the very
early Universe, so here one does not need to explain where the energy
comes from (see {\it e.g.}\ Refs.\ \cite{BhaSig,topdown}). 
One has to explain, however, what kind of 
particles this could be: magnetic monopoles have been advocated
\cite{monopole}, while another community refers to
exotic candidates called ``wimpzillas'' \cite{wimpzilla} --- but there 
are no experimental signals for any of these hypothetical objects.
Moreover, this approach predicts a significant UHECR flux with
$\gamma$ and $\nu$ as primary particles, which is disfavoured by
recent observations: in particular Ref.\ \cite{NoTopDown} established 
above $10^{19}~{\rm eV}$ an upper bound of 2 \% for the photon flux.
The negative impact on top-down scenarios is further discussed in
Refs.\ \cite{NoTopDown2,Hoer}. Ref.\ \cite{Rub} had reported earlier 
measurements  with similar conclusions.
\end{itemize}

\subsection{The Cosmic Microwave Background}

\begin{figure}
\begin{center}
\includegraphics[angle=270,width=.55\linewidth]{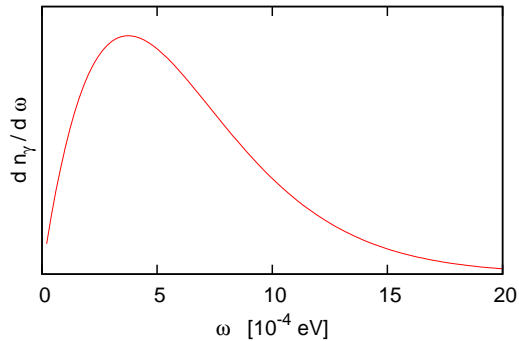} 
\vspace*{-3mm}
\end{center}
\caption{\it The Planck distribution (\ref{planckeq}) of the black body 
radiation at the CMB temperature of $T = 2.725 ~ {\rm K}$. 
The CMB energy density
agrees accurately with this Planck distribution --- it is by far
the most precise Planck radiation that has ever been measured.}
\label{planck}
\end{figure}

We proceed to another highlight of the historic development: in 
1965 A.A.\ Penzias and R.W.\ Wilson discovered (rather accidentally)
the Cosmic Microwave Background (CMB) \cite{PenWil}, which gave rise to 
a break-through for the evidence in favour of the Big Bang scenario. 
It is estimated that the 
CMB emerged after about $380 \, 000$ years. At that time the cooling of
the plasma reached a point where hydrogen atoms were formed.
Now photons could scatter off these electrically neutral objects,
so that the Universe became transparent 
(for reviews, see Refs.\ \cite{CMBrev}). 
The CMB that we observe today
--- after a period  of $13.7(2) ~ {\rm Gyr}$ of further cooling --- 
obeys to a high accuracy Planck's formula for the energy dependent
photon density in black body radiation,  
\be  \label{planckeq}
\frac{d n_{\gamma}}{d \omega} = \frac{1}{\pi^{2}}
\frac{\omega^{2}} {\exp (\omega /(k_{B}T)) -1} \ ,
\ee
where $\omega$ is the photon energy (for $\hbar = 1$)
and $k_{B}$ is Boltzmann's constant. This shape, plotted in Figure
\ref{planck}, is in excellent agreement with the most precise observation,
which was performed by the  Cosmic Background Explorer (COBE) satellite
\cite{COBE}.
Any deviation from the black body formula (\ref{planckeq})
over the wave length range 
$\lambda \simeq (0.5 \dots 5) ~{\rm mm}$ is
below $0.005 \, \%$ of the CMB peak.\footnote{Sizable deviations seem to
occur, however, at much larger wave length, $\lambda \gsim 1 ~{\rm m}$, 
due to an additional {\em radio
background}, the details of which are little known.\\
We add that there is also a Cosmic Neutrino Background, which decoupled
already about $2~{\rm s}$ after the Big Bang, and its present temperature 
is estimated as $\lsim 1.9 ~{\rm K}$. We do not consider it here because 
it has no significant impact on cosmic rays.\label{radio}}
The temperature is identified as $T = 2.725(1) ~ {\rm K}$ 
\cite{particledata}. This implies that the mean photon
energy and wave length are given by $\langle \omega \rangle
\simeq 6 \cdot 10^{-4}~{\rm eV}$ and $\langle \lambda \rangle
\simeq 1.9 ~{\rm mm}$.\footnote{The term ``microwave'' usually
refers to wave lengths in the range of about 
$1~{\rm mm} \dots 10~{\rm cm}$, so it does apply to the CMB.}
The resulting photon density in the Universe
amounts to $\int_{0}^{\infty} d \omega \, (d n_{\gamma}/ d \omega) 
\simeq 410.4(5) ~ {\rm cm}^{-3}$. Interesting further aspects of the
CMB --- which are, however, not directly relevant for our discussion ---
are reviewed for instance in Ref.\ \cite{Durrer}.

\subsection{The Greisen-Zatsepin-Kuz'min cutoff}

In the subsequent year the knowledge about cosmic rays {\em and} about
the CMB led to an epoch-making theoretical prediction, which
was worked out independently by K.\ Greisen at Cornell University 
\cite{Grei}, and by G.T.\ Zatsepin and V.A. Kuz'min 
at the Lebedev Institute \cite{ZK}. 
They expected the flux of cosmic rays to drop abruptly
when the energy exceeds the {\em ``GZK cutoff''\,}\footnote{Since 
this ``cutoff'' is not sharp, it is a bit arbitrary where exactly to put 
it. Eq.\ (\ref{cutoff}) gives a value which is roughly averaged over 
the literature, and which we are going to embed into phenomenological
and theoretical considerations.\label{GZKarb}}
\be  \label{cutoff}
\framebox[1.1\width]{ $ \displaystyle
E_{\rm GZK} \approx 6 \cdot 10^{19} ~ {\rm eV} $ } \ .
\ee
The reason is that protons above this energy interact
with background photons to generate pions. The dominant
channel for this {\em photopion production} follows the scheme
\be  \label{photopion}
p + \gamma \to \Delta (1232~{\rm MeV}) \begin{array}{cl}
\to & p + \pi^{0} \\
\to & n + \pi^{+} \ , \quad n \to p + e^{-} + \bar \nu_{e} \ .
\end{array}
\ee
These two channels cover $99.4 \, \%$ of the $\Delta (1232)$ 
decays \cite{particledata}.
If even more energy than the threshold for this transition is
available, excited proton states 
($p^{*} (1440)$, $p^{*} (1520) \dots $)
and higher $\Delta$ resonances ($\Delta (1600)$, 
$\Delta (1620)$, $\Delta (1700) \dots$)
can contribute to the photopion production as well. In these 
cases one may end up with $p+\pi$ or also with $p+ 2 \pi$;
at even higher energy the production of three pions is possible
too.

It is obvious that a proton with energy 
$E_{p}> E_{\rm GZK}$ looses energy under photopion production, 
until it drops below the threshold for this process.\footnote{In 
the original studies \cite{Grei,ZK}
terms like ``$\Delta$ resonance'' did not occur, but the authors knew
effectively about the photopion production. They inferred this
spectacular conclusion, although Ref.\ \cite{Grei} only consists
of 2 pages without any formula, and Ref.\ \cite{ZK} is just 
a little more extensive. Its original Russian version has been 
translated but only few people have read
(the page number is often quoted incorrectly). Nevertheless 
both works are of course top-cited on a renowned level.} 
We are now going to take a somewhat more quantitative look
at transition (\ref{photopion}) (in natural units, $\hbar = c =1$).

We denote the proton energy $E_{p}$ at the
threshold for the $\Delta (1232)$ resonance as $E_{0}$.
In these considerations one refers to the Friedmann-Robertson-Walker 
metrics (see Appendix B) --- which is
co-moving with the expanding Universe --- as the ``laboratory frame''
(we will call it the ``FRW laboratory frame'').\footnote{In terms
of a CMB decomposition in spherical harmonics, the monopole contribution
sets the temperature of $2.725(1)~{\rm K}$. The dipole term is
frame dependent; the requirement to make it vanish singles out
the maximally isotropic frame \cite{particledata}, 
which we could also refer to in this context.\label{isoCMB}}
We first consider the relativistic invariant
\bea
s &=& (E_{0} + \omega )^{2} - ( \vec p_{p} + \vec p_{\gamma} )^{2} \nn \\
  &=& E_{0}^{2} - \vec p_{p}^{\, 2} + 2 E_{0} \omega - 2
\vec p_{p} \cdot \vec p_{\gamma} 
\overbrace{\simeq}^{\rm head-on} 
m_{p}^{2} + 4 E_{0} \omega \excleq m_{\Delta}^{2} \ . 
\label{Deltacon}
\eea
$\vec p_{p}$ and $\vec p_{\gamma}$ are the 3-momenta of the 
proton and CMB photon. We arrived at the term $4 E_{0}\omega$
by assuming a head-on collision (and $E_{0} \gg m_{p}$). 
The last term refers to a $\Delta$ baryon at rest, 
which sets the energy threshold. 
Thus we obtain an expression for $E_{0}(\omega )$.
As an exceptionally high photon energy we insert
$\omega = 5 \langle \omega \rangle = 3 ~ {\rm meV}$ 
--- photons with even higher energy are 
very rare due to the exponential decay of the Planck
distribution (\ref{planckeq}). This leads to\footnote{In eq.\ 
(\ref{cutoff}) we increased this energy threshold slightly, which is 
consistent with the unlikelihood of exact head-on collisions.}
\be  \label{E0}
E_{0} = \frac{m_{\Delta}^{2} - m_{p}^{2}}{4 \omega}
\vert_{\omega = 5 \langle \omega \rangle} 
\simeq 5.3 \cdot 10^{19} ~ {\rm eV} \ .
\ee

Further kinematic transformations yield a
simple expression for the inelasticity factor $K$, which
represents the relative energy loss of the proton 
under photopion production \cite{Stecker},
\be  \label{Keq}
K := \frac{\Delta E_{p}}{E_{p}} = \frac{1}{2} \Big(
1 - \frac{m_{p}^{2} - m_{\pi}^{2}}{s} \Big) \ .
\ee
This relative loss still refers to the FRW laboratory frame,
but formula (\ref{Keq}) holds for general proton-photon scattering
angles.

Let us now change the perspective and consider the 
Mandelstam variable  $s$ in the rest frame of the proton,
where we denote the photon 4-momentum as 
$(\omega ' ,  \vec p{\, '}_{\!\!\! \gamma})$,
\be  \label{sprot}
s = (m_{p} + \omega ' )^{2} - \vec p{\, '}_{\!\!\! \gamma}^{\, 2} =
m_{p}^{2} + 2 m_{p} \, \omega' \ .
\ee
The photon energy  $\omega '$ is related to
$\omega$ by the Doppler effect,
\be
\omega ' = \gamma \omega \, ( 1 - v_{p} \cos \vartheta ) \ .
\ee
$v_{p}$ and $\vartheta$ are the proton velocity and the
scattering angle in the FRW laboratory frame. The Lorentz factor
$\gamma$ can take a remarkable magnitude;
for instance at the proton threshold energy $E_{p} = E_{0}$ it amounts to
\be  \label{gamma11}
\gamma = \frac{E_{p}}{m_{p}} 
\vert_{E_{p}=E_{0}} \approx 10^{11} \ .
\ee
By averaging $\omega '$ over the angle $\vartheta$ one arrives at 
\cite{Stecker}
\be  \label{omegavan}
\bar \omega \simeq 180 ~ {\rm MeV} \cdot \frac{E_{p}}{E_{0}} \ .
\ee
Hence a proton with $E_{p} \geq E_{0}$ perceives the CMB
photons as quite energetic $\gamma$-radiation.\footnote{The velocity
of the proton in the centre-of-mass frame is given by
$\omega ' / ( \omega ' + m_{p})$. At $E_{p} \approx E_{0}$
it is small on the relativistic scale, 
hence the proton rest frame referred to above
is not that far from the centre-of-mass frame.}

Combining eqs.\ (\ref{Keq}), (\ref{sprot}) and (\ref{omegavan})
we can determine the inelasticity factor at a given
proton energy (and averaged scattering angle),
\bea
K (\bar \omega ) &=&
\frac{1}{2} \Big( 1 - \frac{m_{p}^{2} - m_{\pi}^{2}}
{m_{p} (m_{p} + 2 \bar \omega )} \Big) \nn \\
&=& \left\{ \begin{array}{lr}
0.15 & \qquad {\rm at~~} E_{p} = E_{0} \ , \quad
\bar \omega = 180 ~ {\rm MeV} \ \ \\
0.20 & \quad \quad {\rm at~~} E_{p} = 2 E_{0} \ , \quad
\bar \omega = 300 ~ {\rm MeV} \ .
\end{array} \right. \quad \label{Kval}
\eea
We see that the inelasticity is significant already at the
energy threshold, and beyond it (gradually) rises further.

We are now prepared to tackle the question which ultimately matters
for the GZK cutoff: if a proton with $E_{p} > E_{0}$ travels
through the Universe, how long is the {\em decay time} of its energy~? \\
This question was analysed by F.W.\ Stecker \cite{Stecker}, 
who derived the following formula for the energy decay time $\tau$,
\be  \label{taueq}
\frac{1}{\tau ( E_{p})} = - \frac{k_{B}T}{2 \pi^{2} \gamma^{2}}
\int_{\bar \omega_{0}(E_{p})}^{\infty} d \bar \omega \,
\sigma (\bar \omega ) \, K (\bar \omega ) \, \bar \omega \, \ln
\Big( 1 - e^{-\bar \omega / (2 \gamma k_{B}T)} \Big) \ .
\ee
$\bar \omega_{0}$ is the photon threshold energy for
photopion production in the rest frame of a proton
(which has FRW laboratory energy $E_{p}$).
The product of $\bar \omega$ with the logarithm emerges 
from the Planck distribution (\ref{planckeq}) after integration. 
$\sigma (\bar \omega )$
is the total cross-section for the photopion production,\footnote{Strictly 
speaking it is cleaner to set the lower bound in
this integral to zero and rely on the
strong suppression below $\bar \omega_{0}$ due to $\sigma (\bar \omega )$, 
but the form (\ref{taueq}) is more intuitive.
Moreover eq.\ (\ref{taueq}) simplifies the energy attenuation
to a continuous process --- its discrete nature gives rise
to minor corrections.}
which was explored experimentally already in the 1950's,
so Refs.\ \cite{Grei,ZK,Stecker} could refer to it.
The corresponding experiment with protons at rest,
exposed to a $\gamma$-ray beam of about $200 ~{\rm MeV}$,
had been reported for instance in Ref.\ \cite{KKW}. The 
cross-section varies mildly with the energy, in the magnitude of 
$\sigma \approx 0.1 ~ {\rm mb}$. Its profile was reproduced
in Ref.\ \cite{Stecker}, along with the measured inelasticity
factor $K( \bar \omega )$. 
It is remarkable that this simple experiment at modest energy 
provides relevant information about the fate of UHECRs.
Based on the knowledge about $\sigma$, Figure 2 in
Ref.\ \cite{Stecker} displays the mean free path length for 
the proton, for instance
\be  \label{ellE0}
\ell^{\rm \, free}_{E_{p} \approx 10 E_{0}}
\approx 10 ~ {\rm Mpc} \ .
\ee
A later fit to the experimental
data, which refers directly to the product $\sigma (\bar \omega)
K (\bar \omega)$, is worked out in Ref.\ \cite{sigmaK}.
It includes corrections like 
the higher photopion production channels, which still reduce
the free path length a little.\footnote{The same
is true for the somewhat higher CMB temperature in scattering
processes long ago (although this effect is marginal).
Various CMB temperatures have been considered in Ref.\ \cite{ZK}.}

Refs.\ \cite{Grei,ZK} pointed out already that the 
corresponding {\em energy attenuation for heavier nuclei} is even 
stronger. The main reason here is the {\em photo-disintegration} into 
lighter nuclei due to interactions with CMB photons 
(see Ref.\ \cite{photodesint} for a detailed analysis).
The fragments carry lower energies; the energy per nucleon
remains approximately constant.
Hence we are in fact dealing with an absolute limitation for 
the energy of cosmic rays consisting of nucleons.\footnote{Regarding
non-nuclear rays, we stress that 
neutrinos are not subject to any theoretical energy limit
in the CMB, 
but no UHECR neutrinos have been observed yet.\label{neutrinofn}} 
Numerous works have later reconsidered
the attenuation of protons and heavier nuclei in the CMB in
detail, see {\it e.g.}\ Refs.\ \cite{RevGZK,HMR} and references therein.

\begin{figure}[h!]
\begin{center}
\includegraphics[angle=0,width=.6\linewidth]{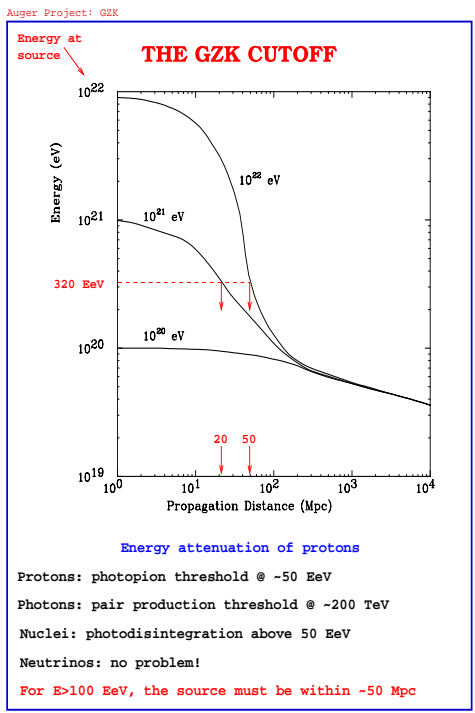} 
\vspace*{-1cm}
\end{center}
\caption{\it The energy attenuation for protonic UHECRs
due to photopion production with CMB photons (plot adapted from 
the Pierre Auger Collaboration \cite{PAO}). Note that $1 ~{\rm EeV} =
10^{18} ~{\rm eV}$ (see Appendix B).}
\label{GZK.PA}
\end{figure}
At last one might object that protons may still travel very long distances
with energies above $E_{\rm GZK}$ if they start at a much higher 
energy. However, at $E_{p} \gg E_{\rm GZK}$ the attenuation
is strongly intensified. (For instance the value of the lower integral 
bound $\bar \omega_{0}$ in eq.\ (\ref{taueq}) decreases; once the 
integral captures the peak in Figure \ref{planck}, many more
photons can contribute to the photopion production).
As a result, extremely high energies decrease rapidly, and
the final super-GZK path does not exceed the 
corresponding path length at $ E_{p} \approx E_{0}$
drastically. This feature is shown in a plot by the
Pierre Auger Collaboration, which we reproduce in Figure \ref{GZK.PA}.

It seems natural to assume a homogeneous distribution of
UHECR sources in the Universe, both for bottom-up and for
top-down scenarios (cf.\ Subsection 1.1). Then one could
expect a pile-up of primary particles just below
$E_{\rm GZK}$. As a correction one should take into 
account another process, which dominates 
somewhat below $E_{\rm GZK}$, and which
enables the CMB to reduce the proton energy further:
in the energy range $\, 5 \cdot 10^{17}
~ {\rm eV} < E < E_{\rm GZK}\,$ the 
{\em electron/positron pair production}
\be
p + \gamma \to p + e^{+} + e^{-} 
\ee
causes this effect, though the inelasticity is much smaller than in
the case of photopion production (consider the
ratio in eq.\ (\ref{Kval})).

\begin{figure}[h!]
\begin{center}
\hspace*{-2mm}
\includegraphics[angle=270,width=.5\linewidth]{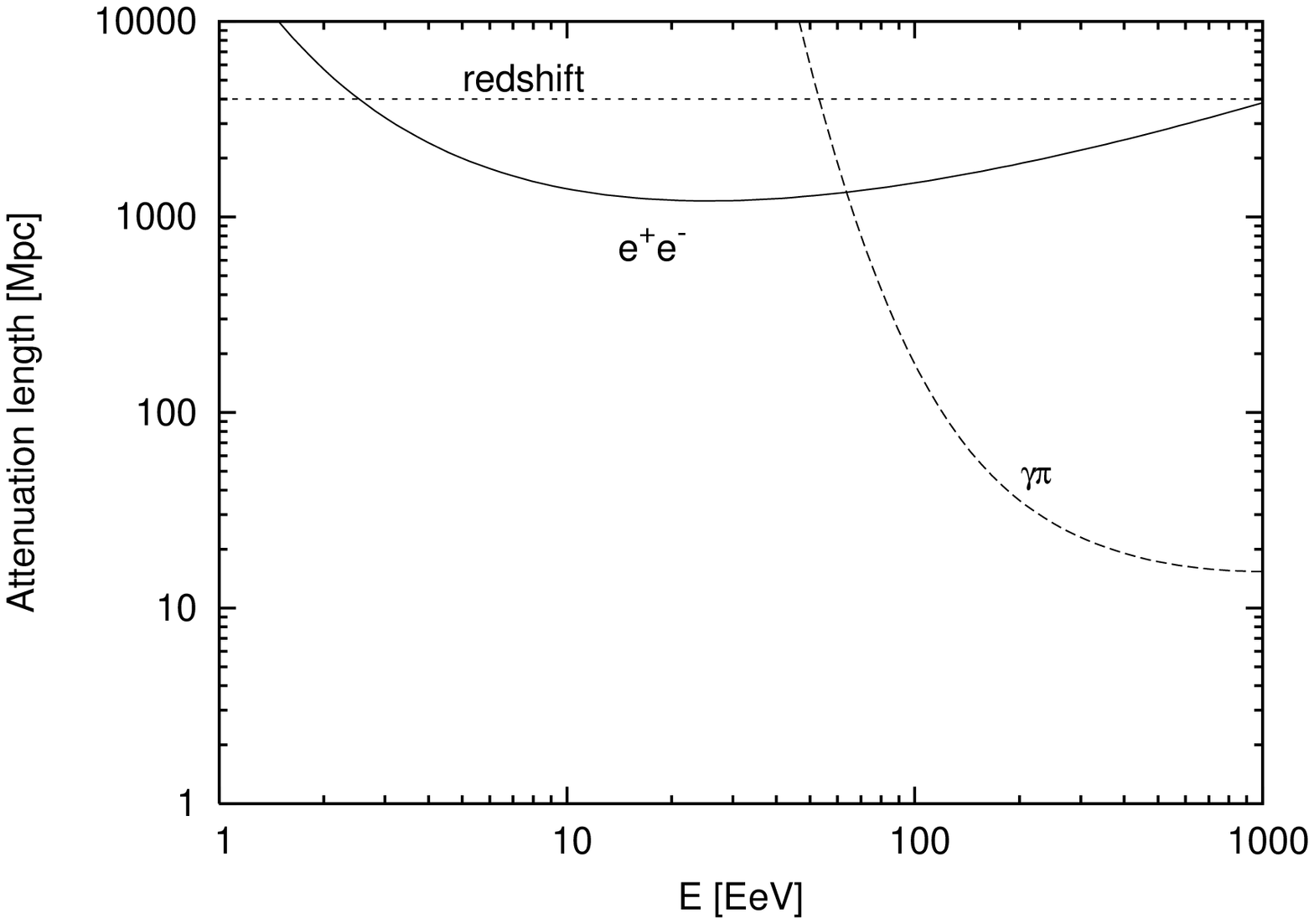}
\hspace*{-2mm}
\includegraphics[angle=270,width=.5\linewidth]{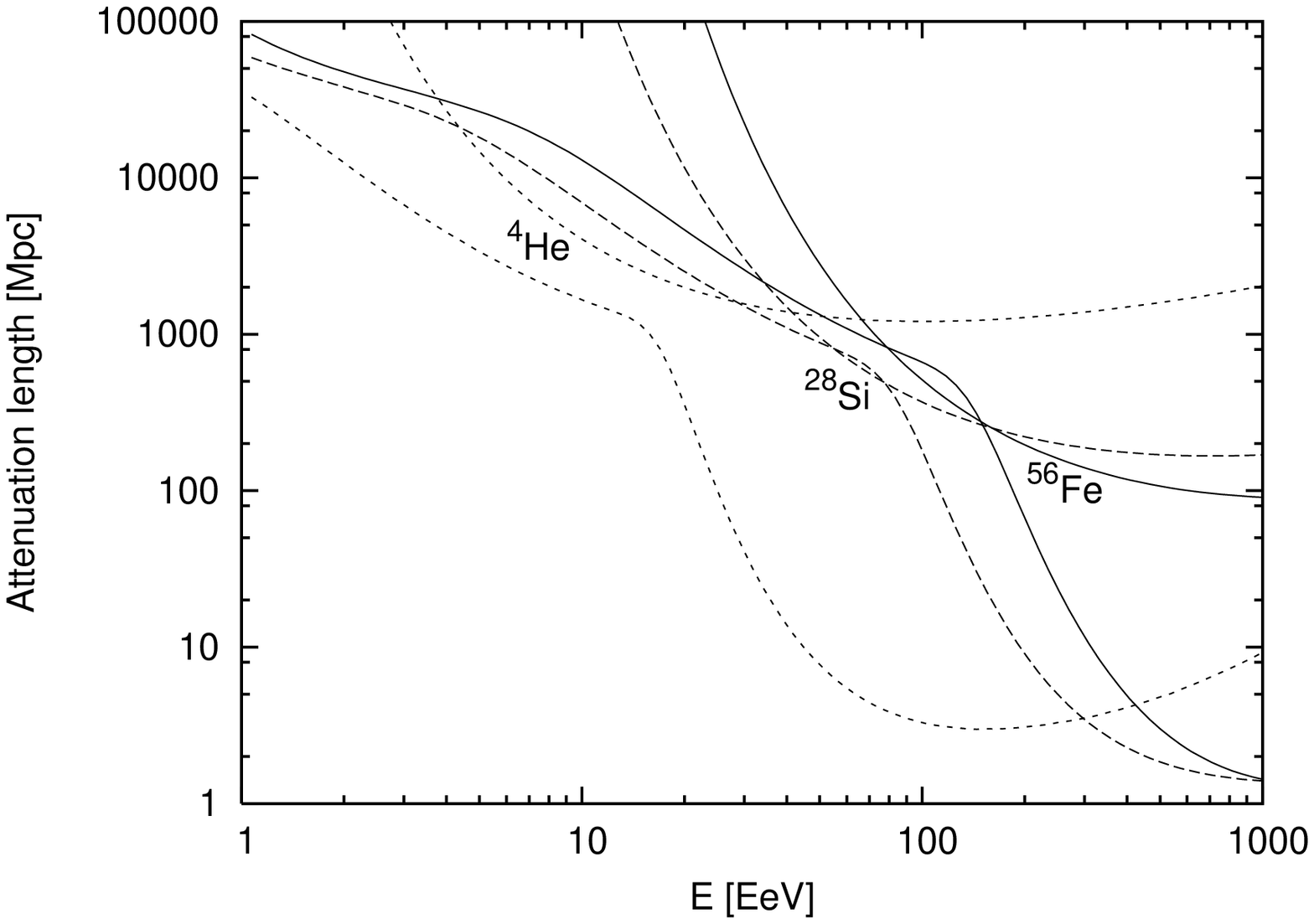} 
\end{center}
\caption{\it Illustration of the energy attenuation of UHECRs
due to photopion production with CMB photons for protons (on the left), 
and due to photo-disintegration of heavier nuclei (on the right).
In both cases also $e^{-} \, / \, e^{+}$ pair production is considered
(upper curves at high energy).
The plot on the left even captures the redshift, which is, however,
not really important at the ultra high energies that we are interested in.
These plots are adapted from Ref.\ \cite{HMR}.}
\label{attenplot}
\end{figure}

Ref.\ \cite{HMR} presents an updated overview, which incorporates
all relevant effects. This leads to Figure \ref{attenplot} 
for the {\em attenuation lengths} 
for various primary nuclei. 
For the proton it
is composed as $\ell^{-1} \simeq \ell^{-1}_{p \pi} + \ell^{-1}_{ee}$,
which refers to photopion production and $e^{-}\, / \, e^{+}$ pair 
creation (at even lower energy adding also the inverse 
attenuation length due to the redshift becomes significant).

All in all, one concludes the following:
if protons --- or heavier nuclei ---
hit our atmosphere with energies $E > E_{\rm GZK}$, they are
expected to come from a distance of maximally
\be  \label{ellmax}
\framebox[1.1\width]{ $ \displaystyle
\ell_{\rm max} \approx 100 ~ {\rm Mpc} $ } \ .
\ee
If we rise the ``cutoff'' to $10^{20}~{\rm eV}$ (cf.\ footnote \ref{GZKarb})
the range decreases to $\ell_{\rm max} \approx 50 ~ {\rm Mpc}$.
For instance this distance reaches out to the Virgo cluster of 
galaxies (its centre is about $20 ~ {\rm Mpc}$ from here). 
$\ell_{\rm max}$ is a large distance compared to our galaxy (or Milky 
Way) --- our galactic plane has a diameter of about $30 ~ {\rm kpc}$. 
But $\ell_{\rm max}$ is short
compared to the (co-moving) radius of the visible
Universe. 
On that scale the sources of super-GZK radiation should be {\em nearby},
\be
(R_{\rm Milky~Way} \simeq 15 ~ {\rm kpc}) \ll \ell_{\rm max} \ll
(R_{\rm visible~Universe} \simeq 14 ~ {\rm Gpc}) \ .
\ee


Despite the $e^{+} / e^{-}$ pair production,
some pile-up seems to occur at energies $E \lsim E_{\rm GZK}$
(cf.\ footnote \ref{Epeak}).
The status of the subsequent dip is discussed in Refs.\ \cite{ee-pairdip}.
As we mentioned in footnote \ref{knee2}, a ``second knee'' in the
cosmic flux was observed by various collaborations around
$10^{17.5}~{\rm eV}$. As a possible interpretation it could be the 
pile-up at the lower end of the  pair creation threshold \cite{GMT}.

However, in this review we want to focus on the issue of 
super-GZK energies in cosmic rays.
Not even in our vicinity --- given by the range (\ref{ellmax}) ---
we know of any acceleration mechanism, which could come near
such tremendous energies. Therefore the detection of
{\em super-GZK events} could represent a mystery --- and perhaps 
a hint for new physics. So let us now summarise the actual 
observations in this respect.

\subsection{Observations of super-GZK events}

Even before the GZK cutoff was established from the theoretical side,
one spectacular event had been reported by the Volcano Ranch
Observatory in the desert of New Mexico \cite{Linsley}. 
The energy of the cosmic particle that triggered this event was estimated 
at $10^{20}~{\rm eV}$, which exceeds already the GZK cutoff (\ref{cutoff}). 
Refs.\ \cite{Grei,ZK} both quoted this report and commented
that they do not expect any events at even higher energies (Greisen
added that he found even that event ``surprising'').
In fact, identifying the energy of the primary particle is
a delicate issue --- which we will address below --- and the
corresponding techniques were at an early stage. 

In any case this initiated a long-standing and 
controversial challenge to verify the validity of the GZK cutoff.
In 1971 yet another super-GZK event was 
reported from an air shower detected near Tokyo 
\cite{Suga}.\footnote{For that event the energy was estimated 
even as $4 \cdot 10^{21}~{\rm eV}$, but it is not often quoted in the
literature.}
This inspired a large-scale installation in Akeno (170 km west of
Tokyo), which is known as {\em AGASA} (Akeno Giant Air Shower Array);
a historic account is given in Ref.\ \cite{Sato}.

Meanwhile also the {\em Fly's Eye} project in Utah went into
operation, and in 1991 it reported an event with $3 \cdot
10^{20}~{\rm eV}$ primary particle energy \cite{Bird93}. Up to Ref.\ 
\cite{Suga}, this is the highest cosmic ray energy ever reported.
In macroscopic units it corresponds to 
$48~{\rm J}$.\footnote{It is popular to
compare this to the kinetic energy of a tennis balls. 
Its mass is typically $57.6 ~{\rm g}$, so that
these $48~{\rm J}$ are the kinetic energy for a speed of
$147 ~{\rm km/h}$, which is somewhat below the
speed attained in a professional game.}

\begin{figure}[h!]
\begin{center}
\vspace*{-0.5cm}
\includegraphics[angle=0,width=.5\linewidth]{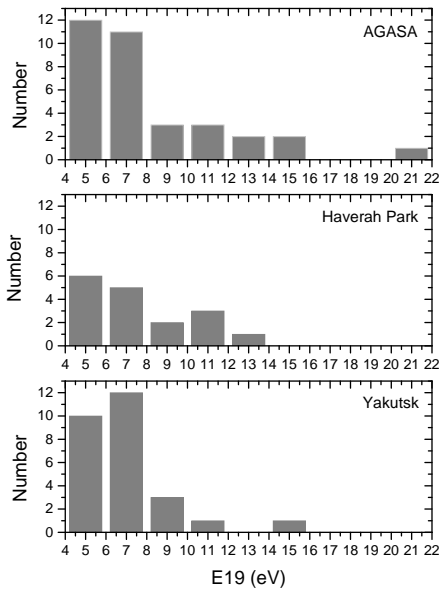} 
\end{center}
\vspace*{-1cm}
\caption{\it The statistics of GZK and super-GZK events 
reported by the observatories
AGASA, Haverah Park and Yakutsk, against the energy
in units of $10^{19}~{\rm eV}$ (Figure adapted 
from Ref.\ \cite{Niteroi}).}
\label{superGZKevents}
\end{figure}

Up to the 21$^{\rm st}$ century numerous new super-GZK events were
detected, in particular by AGASA. According to the data of this
collaboration, the spectrum --- {\it i.e.}\ the flux of
UHECR rays --- continues to follow the power-like decrease
in the energy that we saw in Figure \ref{fluxfig}, 
with hardly any extra suppression as $E_{\rm GZK}$
is exceeded. This contradiction to the theoretical expectation
triggered an avalanche of speculations. 
The AGASA results have been considered essentially consistent with 
the data of further air shower detectors in {\em Yakutsk} 
(Russia, at Lake Baikal) \cite{Yakutsk} and in {\em Haverah Park} 
(England, near Leeds) \cite{Haverah}. An overview of 
their super-GZK statistics is given in Figure \ref{superGZKevents}.

This observation {\em disagrees}, however, with the data of
the {\em HiRes} (High Resolution Fly's Eye) observatory, which 
concludes that the cosmic rays do respect the GZK
cutoff \cite{HiResPap}. HiRes is the successor experiment of 
Fly's Eye since 1994, with a refined technology.\footnote{Initially
the HiRes results seemed to agree as well,
but here and in the following we refer to the results presented
later by this Collaboration.}

\begin{figure}
\hspace{-4mm}
\includegraphics[angle=0,width=.55\linewidth]{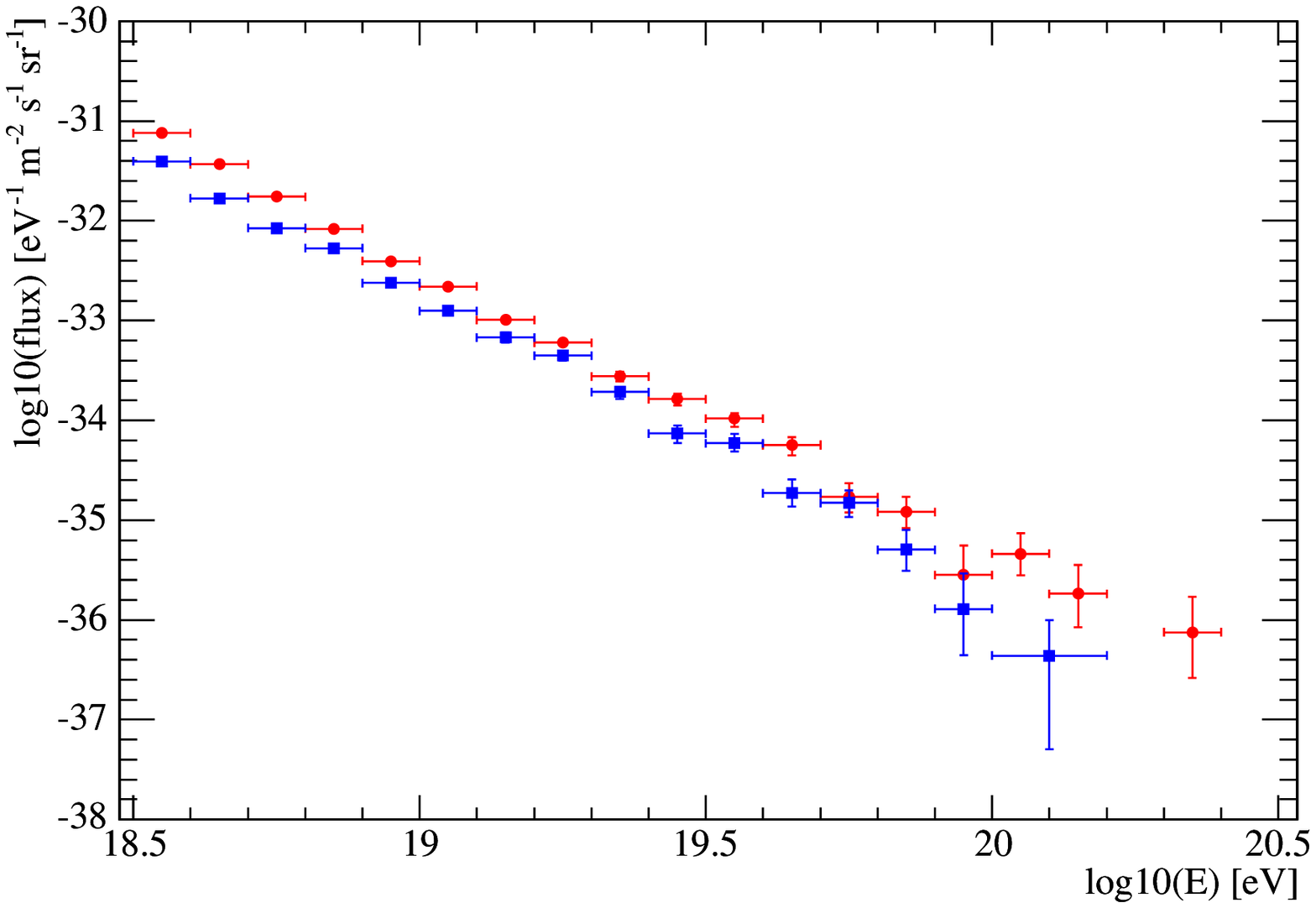} 
\hspace{-6mm}
\includegraphics[angle=0,width=.55\linewidth]{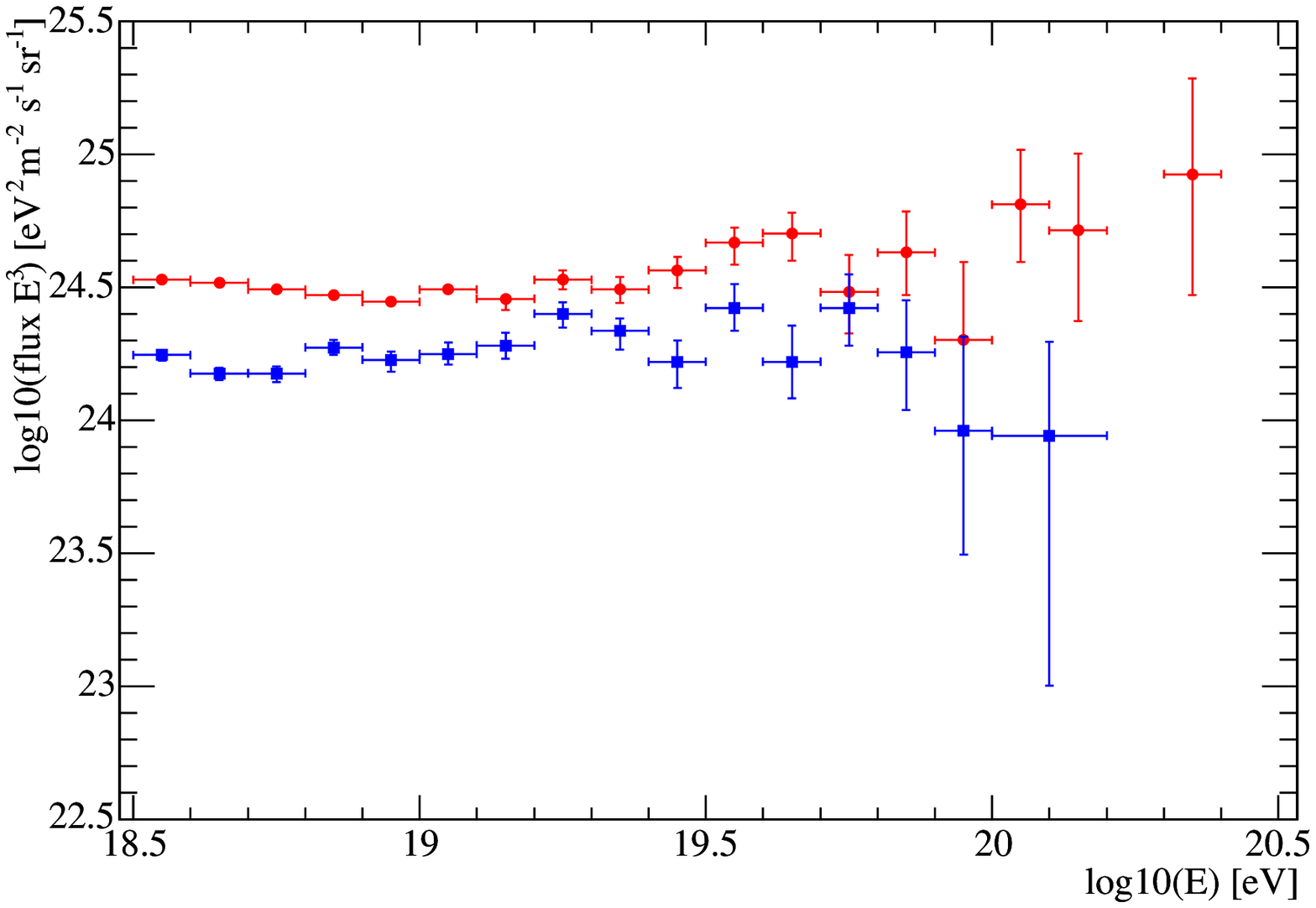}
\caption{\it A direct comparison of the cosmic flux at ultra high
energy as measured by the Observatories AGASA (upper symbols)
and HiRes (lower symbols). The plot
on the left shows the flux as a function of the energy $E$. On the
right it is rescaled by a factor $E^{3}$ (cf.\ Figure \ref{fluxfig}), 
which renders it approximately constant up to $E_{\rm GZK}$.
We see that the continuation beyond $E_{\rm GZK}$ is controversial
(plots adapted from Ref.\ \cite{DMBO06}).}
\label{AGASAvsHiRes}
\vspace*{-2mm}
\end{figure}

A comparison of the cosmic ray spectra as determined by
AGASA (1990 - 2004) and by HiRes (1997 - 2006)
is shown in Figure \ref{AGASAvsHiRes}.
Commented overviews over the results of these collaborations
are included in Ref.\ \cite{RevGZK,Kampert}.
Figure \ref{superGZKevents} also shows that 
the overall statistics of these UHECR observations is modest.
Note that the flux above $10^{12}~{\rm eV}$ is around
$10$ primary particles per minute and ${\rm m}^{2}$, 
but above $10^{18.5}~{\rm eV}$ it drops to
$O(1) ~ {\rm particle}/({\rm km}^{2} \, \, {\rm year})$ 
(as indicated in Figure \ref{fluxfig}), so the
search for UHECR takes patience.
One may question if there is really
a significant contradiction between HiRes and the other
groups mentioned above. In fact, an analysis in Ref.\ \cite{MBO03}
concludes that this discrepancy between AGASA and HiRes might also be 
explained by statistical fluctuations, if the energy scale is corrected 
in each case by $15 \, \%$ 
(which is well below the systematic uncertainty).

Nevertheless a large community
assumed that a contradiction is likely and wondered about
possible reasons. An obvious suspicion is that this may
actually be a discrepancy between {\em different methods.}

\begin{figure}
\vspace*{-2mm}
\begin{center}
\includegraphics[angle=0,width=1.\linewidth]{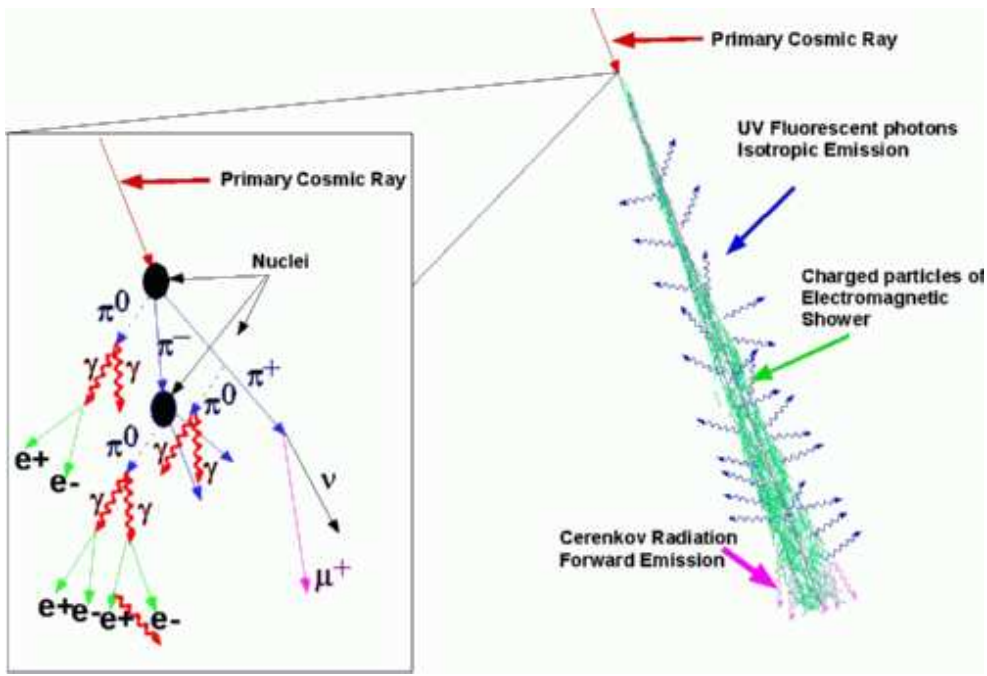}
\caption{\it The evolution of an air shower, as illustrated by the
HiRes Collaboration \cite{hires}. We recognise the fluorescence 
light (UV or blue), the \v Cerenkov radiation in the air,
and the production of light mesons, which rapidly decay
into leptons and photons (see zoom on the left).}  
\label{airshower}
\end{center}
\vspace*{-8mm}
\end{figure}

\begin{itemize}

\item To sketch these two methods, we first address
the air shower, which is illustrated in Figure \ref{airshower}.
At an early stage the huge energy transfer of
an UHECR primary particle onto the molecules in the atmosphere
generates --- among other effects --- a large number of light mesons
(pions, kaons $\dots$), which rapidly undergo leptonic
(or photonic) decays. Of specific interest for the observer are the
muons emerging in this way: thanks to the strong
time dilation they often survive the path of about
25 km to the surface of the Earth (although their mean
life time is only $2 \cdot 10^{-6}~{\rm s}$). Another
consequence of the extremely high speed is that the air
shower is confined to a narrow cone --- typically the
directions of motion of the secondary particles deviate by 
less than $1^{\circ}$ from the primary particle direction
in the shower maximum (a few hundred meters from the core).
The method applied by AGASA, Yakutsk and Haverah Park is
an array of numerous \v Cerenkov detectors --- spread over a
large area on ground --- which identify the
trajectories and energies of highly energetic secondary 
particles. They use tanks with pure water, where 
\v Cerenkov radiation is amplified by photomultipliers.
Detecting in particular a set of leptons $\mu^{\pm}, \, e^{\pm}$
(and photons) belonging to the
same shower --- with precise time of arrival, speed
and direction --- provides valuable information
about the shower. Its evolution in the atmosphere, and
ultimately the primary particle energy, are reconstructed
by means of sophisticated numerical methods. (Of course,
the information recorded on Earth is still insufficient
for a unique reconstruction, so that maximal likelihood
methods\footnote{A synopsis of this method is given in 
Ref.\ \cite{particledata}.}
have to be applied to trace back the probable scenario.)

\item The method used by HiRes --- and previously by Fly's Eye 
--- observes the showers 
in the atmosphere, before they arrive at Earth. The showers
excite $N_{2}$ molecules of the air, which subsequently 
emit UV or bluish light when decaying to their ground state
(with wave lengths $\lambda \approx (220 \dots 440) ~{\rm nm})$.
Although this light is weak, it is visible from Earth
to powerful telescopes, along with light collecting mirrors,
at least in nights with hardly any moonshine or clouds. 
(These telescopes are structured 
in a form similar to the compound eye of an insect.)
In 1976 Volcano Ranch achieved the first successful observation,
which was confirmed by the detection of the same air shower on ground.
This {\em air fluorescence light\,}\footnote{We adapt this term from
the literature, although it actually suggests that the primary 
particle must be a photon. The appropriate term for this effect
would be ``scintillation'', or more generally ``luminescence''.}
is emitted isotropically. 
It is also illustrated in Figure \ref{airshower},
along with the \v Cerenkov radiation in the air as a further
effect. The direct nature of this
observation is a valuable virtue, but an obvious
disadvantage is the limitation of the detection time
to $\approx 10 \, \%$. In this method, the observable
height grows with the energy of the primary particle,
which complicates the interpretation of the data \cite{Kampert}.

\end{itemize}

We add a couple of qualitative remarks, without going into 
details. An obvious question is how protons
as primary particles can be distinguished from heavier
nuclei. For the latter the air showers arrive at the maximal
number of particles at a higher point. The depth $D_{\rm max}$ in 
the atmosphere to that point can be parametrised as \cite{Matthews,Hoer}
\be  \label{Dmax}
[ D_{\rm max}({\rm proton}) - D_{\rm max}(A) ] \propto \ln A \ ,
\ee
where $A$ is the atomic mass number.
This point can approximately be identified 
by either of these two methods: for the fluorescence method this
observation is quite direct, and on ground it is characterised
by the electron/muon ratio. Further criteria --- such as the time
profile of the signal and the curvature of the shower front ---
are reviewed in detail in Refs.\ \cite{LaPlata},
and summarised in Ref.\ \cite{GMT}.

In particular the Fly's Eye record of $3 \cdot 10^{20}~{\rm eV}$ 
seemed to originate from a quite heavy primary nucleus, such as oxygen.
However, the identification is not easy at all, and in practice
the criteria are not always consistent. Only cosmic
$\gamma$-rays can be distinguished quite clearly from other
primary particles (they penetrate much deeper into the
atmosphere, hence the air shower maximum is closer to Earth).

In the course of a long flight through the CMB,
heavy nuclei are expected to 
break apart (photo-disintegration), as we mentioned 
before.
Therefore a dominance of protons in the UHECRs suggests that the
primary particles come from relatively far distances, and vice versa.\\

The status of UHECR observations to this point is reviewed in
detail in Refs.\ \cite{Rev2002,RevGZK}.
In light of the dilemma between the results obtained
with these two techniques, the {\em Pierre Auger Collaboration}
designed a new project in Argentina (near Malag\"{u}e, province of
Mendoza) with the goal to 
clarify the situation \cite{PAO}. Its planning started in 1992 and it 
is in stable operation since January 2004, while part of the
equipment has still been installed. The concept of
the Pierre Auger project is the combination of {\em both}
techniques described above:
\begin{itemize}
\item On Earth it involves 1600 detectors with 12 tonnes of
water, where three photomultipliers monitor the
\v Cerenkov light caused by air showers. 
The installation of these tanks has been terminated in 2008.  
They are distributed over an area of 3000 km$^{2}$ on a triangular
grid of spacing $1.5~{\rm km}$
(for comparison: air showers due to UHECRs have diameters of about 
$6~{\rm km}$ at the terrestrial surface).
\item In addition 24 telescopes are searching for fluorescence 
light from 4 well-separated sites
--- all of them have already been operating since 2004. 
\end{itemize}
The \v Cerenkov array is suitable for collecting large
statistics, while the telescopes are important for
a reliable calibration of the energy measurement.
This can be achieved best based on the data from air showers, 
which are observed in both ways, {\em hybrid events},
by evaluating a variety of correlations \cite{Roth}. With a single
detection technique the energy calibration has 
been notoriously problematic.

\begin{figure}
\begin{center}
\includegraphics[angle=0,width=.8\linewidth]{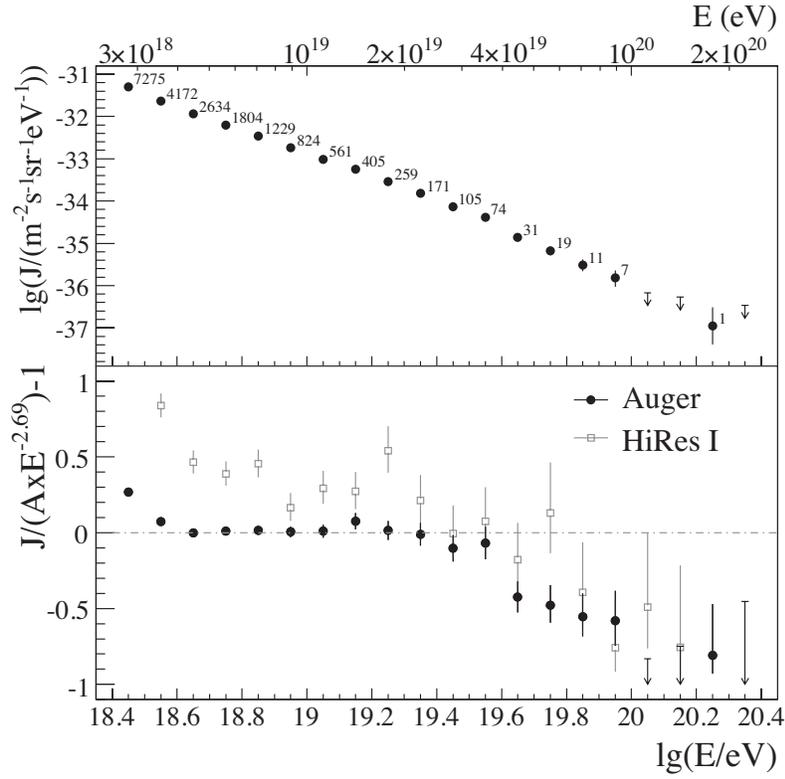}
\end{center}
\caption{\it The UHECR spectrum according to the data by the
Pierre Auger Collaboration collected from January 2004 until August 
2007 (plots adapted from Ref.\ \cite{PAflux}). The lower panel
reveals a clear reduction of the rescaled flux beyond $E_{\rm GZK}$, 
roughly consistent with the HiRes data and the predicted GZK cutoff. 
Nevertheless there is a considerable
number of new super-GZK events (the upper panel displays
the number of events).}
\label{PAflux}
\end{figure}

In Refs.\ \cite{Roth,Yamamoto,PAflux} the Pierre Auger Collaboration 
presented data collected until the middle of 2007, which we reproduce 
in Figure \ref{PAflux}. The exposure to this point already exceeds
the total exposure which was accumulated by HiRes and AGASA by
about a factor of 2 resp.\ 4.
The systematic uncertainty in the energy
measurement is estimated around $ 22 \, \%$ and the statistical
error around $6 \, \%$ (which is rather harmless in this context).
The vertical arrows in this plot mark the upper limits for the 
energy bins with $84 \, \%$ C.L. \cite{PAflux}, based on the
Feldman-Cousins method for Poisson distributions \cite{FeldCou}.
The small labels indicate the number of 
events detected in the corresponding bin, and the error bars 
represent the statistical uncertainty of the flux.
Below $E_{\rm GZK}$ (as given in eq.\ (\ref{cutoff}))
we recognise the power-like behaviour of the energy spectrum. 
In the intervals $4 \cdot 10^{18} \dots 4 \cdot 10^{19}~{\rm eV}$,
and beyond $4 \cdot 10^{19} ~{\rm eV}$
its slope was measured $\propto E^{-2.69(6)}$ resp.\
$\propto E^{-4.2(2)}$, which rules out a single power law
by 6 standard deviations. Hence for $E > E_{\rm GZK}$ the 
flux drops clearly below the extrapolated power law.
Nevertheless a considerable number of
new super-GZK events have been observed.\footnote{If one
rescales the flux with the factor $E^{3}$ --- as 
in Figure \ref{AGASAvsHiRes} --- the shape turns into a small
peak below $E_{\rm GZK}$ (which could be interpreted as a pile-up,
as we mentioned in Subsection 1.3), and the suppression beyond 
the threshold appears weaker.\label{Epeak}}
Confronted with the fluxes measured earlier by AGASA and HiRes,
these new data are closer to the latter.
They are certainly consistent with an extra damping once the energy
threshold for the $\Delta$ resonance is exceeded.
On the other hand, the sizable number of super-GZK
cosmic rays asks for an explanation and keeps the
door open for speculations. An interesting interpretation
of the UHECR arrival directions was published by the 
Pierre Auger Collaboration in November 2007, see Appendix A.

In this review we focus on Lorentz Invariance Violation (LIV)
as one attempt to explain a possible excess of
UHECRs compared to the theoretical prediction.
We should add, however, that the ``boring'' outcome
of full consistency with the GZK cutoff is by no means 
disproved --- recently that scenario has actually been boosted,
see again Appendix A.

%% file: LIV2.tex
So far we have assumed Lorentz Invariance (LI) to hold.
If fact, it played a key r\^{o}le in the derivation
of the GZK cutoff: the transformation of the scattering
process of an UHECR proton and a CMB photon to the rest frame 
of the proton established the link to the experimentally known
cross-section. Also the determination of the inelasticity 
factor for the proton under photopion production depends on
LI, since it refers again to the proton rest frame (regardless
whether one relies on theory, as in eq.\ (\ref{Kval}),
or on laboratory experiments). The same holds for the photo-disintegration
of heavier nuclei, and for 
electron/positron 
pair creation.

LI was therefore crucial for the analysis
of the energy attenuation of super-GZK cosmic rays.
Here we relied on Lorentz transformations with extreme
Lorentz factors of the magnitude 
\be
\gamma = \frac{1}{\sqrt{1 - v^{2}}} \, \gsim \, 10^{11} \ , 
\ee
see eq.\ (\ref{gamma11}) (we still set $c=1$).
There are no direct experimental tests if LI
still holds without any modifications for such extreme boosts;
accelerator experiments are limited to 
Lorentz factors $\gamma \lsim O(10^{5})$. 
If the validity of the GZK cutoff will ultimately be confirmed,
then this observation may be considered an indirect piece of
evidence for LI under tremendous boosts. 
If, on the other hand, the final analysis 
reveals a mysterious excess of super-GZK events, 
it might indicate new physics. 
In that case, one point 
to question about the standard picture is LI --- so let us 
take a closer look at it. \\

LI is a central characteristic of relativity:
\begin{itemize}

\item In {\em Special Relativity Theory} (SRT) LI holds
as a {\em global} symmetry. \\
The transformation rules were derived by H.A.\ Lorentz
(1899 and 1904), H.\ Poincar\'{e} declared them a Law of Nature,
A.\ Einstein provided a consistent physical picture
(1905) and H.\ Minkowski embedded it into the 
geometry of a 4d space (1907).  

\item In {\em General Relativity Theory} (GRT) LI holds as a {\em local
symmetry} (A.\ Einstein, around 1915). 
In each space-time point the frame can be chosen
so that locally the metrics takes the Minkowski 
form $g = {\rm diag} \, (1,-1,-1,-1)$,
which holds globally in SRT.\\
It is therefore tempting --- and possible --- to write GRT
in the terminology of a gauge theory (see for instance
Ref.\ \cite{Ramond}, Section 6.4). However, as a quantum 
field theory this formulation is not renormalisable: the renormalisation
group flow does not lead to a conformal field theory ({\it i.e.}\ to scale
invariance) at high energy, as reviewed recently in Ref.\ \cite{Shomer}. 
Therefore our established description of particle physics 
(local quantum field theory) is incompatible with GRT, 
which applies to very low energy and 
describes long-range gravity successfully.

\end{itemize}

Hence LI plays an essential r\^{o}le in both, SRT and GRT, although
this r\^{o}le is not the same. In this review we mostly address
high particle energies up to the maximum that has been observed,
hence SRT is in general the appropriate framework.
At specific points, however, we are also going to comment on conceivable
connections to GRT --- this becomes relevant in view of speculations
up to the Planck scale.\footnote{For approaches to quantum gravity, 
which discuss in particular the r\^{o}le of LI and its possible
violation, see {\it e.g.}\ Refs.\ \cite{SMEgrav,Alfaro2,SSBinSME,Alfaro1}.}

In general it is untypical for global symmetries to hold
exactly, so in SRT one could puzzle why this should be the
case for LI (and the related CPT invariance, 
see Subsection 2.1).\footnote{In this spirit, the Standard Model 
of particle physics
appears more natural in the now established form, which incorporates 
neutrino masses, and therefore only approximate chiral symmetry.}
On the other hand, gauge symmetries are exact, hence in this
respect the GRT picture appears helpful (as long as no violation
of LI is observed).

Nevertheless for our further discussion the framework of SRT is 
essential, since it allows us to apply particle physics as 
described by quantum field theory. In field theory, the manifestation
of LI as a global symmetry is well established. Some field $\Phi (x)$, which
may represent a scalar, a spinor, a 4-vector or some tensor, transforms
in a (finite dimensional) representation $D$ of the Lorentz group $SO(1,3)$,
\bea
\Phi (x) & \longrightarrow & 
U(\Lambda)^{\dagger} \Phi (x) U(\Lambda) =   
D(\Lambda ) \Phi (\Lambda^{-1}x) \nn \\
&& U ~ {\rm unitary} \ , \quad \Lambda \in SO(1,3) \ ,
\eea
see for instance Ref.\ \cite{Ramond}.
In particular scalar quantities --- like the Lagrangian of a
relativistic field theory --- are Lorentz invariant ($D(\Lambda ) =1$).

In contrast to the rotation group,
the Lorentz group is {\em non-compact.} The extrapolation of
LI to arbitrarily high energy leads to the generic UV divergences 
in quantum field theories. The formulation on non-commutative 
spaces --- to be addressed in Subsection 3.3 --- was originally
motivated as an attempt to weaken these UV divergences. Further
early suggestions to deviate from LI in quantum field theories
include Refs.\ \cite{LIVprehisto}.

\subsection{Link to the CPT Theorem}

According to the {\em CPT Theorem,} particle physics is invariant
if three discrete transformations are performed simultaneously:\\
C : flips particles into the corresponding anti-particles and vice versa.\\
\hspace*{6mm} This changes the signs of all couplings to gauge fields, 
hence C is \\ 
\hspace*{6mm} denoted as {\em charge conjugation}.\\
P : {\em parity}, space reflection, $\vec x \to - \vec x \,$. \\
T : inversion of the time direction, $t \to -t \,$.

\noindent
The CPT Theorem states that, assuming {\em locality and LI},
CPT invariance is inevitable. This was recognised first
by W.\ Pauli and G.\ L\"{u}ders \cite{GLud} and put on
a rigorous basis by R.\ Jost \cite{Jost}. \\
The basic idea is an analytic continuation of the Lorentz
group to imaginary time (Wick rotation), so one arrives at a reflection 
at the origin of 4d Euclidean space. 
This reflection captures P and T, and the Wick rotation 
implies that C is included as well. The combination amounts to an
anti-unitary symmetry transformation.
Applying the CPT transformation twice, one also
obtains a derivation of the Spin-Statistics Theorem.
(A historically and mathematically precise account is 
given in Ref.\ \cite{SpinCPT}).

Only recently O.W.\ Greenberg turned the consideration around
and proved the following Theorem \cite{Greenberg}:
%
{\em Any breaking of CPT invariance
necessarily entails a Lorentz Invariance Violation (LIV).}

This Theorem is very general, it applies whenever CPT invariance
breaks in any Wightman function.

On the other hand, if we only assume LIV, then CPT
symmetry may be broken or not, {\it i.e.}\
there are many ways to break LI while leaving CPT symmetry intact. 
The leading LIV terms are linear in the energy resp.\ momenta,
and these terms break CPT too; they are CPT odd.
If we only admit LIV terms starting quadratically in the energy, 
then CPT can be preserved \cite{ColGla}, so it is natural to assume
\be  \label{CPTprop}
{\rm LIV}\vert_{\rm CPT~odd} \propto E \ , \quad
{\rm LIV}\vert_{\rm CPT~even} \propto E^{2} \ .
\ee

\subsection{Status of experimental LI tests}

According to Greenberg's Theorem, LI tests can be performed 
indirectly by verifying CPT invariance. 
Experiments probe in particular the CPT prediction of identical
masses for particles and their anti-particles. 
This holds for instance for $e^{+}$, $e^{-}$
up to $8 \cdot 10^{-9}$, and
for $p$, $\bar p$ up to $ 10^{-8}$ (in both cases this is 
the bound for the relative deviation with $90 \, \%$ C.L.)
\cite{particledata}.
On-going tests investigate the strangeness driven oscillation
$K_{0} \leftrightarrow \bar K_{0}$
and conclude \cite{CPTtest}
\be
| M_{K_{0}} - \bar M_{K_{0}} | < 5.1 \cdot 10^{-10} ~ {\rm eV}
\qquad (95 \, \% ~ {\rm C.L.}) \ ,
\ee
which corresponds to a remarkable relative precision of $O(10^{-18})$.

Similar high precision LI tests via CPT measure the magnetic moments 
of elementary particles and anti-particles \cite{magmo}. 
A further indirect method to test LI in part deals with the
equivalence principle \cite{Fischbach}.
However, direct tests are even more powerful, and we will explain below 
why we are more interested in {\em CPT-conserving} LIV parameters.

The status of numerous experimental LI tests by means of a variety
of methods has been reviewed extensively in 
Refs.\ \cite{Matt,JLM,LIVboundsKos}.
%
%
Atomic physics gives access to excellent precisions
for specific LIV parameters. Such parameters describe, for instance,
a particle spin coupling to a conceivable
``tensor background field''. We will come back to these objects in 
the following Subsection. Here we anticipate that a large number of
such parameters can be introduced as additional couplings in the
Standard Model. In light of the discussion in Subsection 2.1,
we stress that these LIV terms are local, but they can be
even or odd under CPT transformation.
According to Ref.\ \cite{Matt2} they are all bounded by $O(10^{-27})$,
in a form which factors out the Planck scale (\ref{planckscale}).


These precisions are impressive, and all results are in agreement with 
LI --- {\em no LIV has ever been observed.} One may wonder if even higher 
precision is still of interest, or if it is already sufficient
for all phenomenological purposes. In the following we would 
like to argue that this is not necessarily the case.
The precision achieved so far may be insufficient with respect to
scenarios, which could drastically affect physics on the Planck
sale, or even the propagation of cosmic rays.

\begin{itemize}

\item We first address the CPT conserving class of LIVs, 
which exclude terms of $O(E)$ \cite{ColGla}, so we can assume the violation 
effect to be $ \propto E^{2}$, as in relation (\ref{CPTprop}). Let us
consider the case where LI is violated dramatically on the Planck
scale, where one expects both, gravity effects and quantum effects 
to be strong.
For instance the magnitude of (dimensionless) LIV effects 
at such energies could be
\be
{\rm LIV} \, (E = M_{\rm Planck}) = O(1) \ ,
\ee
where $M_{\rm Planck}$ is the Planck mass, {\it i.e.}\ the energy
scale set by the gravitational constant $G$,
\be  \label{planckscale}
M_{\rm Planck} = 1 / \sqrt{G} 
\simeq 1.2 \cdot 10^{28}~{\rm eV} \simeq 22 ~ \mu {\rm g} \ .
\ee
Accelerator experiments reach energies $E \leq O(10^{13})~{\rm eV}$, so
they could only be confronted with such LIV effects
$< O(10^{-30})$. It is therefore conceivable that a CPT conserving
LIV has been overlooked up to now, although it plays an essential
r\^{o}le on the Planck scale.

\item We move on to the case where LI {\em and} CPT
are violated. Then we expect LIV effects
$ \propto E$. Let us assume that even a surprisingly strong LIV
parameter of this kind, say with a magnitude of $O(10^{-23})$,
has been overlooked. It would be amplified to $O(10^{-8})$ on the 
Planck scale, so even there it would be of minor importance.

\end{itemize}

Therefore the {\em CPT conserving LIV terms} are more interesting. 
In addition to this estimate of magnitudes one might also argue that
breaking one fundamental law of standard quantum field theory 
--- which has been a principal pillar of physics since
the $20^{\rm th}$ century --- 
is a very daring step already. Hence it appears reasonable to study
this step in detail, rather than damaging yet another highly
established principle on top of it.

Cosmic rays represent indeed a unique opportunity to 
observe effects not too far below the Planck scale. The point
is not only that they carry the {\em highest particle energies} 
in the Universe, but --- even more importantly
--- their extremely {\em long flight,} which could accumulate 
tiny effects of new physics also at low or moderate energy (we will 
sketch examples in Subsection 3.4). If there is any hope to find 
experimental hints related to theoretical approaches like string 
theory\footnote{String theories start from a higher
dimension and first need a spontaneous LIV in that space to explain
why we experience only 4 extended dimensions. This requires a 
non-perturbative mechanism, which is not well understood.
One attempt to formulate string theory beyond perturbation theory
is the 10d IIB matrix model (or IKKT model) \cite{IKKT}; 
a possible spontaneous splitting into extended and compact
directions is discussed in Refs.\ \cite{10dIIB}, and for its
4d counterpart in Refs.\ \cite{4dIIB}. \\ As a different aspect,
LIV effects on the Casimir force in a 5d brane world are 
discussed in Ref.\ \cite{Obousy}.\label{IIB}}
or (loop) quantum gravity,\footnote{Conceivable loop quantum gravity 
effects on cosmic rays are discussed in Refs.\ \cite{AlfPal}.
So far we can restrict parameters in specific quantum gravity approaches
based on the bounds on LIV parameters \cite{UNAM}.}
then it is likely to involve cosmic rays (although also
LHC might have a chance \cite{Lust}).\footnote{We mention 
at the side-line that the scattering 
of UHECR primary particles on atmospheric molecules takes place
at higher energies in the centre-of-mass frame than any scattering
in accelerator experiments. In particular the proton-proton 
collisions at LHC (with 7 TeV per beam) will be equivalent 
in the centre-of-mass energy to the scattering of a $10^{17} ~ {\rm eV}$
proton on a fixed target. This happens about $2.5 \cdot 10^{5}$ times
per second in the terrestrial atmosphere, in addition to the cosmic
rays hitting other astronomical bodies, which obviously invalidates
popular concerns about the safety of LHC \cite{LHCsafty}. 
Here, however, we are interested
in interactions of UHECRs with CMB photons, which
involve modest centre-of-mass energies, see Subsection 1.3.}
These hypothetical theories tend to install new fields in the
vacuum, in addition to the Higgs field of the Standard Model.
We are now going to glimpse at the mechanism how such
new background fields could lead to spontaneous LIV.

\subsection{Standard Model Extension (SME)}

A systematic approach to add local LIV terms to the
Standard Model has been worked out by A.\ Kosteleck\'y and
collaborators starting 10 years ago \cite{ColKos}.
This collaboration provides a kind of encyclopedia of such terms,
which is reviewed --- along with its experimental bounds ---
in Ref.\ \cite{Bluhm}.
The original motivation emerged from string theory \cite{KosSam}, 
which is, however, not needed for the resulting {\em Standard
Model Extension} (SME).

We take a look at a prototype of a LIV term in this extension. 
Consider (in a short-hand notation) the Lagrangian
of some fermion field $\psi (x)$, which may represent a
quark or a lepton,
\be  \label{LSME}
{\cal L} = {\rm i} \bar \psi \gamma_{\mu} \partial^{\mu} \psi
- g \bar \psi \phi \psi - {\rm i} g' G_{\mu \nu} \bar \psi
\gamma^{\mu} \partial^{\nu} \psi + \dots 
\ee
We start with the free kinetic term, followed by 
the Yukawa coupling $g$ to the Higgs field $\phi$.
The Standard Model assumes $\phi$ to undergo
Spontaneous Symmetry Breaking (SSB), so that one of its
components takes a non-vanishing expectation value, 
say $\langle \phi_{0}\rangle > 0$. This leads to the fermion 
mass $g \langle \phi_{0} \rangle$. A mass term cannot be
written into the Lagrangian directly if we want to
keep the freedom to couple the left- and the right-handed
chirality components of the fermion field independently
to a gauge field; this is the situation in the
electroweak sector of the Standard Model.\footnote{On the other
hand, a fermion mass term $ - m \bar \psi \psi$ is allowed
in a vector theory like QCD, where both chirality components
of a quark couple to the gluons in the same manner.}

Let us now consider the third term in the Lagrangian (\ref{LSME}),
which is one of the possible extensions beyond the
Standard Model. The Yukawa-type coupling $g' \in \R$ is
another free parameter, which couples the fermion
to a new background field of the Higgs-type, albeit with
a tensor structure. Again we assume SSB, which could, 
for instance, lead to
\be  \label{G00}
\langle G_{00} \rangle > 0 \ , \quad 
\langle G_{\mu \nu} \rangle = 0 \quad {\rm (otherwise)~.}
\ee
This additional term will obviously distort the fermion
dispersion relation compared to the Standard Model, which
amounts to a spontaneous LIV.

In contrast to the Higgs field we are now confronted with the
question if new background fields --- which may have a vector
or a tensor structure --- are Lorentz transformed as well.
For a simple change of the observer's inertial frame, all fields ---
including the background fields --- are transformed, and ${\cal L}$
remains invariant. However, the question if particles, which are
described by fields like $\psi$, really perceive LI depends on a 
transformation of these fields only in a
constant background \cite{Kosphilo,Matt}.
Kosteleck\'y {\it et al.} denote this as an
``active'' (or ``particle'') Lorentz transformation, and in this
respect LI may break {\em spontaneously}, 
as in eq.\ (\ref{G00}).\footnote{Here one 
re-introduces a distinction, which one had hoped to
overcome since the historic work by Einstein.}

In this way we can construct a non-standard dispersion relation
for any particle, depending on its coupling to the tensor
field $G_{\mu \nu}$ or further new background fields. 
The inclusion of gauge interactions
proceeds in the familiar way (one promotes some global
symmetries to local ones by means of covariant derivatives).

Of course the LIV parameter $g'$ is just one example ---
the mass term and kinetic term in the Lagrangian (\ref{LSME})
could both be extended by including a background field
for each element of the Clifford algebra (further terms 
are written down in Section 3.5).
Another example, which was considered earlier \cite{CFJ},
is the addition of an extra gauge term of the
Chern-Simons type to modify QED,\footnote{In non-Abelian
gauge theory, LIV through Chern-Simons-type terms is studied
in Refs.\ \cite{Petrov2}.}
\be  \label{CSterm}
{\cal L}_{\rm CS} = - \frac{1}{4} \epsilon^{\mu \nu \rho \sigma}
p_{\mu} A_{\nu} F_{\rho \sigma} 
= \frac{\mu}{2} \vec B \cdot \vec A \ , \quad
( p_{\mu} p^{\mu} = \mu^{2} >0) \ .
\ee
Here $p_{\mu}$ is a vector of dimension mass (and 
$\vec B = \nabla \times \vec A$). Each
component has vanishing covariant derivatives in all 
frames. This vector introduces a ``preferred direction'', 
where also all the ordinary derivatives $\partial_{\nu}p_{\mu}$ 
vanish --- Ref.\ \cite{CFJ} associates it with the preferred frame in 
a galaxy. It identifies an astrophysical bound of
$\mu < O(10^{-33}) ~ {\rm eV}$ based on radio galaxies (the method
will be sketched in Subsection 3.5).
Recently there have been attempts to achieve even higher precision
for this bound based on the rotation of CMB photon polarisation 
vectors \cite{IHEP}.

Note that ${\cal L}_{\rm CS}$ breaks also CPT invariance
spontaneously. Examples for CPT conserving LIV terms in the
fermionic and gauge part of the extended Lagrangian are
the LIV term in eq.\ (\ref{LSME}), or
$K_{\mu \nu \rho \sigma} F^{\mu \nu} F^{\rho \sigma}$
(where $K$ is another background tensor field).
Kosteleck\'y {\it et al.}\ have identified more than 100 LIV
parameters in this way, including CPT breaking terms \cite{ColKosCPT}.
The resulting model preserves a number of properties of the
Standard Model, like energy and momentum conservation, gauge
invariance and locality. For massive fermions at low energy, and
small LIV parameters, also energy positivity and causality are safe
\cite{stabcaus}. 
In the framework of SRT the photon is identified with
the Nambu-Goldstone boson of the SSB of Lorentz symmetry.
That collaboration also discussed various scenarios for an 
interpretation in the GRT framework \cite{ColKos,SSBinSME,Bluhm}.

In this parametrisation, the current status is summarised
in Ref.\ \cite{Matt2} as follows: all (dimensionless) LIV coefficients
are $< 10^{-27}$ (as we mentioned before). 
For those which break CPT the upper bound is as tiny as
$10^{-46}$, which further motivates our focus on CPT even terms. 
A detailed overview of the various bounds is given in 
Ref.\ \cite{LIVboundsKos}.

An obvious question is why LIV should become manifest only
at huge energies, although SSB is typically a low energy
phenomenon. In principle even a LIV detected a low energy
would not necessarily imply LIV at high energies. 
Of course we are interested in the opposite
situation, where LIV is significant only at high energies, so we
have to assume the SSB to persist over all energies that we 
consider. This setting requires {\em tiny} LIV parameters
multiplying momenta of the fields (as in the examples above). 
Then the deviation from LI is only visible at huge momenta
and a contradiction to known phenomenology can be avoided.

Although consistent, this scenario implies a severe {\em fine tuning
problem.} On tree level one might hope for a somehow natural
suppression of the LIV parameters by a factor
$m_{\rm ew} / M_{\rm Planck} \sim 10^{-17}$ (where $m_{\rm ew}$ is the
electroweak mass scale) \cite{KosPot}. If one includes
1-loop radiative corrections, however, this effect is lost and the 
magnitude of LIV effects is just given by $O(g_{\rm ew}^{2})$ 
($g_{\rm ew}$ being an 
electroweak Yukawa coupling) \cite{LIVfinetune}, which is far from 
the suppression needed. If we consider LIV nevertheless, we have 
to add this problem to the list of unsolved
hierarchy problems, like the small vacuum angle $\theta$ in QCD,
the small particle masses (on Planck scale) or the small cosmological 
constant. On the other hand, the Standard Models of particle physics
and cosmology are well established, despite
the presence of these unsolved problems.

\subsection{Maximal Attainable Velocities (MAVs)}

Within the same framework we now switch to the pragmatic
perspective of Ref.\ \cite{ColGla}. We saw in the
previous Subsection that LIV can be arranged for. 
Now we assume that this mechanism has been at work,
and we proceed to an {\em effective Lagrangian,} which includes 
explicit LIV terms. Such an effective Lagrangian may also
capture hadrons as composite particles 
(like Chiral Perturbation Theory).

Following Ref.\ \cite{ColGla} we leave, however,
other Standard Model properties intact as far as possible.
We therefore require CPT and gauge invariance to persist,
and $SO(3)$ rotation symmetry to hold in a ``preferred 
frame''.\footnote{In contrast to Ref.\ \cite{CFJ}, the following
considerations on cosmic rays can hardly employ the
preferred frame in a galaxy, because the primary particle
paths extend over much larger distances. One might refer
to the frame which renders the CMB maximally isotropic,
cf.\ footnote \ref{isoCMB}.}
Moreover we do not break power counting renormalisability, {\it i.e.}
we only add terms of mass dimension $\leq 4$.

Let us outline which terms are admitted by these conditions.
For a bosonic field $\vec \Phi$ we can add a purely spatial
kinetic term
\be
{\cal L}_{\rm eff} = 
\dots + \frac{1}{2} \sum_{i=1}^{3} \sum_{a,b} (\partial_{i}
\phi^{a}) \varepsilon_{ab} (\partial_{i} \phi^{b}) \ , \quad
(\varepsilon_{ab} = \varepsilon_{ba}) \ .
\ee
(Equivalently we could add a term with only time derivatives
instead). For a fermion field we insert a
spatial kinetic term as well, where the LIV parameters can
be distinct for positive or negative chirality,
\be  \label{LIVchiral}
{\cal L}_{\rm eff} = 
\dots + {\rm i} \bar \psi \, \vec \gamma \cdot \vec \partial 
\ [ \, \varepsilon_{+} (1 + \gamma_{5}) + \varepsilon_{-}
(1 - \gamma_{5} ) \, ] \ \psi \ .
\ee
Gauge interactions require covariant derivatives also in these
non-standard terms. For the pure gauge part we first consider the
Abelian gauge group $U(1)$, where the electric and the magnetic
field are given by $E^{i} = -F^{0i}$, $B^{i} = \frac{1}{2}
\epsilon^{ijk} F_{jk}$. The above conditions allow for
three independent terms,\footnote{Without insisting on CPT
invariance, the Chern-Simons term (\ref{CSterm}) would be added
to this list.}
\be
\vec E^{2} - \vec B^{2} \ , \quad \vec E \cdot \vec B \ , 
\quad \vec B^{\, 2} \ .
\ee
The first two among these terms are LI, so we work with
the third one. Also in Yang-Mills gauge theories Ref.\ \cite{ColGla} 
uses the corresponding term
\be  \label{Bsquare}
{\cal L}_{\rm eff} = \dots + g'' \sum_{a} \vec B^{a} \vec B^{a} \ , 
\ee
where the index $a$ runs over the generators of the Abelian
factors in the gauge group.

In this way one obtains a quasi-Standard Model with 46 LIV
parameters. In view of the above list the large number may
come as a surprise at first sight, but it can be understood
by the numerous new fermion generation mixings (for one case,
the phenomenological impact on neutrino oscillation will
be discussed in Subsection 2.5.4). 
Coleman and Glashow verified that all these parameters
preserve the vanishing gauge anomaly,
so that the model remains gauge invariant on quantum level. \\

We proceed to another prototype illustration and
consider a neutral scalar field \cite{ColGla}. In some 
background the renormalised propagator $G$ is written as
\be
- {\rm i} G^{-1} = (p^{2} - m_{0}^{2}) f(p^{2}) +
\varepsilon \, \vec p^{\, 2} g(p^{2}) \ .
\ee
Geometrically we assume the usual Minkowski space with $c=1$,
$p^{2} = E^{2} - \vec p^{\, 2}$, and $m_{0}$ is the renormalised
mass at $\varepsilon =0$. The functions $f$ and $g$ are not
specified, but $f$ obeys the conventional normalisation, and
$g$ is adapted to it, $f (m_{0}^{2}) =  g (m_{0}^{2}) =1$
(both functions are smooth in this point).

Let us now turn on a tiny LIV parameter $\varepsilon$ and consider its 
leading order. The poles in the propagator are shifted to 
\bea
E^{2} & \simeq & \vec p^{\, 2} + m_{0}^{2} - \varepsilon \vec p^{\, 2}
\simeq \vec p^{\, 2} c_{\rm P}^{2} + m^{2} c_{\rm P}^{4} \nn \\
{\rm with} && m = \frac{m_{0}}{1 + \varepsilon} \ , \quad
c_{\rm P} = 1 - \varepsilon \qquad ({\rm to ~} O(\varepsilon )) \ .
\label{MAV}
\eea
The renormalised mass is modified, $m_{0} \to m$. However, we are more 
interested in the feature of the dispersion relation (\ref{MAV}):
the parameter $c_{\rm P}$ takes the r\^{o}le, which is
usually assigned to $c$. From the group velocity
\be
\frac{\partial E}{\partial | \vec p |} = \frac{| \vec p \, |}
{\sqrt{ \vec p^{\, 2} + m^{2} c_{\rm P}^{2} }} \ c_{\rm P}
\ee
it is obvious that this is the speed that the particle approaches
asymptotically at large $| \vec p \, |$.

Following this pattern, each particle type P can pick up its
own {\em Maximal Attainable Velocity (MAV)} $c_{\rm P}$; it might
slightly deviate from the speed $c$, which establishes the Minkowski
metrics. A tiny value of the dimensionless parameter $\varepsilon$
justifies the above linear approximations, and it is compatible with
the desired setting where LIVs only become noticeable at huge momenta.
We mentioned before that this imposes a new hierarchy problem (which
we have to live with) and that we are referring here to an effective 
approach, which captures composite particles, unlike Subsection 2.3.

\subsection{Applications of distinct MAVs}

Let us now address some applications that we are led to
if we assume certain particles to posses
MAVs distinct from $c$. We still follow Ref.\ \cite{ColGla}
and focus on applications of potential interest in cosmic
ray propagation.\footnote{It has also been suggested to test
MAVs in air showers \cite{MAVtestinEAS}.}

\subsubsection{Decay at ultra high energy}

We consider the possibility that some particle with
index $0$ decays into a set of particles with indices 
$a = 1,2 \dots$. At ultra high energy all masses are
negligible, but we assume individual MAVs for the
particles involved, $c_{0},\ c_{a}$. We further define
$c_{\rm min}$ as the minimal MAV among the decay products.
Then the energetic decay condition reads
\bea
|\vec p^{\, (0)}| \, c_{0} & = & \sum_{a} 
|\vec p^{\, (a)}| \, c_{a} \geq c_{\rm min} 
\sum_{a} |\vec p^{\, (a)}| \geq c_{\rm min} \, |\vec p^{\, (0)}| \nn \\
\Rightarrow && c_{0} \geq c_{\rm min} 
:= \ ^{\rm min}_{~a} \ c_{a} \ .  \label{decaycon}
\eea
The interesting observation is that tiny differences in the
MAVs can be arranged such that a usual decay cannot take place 
anymore if the primary particle carries ultra high energy.
This happens when the condition (\ref{decaycon}) is violated,
{\it i.e.}\ if $c_{0}$ is smaller than any of the $c_{a}$.

On the other hand, new decays could be allowed at high energy,
such as the photon decay $\gamma \to e^{+} + e^{-}$, 
see Subsections 2.5.2 and 3.5, or the radiative muon decay
$\mu^{-} \to e^{-} + \gamma$. A further example is the inverse 
$\beta$-decay $p \to n + e^{+} + \nu_{e}$, which tightens the 
bounds on specific LIV parameters \cite{Altschulbeta}
based on the new results for UHECR that we review in Appendix A.

\subsubsection{Vacuum \v Cerenkov radiation}

Next we consider a particle with a MAV that slightly exceeds the
speed of light, {\it i.e.}\ it exceeds the MAV of the photon in vacuum,
\be  \label{espdef}
\frac{c_{\rm P}}{c_{\gamma}} = 1 + \varepsilon > 1 \ .
\ee
This particle can be accelerated to a speed $v > c_{\gamma}$.
Then it will emit {\em vacuum \v Cerenkov radiation},
so it slows down intensively.
The energy required for vacuum \v Cerenkov radiation has to fulfil
\be
E > \frac{m}{\sqrt{1 - (c_{\gamma}/c_{\rm P})^{2}}} \simeq
\frac{m}{\sqrt{2 \varepsilon}} \ ,  \label{vacC}
\ee
where $m$ is the particle mass, and we consider the leading order
in $\varepsilon$. \\

As we discussed in Section 1, we know from cosmic rays
that {\em protons} support an energy up to $O(10^{20})~{\rm eV}$
over a considerable path, hence this energy does not fulfil
inequality (\ref{vacC}). Coleman and Glashow infer
\be  \label{epsprot}
\varepsilon_{p} < \frac{m_{p}^{2}}{2 E^{2}} 
\vert_{E = 10^{20} \, {\rm eV}} \approx 5 \cdot 10^{-23} \ ,
\ee
where $m_{p}$ and $\varepsilon_{p}$ are the mass and the 
$\varepsilon$-parameter (see eq.\ (\ref{espdef})) of the proton.

Cosmic {\em electrons and positrons} are observed only up to 
$E \approx 1 ~{\rm TeV}$, hence the \v Cerenkov constraint on 
their $\varepsilon$-parameter is much less stringent,
$ \varepsilon_{e} < 10^{-13} \, .$
On the other hand, the absence of the photon decay $\gamma \to e^{-}
+ e^{+}$ constrains $\varepsilon_{e}$ from below \cite{ColGlaPLB}.
In total one arrives at
\be  \label{epsele}
- 10^{-5} < \Big( \frac{c_{e}}{c_{\gamma}} -1 \Big) < 10^{-13} \ .
\ee
%

\subsubsection{Impact on the GZK cutoff}

\begin{figure}[h!]
\begin{center}
\includegraphics[angle=270,width=.55\linewidth]{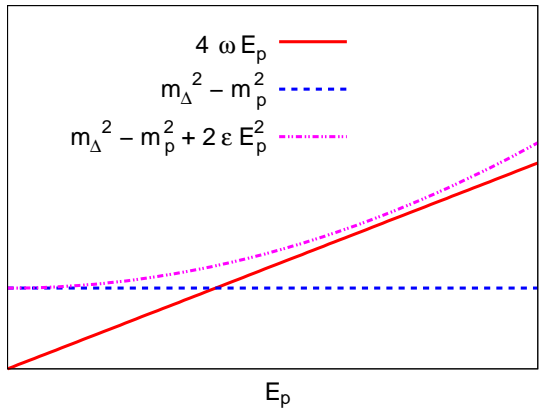} 
\vspace*{-2mm}
\end{center}
\caption{\it An illustration of the threshold for 
the term $ 4 \omega E_{p}$ to enable 
photopion production through the $\Delta (1232)$ resonance:
in the standard situation (all MAVs are $1$) this threshold
is constant, $m_{\Delta}^{2} - m_{p}^{2} 
\simeq (807 ~ {\rm MeV})^{2}$. For $\varepsilon > 0$ the threshold
squared picks up a contribution $\propto E^{2}_{p}$, 
see eq.\ (\ref{epsGZK}), which requires
a higher proton energy, and $\varepsilon > \varepsilon_{c}$ 
(illustrated by the dashed-dotted line)
closes this channel for photopion production.}
\label{noGZK}
\end{figure}

In this generalised framework we now reconsider the head-on
collision $ p + \gamma \to \Delta (1232)$ of relation 
(\ref{photopion}). For the MAVs we could check all kinds of
scenarios --- they are all conceivable in the effective
ansatz of Subsection 2.4.
Here we pick out the particularly interesting option that only 
the proton has a slightly slower MAV,
\be
c_{\gamma} = c_{\Delta} = 1 \ , \quad c_{p} = 1 - \varepsilon < 1 \ ,
\ee
(in the notation of eq.\ (\ref{epsprot}): 
$\varepsilon = - \varepsilon_{p}$). The condition for a $\Delta$ 
resonance in the scattering on a CMB photon was written down
before in eq.\ (\ref{Deltacon}). The following requirement still holds,
\be
m_{\Delta}^{2} \leq (E_{p} + \omega )^{2} - (p - \omega )^{2} 
\simeq E_{p}^{2} - p^{2} + 2 \omega (E_{p} + p) \ ,
\ee
where $E_{p}, \, p = | \vec p \, |$ are the proton energy and 
momentum, and $\omega$ is the photon energy, all in the FRW laboratory 
frame. The evaluation of the right-hand-side is now modified due to
the non-standard dispersion relation of the proton,
\be
E_{p}^{2} - p^{2} (1 - \varepsilon )^{2} = m_{p}^{2} (1 - \varepsilon )^{4} 
\quad \to \quad E_{p}^{2} - p^{2} \simeq m_{p}^{2} - 
2 \varepsilon E_{p}^{2} \ ,
\ee
where we assumed $E_{p} \gg m_{p}$ and $|\varepsilon | \ll 1$.
If we additionally make use of $E_{p} \gg \omega$, we arrive at
a new energy threshold condition, where one term 
is added compared to eq.\ (\ref{Deltacon}),
\be  \label{epsGZK}
m_{\Delta}^{2} - m_{p}^{2} + {\underline{2 \varepsilon E_{p}^{2}}} 
\leq 4 \omega E_{p} \ .
\ee
Once more, the extra term is suppressed by the factor $\varepsilon$,
but it can become powerful at very large energies. 
The consequence of this modification is illustrated in Figure
\ref{noGZK}: the right-hand-side increases linearly with the
proton energy $E_{p}$ and reaches the standard threshold
(for $\varepsilon =0$) at the value $E_{0}$ given in eq.\ (\ref{E0}).
A small $\varepsilon > 0$ leads to an increased threshold energy,
and above a critical value $\varepsilon_{c}$ the $\Delta$ resonance
is avoided, 
\be  \label{epsc1}
\varepsilon_{c} = \frac{\omega}{2 E_{0}} \simeq
\frac{2 \omega^{2}}{m_{\Delta}^{2} - m_{p}^{2}} \vert_{\omega =
5 \, \langle \omega \rangle = 
3 \, {\rm meV}} \simeq 3 \cdot 10^{-23} \ .
\ee
Hence the critical parameter $\varepsilon_{c}$ is tiny indeed,
and this is all it takes to remove the GZK cutoff, at least in 
view of the dominant channel of photopion production.\footnote{The
idea that LIV could invalidate the GZK cutoff was first
expressed in Refs.\ \cite{SatoTati}. Later it has also been 
discussed in the SME framework \cite{Mofa}.}
(This is of course compatible with the condition (\ref{epsprot}),
which only requires $\varepsilon > - 5 \cdot 10^{-23}$.)
The experimental bounds for the LIV parameters in the
fundamental SME Lagrangian are below $10^{-27}$ 
(cf.\ Subsection 2.3). However, it is not obvious
how this translates into bounds on
effective MAV differences (remember that here we
started from a renormalised propagator).\footnote{A possible 
transition from SME parameters to hadrons by means of a parton model 
was discussed in Ref.\ \cite{GaMo}\label{partonfn}.}

Let us stress once more that this is not in contradiction
to the fact that this $\Delta$ resonance is observed at low
energy --- in that case the masses contribute significantly
to the energies and the above approximations are not valid
any longer. 

For $\varepsilon > \varepsilon_{c}$ the dominant
candidate for a photopion production channel is an
excited proton state,
\be
p + \gamma \ \to \ p^{*}(1435) \ \to \ p + \pi^{0} \ .
\ee
But at ultra high energy $p^{*}$ can only decay
if $c_{p} \geq c_{\pi} $,
according to condition (\ref{decaycon}).\footnote{Bounds on
$c_{\pi}$ based on UHECR are discussed in 
Refs.\ \cite{Alfaro1,AltschulMEVpi}.} With a suitable choice of
the MAVs we could close that channel too, and so on.

Just before Ref.\ \cite{ColGla} appeared, 
Ref.\ \cite{FarBier} attracted attention
to the observation that the top 5 super-GZK events known
at that time all arrived from the direction of a 
quasar (cf.\ footnote \ref{quasar}).
This inspired Coleman and Glashow to suggest an unconventional
hypothesis: the super-GZK primary particles could
be {\em neutrons.} Assuming particle specific MAVs,
the following arguments yield a consistent picture
(see also Ref.\ \cite{DubTin}):
\begin{itemize}
\item For $c_{n} < c_{\rm min}$ (where $c_{\rm min}$ refers to the
$\beta$-decay products in the sense of relation (\ref{decaycon})) the
decay of the neutron can be avoided.
\item Since we already assumed a slightly reduced neutron MAV,
we might add $c_{n} < c_{\Delta}$ to protect the neutron
from the GZK cutoff.
\item Unlike the proton, the neutron is hardly deflect by interstellar
magnetic fields, hence the scenario of an approximately straight path
--- which is required to make sense of the observation in 
Ref.\ \cite{FarBier} for quasars at large distances --- 
becomes realistic.
\end{itemize}

This suggestion is currently not discussed anymore, but its
consistency is beautiful enough to be reviewed nevertheless.
Nowadays $O(100)$ super-GZK events have been reported (see Section 1), 
and the quasar hypothesis is out of fashion as well.
However, the hypothesis of clustered UHECR directions
has recently been revitalised, see Appendix A. Also
the idea of electrically neutral primary particles
is still considered, see {\it e.g.}\ Refs.\ 
\cite{neutralprimary}.\footnote{However, neutral 
primary particles are not required for the hypothesis discussed 
in Appendix A, since it assumes {\em nearby} UHECR sources.}
They include 
pions $\pi^{0}$ \cite{Alfaro1},
photons \cite{neutralprimarygamma} (though
Ref.\ \cite{NoTopDown} puts a narrow bound on their UHECR flux), 
neutrinos \cite{neutralprimarynu} and hadrons of $(u,d,s)$ quarks, 
which could be stable in SUSY theories \cite{neutralprimarySUSY}.

\subsubsection{Impact on neutrino oscillation}

If the MAVs can deviate from $c$, there are {\em three} bases for
the neutrino states:
\begin{itemize}
\item the flavour basis with the states
$\, | \nu_{e} \rangle , \ |\nu_{\mu}\rangle , \ |\nu_{\tau}\rangle $,
\item the basis of the neutrino masses,
\item the basis of the MAVs.
\end{itemize}

In this framework neutrino oscillation could occur even at vanishing 
neutrino masses, due to a flavour mixing in the MAV basis.\footnote{A
discussion in the SME terminology is given in Ref.\ \cite{KosMew04}.}
However, in light of experimental data this scenario
has soon been discarded \cite{numixnopureMAV,K2Ka}, so we start here from 
the usual point of view that the observed neutrino oscillation is evidence
for non-vanishing neutrino masses.

We do consider, however, the possibility that the mass driven
oscillation could still be modified as a sub-leading effect
by additional MAV mixing. This scenario was investigated
\cite{MACRO} based on the data by the MACRO Collaboration in 
Gran Sasso (Italy). The study can be simplified by focusing on the
oscillation $\nu_{\mu} \leftrightarrow \nu_{\tau}$, which is
most relevant for the observation of cosmic neutrinos.
In this 2-flavour picture the two parameters
\bea
\Delta v & := & | {\rm MAV}(\nu_{1}) - {\rm MAV}(\nu_{2}) | \nn \\
\theta_{v} &:=& {\rm mixing~angle~of}~| \nu_{\mu} \rangle ~
{\rm and}~ ~| \nu_{\tau} \rangle ~{\rm in~the~MAV~basis} \qquad
\label{mixpar}
\eea
modify the life time of $\nu_{\mu}$.

\begin{figure}
\begin{center}
\includegraphics*[angle=0,width=.6\linewidth]{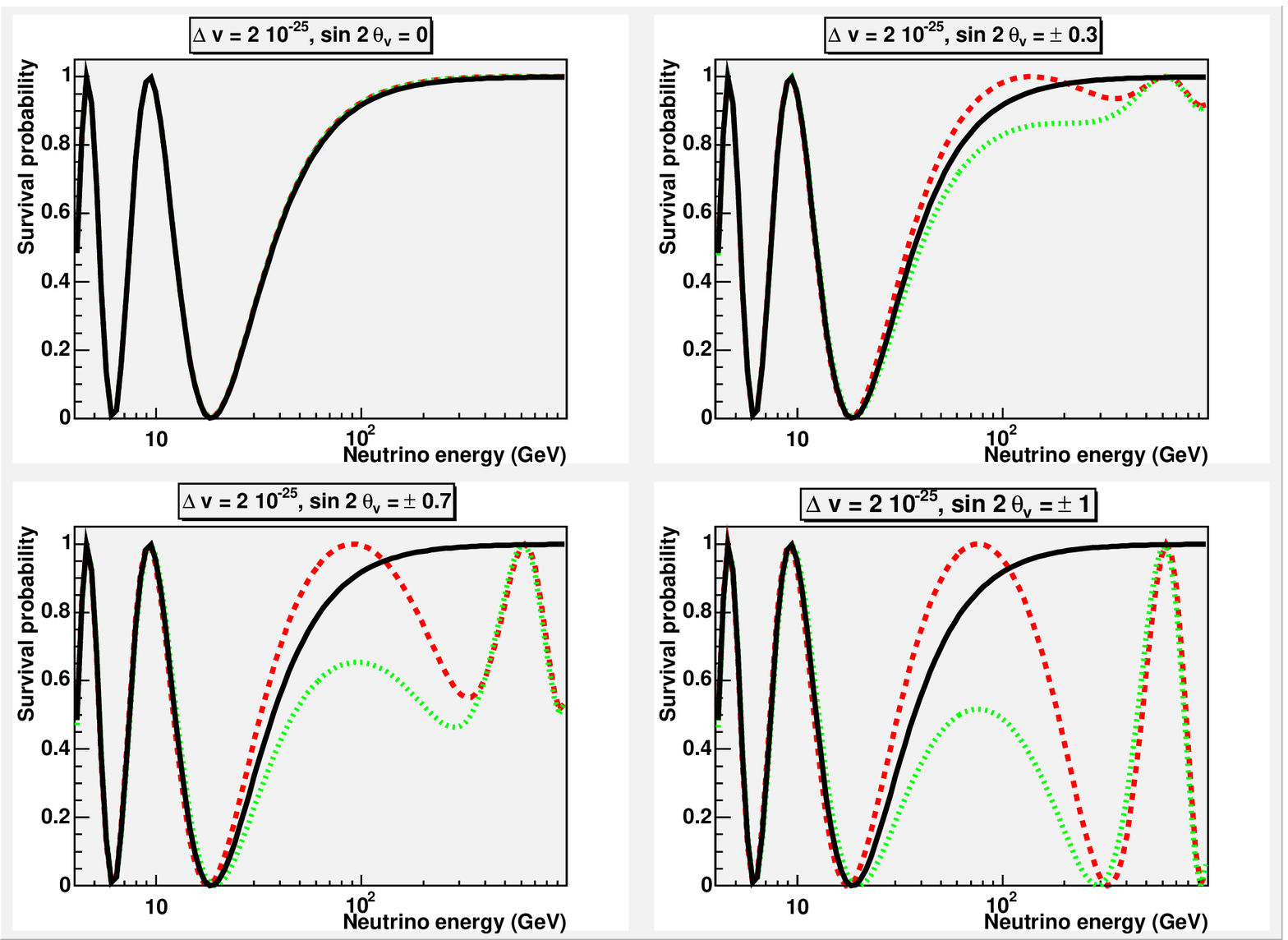}
\end{center}
\vspace*{-2mm}
\caption{\it The survival probability for a neutrino $\nu_{\mu}$
over a distance of $10^{7}~{\rm m}$, for a MAV difference 
$\Delta v = 2 \cdot 10^{-25}$ (see eq.\ (\ref{mixpar})).
The plot is adapted from Ref.\ \cite{MACRO}. It shows the curves
for mixing angles $\theta_{v} =0$ (only mixing due to the
neutrino masses) and $\theta_{v} = \pm \pi /4$ (red dashed line
and green dotted line), where $\theta_{v}$ parameterises a mixing
with $\nu_{\tau}$ in the MAV basis. We see a strong effect on 
neutrinos of $O(10^{11})~{\rm eV}$.}
\vspace*{-2mm}
\label{survive}
\end{figure}

As an example, Figure \ref{survive} shows the survival probability of
$\nu_{\mu}$ over a distance of $10^{7}~{\rm m}$ (roughly
a path across the Earth passing near the centre)
at $\Delta v = 2 \cdot 10^{-25}$ (still in natural units, $c=1$).
The three curves refer to the MAV mixing angles with
$\sin 2 \theta_{v} = 0 \, , \ \pm 1\,$. We see that the effect
of such a mixing is very significant for neutrino energies of
$O(100)~{\rm GeV}$. If we assume neutrino masses $m \lsim 1 ~{\rm eV}$,
this corresponds to a Lorentz factor $\gamma \gsim 10^{11}$, similar
to the primary proton in a UHECR (see eq.\ (\ref{gamma11})).
Cosmic neutrinos in this energy range occur,\footnote{One suspects
that there are cosmic neutrinos at much higher energies as well ---
they are not restricted by any GZK-type energy cutoff (cf.\ 
footnote \ref{neutrinofn}) nor deviated by magnetic fields, so
they could open new prospects to explore the Universe.
Several projects are going to search for them systematically,
see Section 4.}
and their behaviour is again suitable for a sensitive test of LI at 
ultra high $\gamma$-factors.

The MACRO Collaboration detected upward directed muons 
generated by neutrinos in the $| \nu_{\mu} \rangle$ state
through the scattering on a nucleon,
\be
\nu_{\mu} + n \to \mu^{-} + p \qquad {\rm or} \qquad
\nu_{\mu} + p \to \mu^{+} + n \ .
\ee
By monitoring multi-Coulomb scattering they reconstructed 
the energy $E_{\mu}$ and finally $E_{\nu_{\mu}}$. They observed
58 events with $E_{\nu_{\mu}} \gsim 130 ~ {\rm GeV}$ and compared 
this flux to the one of $\nu_{\mu}$ at low energies.
It turned out that the oscillation curve obtained solely by mass 
mixing (corresponding to the curve for $\sin 2 \theta_{v} = 0$ in
Figure \ref{survive}) works very well. Including
the additional parameters (\ref{mixpar}) does not improve the
fits to the data. Therefore the scenario of additional 
$\nu_{\mu} \leftrightarrow \nu_{\tau}$ mixing due to different 
MAVs is not supported, and Ref.\ \cite{MACRO} concludes
\be
\Delta v < 6 \cdot 10^{-24} \ .
\ee
Of course the exact bound on $\Delta v$ depends on the mixing angle ---
it becomes even more stringent when $| \sin 2 \theta_{v}|$
approaches $1$. \\ 

This result is in agreement with similar analyses
based on the Super-Kamiokande K2K data \cite{K2Ka,K2Kb}.
Ref.\ \cite{K2Ka} introduced a parameter $n$ to distinguish
scenarios for the origin of neutrino mixing as follows:
\be
n = \left\{ \begin{array}{cl}
-1 & ~~{\rm mass~mixing~(standard)} \\
~0 & ~~{\rm energy~independent~mixing} \\
~1 & ~~{\rm LIV~resp.~violated~equivalence~principle.}
\end{array} \right.
\ee
The fit to the K2K data led to $n=-0.9(4)$, supporting the
standard picture and strongly constraining the alternatives.
Similarly Ref.\ \cite{K2Kb} generalised the scenario corresponding
to $n=1$ by admitting vector and tensor background fields --- as in
Subsection 2.3 --- and arrived qualitatively at the same conclusion.

Regarding the SME mixing parameters, a phenomenological test 
with $\bar \nu_{e} \leftrightarrow \bar \nu_{\mu}$ data has been
presented in Ref.\ \cite{Auerbach}. On the theoretical side the
$\nu_{\mu} \leftrightarrow \nu_{\tau}$ oscillation of {\em Majorana 
neutrinos} is discussed in Ref.\ \cite{XiaoMa}.

%% file: gammaray2.tex
In this section we are going to consider the photons themselves
as they appear in cosmic rays --- so far they entered our
discussion only in the background.
Their identification as primary particles
is on relatively safe grounds (cf.\ Section 1).

\subsection{Another puzzle for high energy cosmic rays ?}

The highest energies in cosmic $\gamma$-rays were observed around 
$E_{\gamma} \gsim 50 ~ {\rm TeV}$ (see {\it e.g.}\ Refs.\ \cite{crab}).
They originate from the {\em Crab Nebula,} which is the remnant of a 
supernova --- recorded by Chinese and Arab astronomers in the year
1054 --- at a distance of $2~{\rm kpc}$. 

The sources of the strongest $\gamma$-rays reaching us 
from {\em outside} our galaxy are {\em blazars}. Blazars are 
a subset of the AGN, see footnote \ref{AGNf}.\footnote{Appendix A
addresses the hypothesis that AGN could emit UHECRs as well.}
A few hundred blazars are known. A prominent
example is named {\em Markarian 501}; in 1997 the 
High Energy Gamma Ray Astronomy (HEGRA) satellite
detected $\gamma$-rays with $E_{\gamma} \gsim 20 ~ {\rm TeV}$
from an outburst that this blazar had emitted \cite{HEGRA}. 
It is located at a distance 
of $157~{\rm Mpc}$ --- this can be determined 
from the redshift ($z \simeq 0.034$). 
Here also the direction of the source is known, 
in contrast to far-travelling charged 
cosmic rays (gravitational deflections are small).
Another example, which is noteworthy in this
context, is {\em Markarian 421}, at a distance of 110~Mpc, which 
emitted photons that arrived with $\sim 10 ~ {\rm TeV}$ \cite{Mkn421}.

It was suggested that
the observation of these highly energetic $\gamma$-rays
could pose a new puzzle, which is similar to
the GZK cutoff \cite{Camel}. Again the question is why the CMB can be
transparent for rays of such high energies. In this case,
one expects electron/positron pair creation of an
UV photon in the ray and an IR photon in the background,
\be  \label{gammagamma}
\gamma_{\rm UV}(E) + \gamma_{\rm IR}(\omega ) 
\ \to \ e^{+} + e^{-} \ , 
\ee
if sufficient energy is involved. This pair creation is followed 
by inverse  Compton scattering on further CMB photons, triggering a 
cascade. In the centre-of-mass frame both photons in the transition
(\ref{gammagamma}) have the energy 
\be
\bar \omega = \frac{E}{\gamma} = \omega \gamma \ ,
\ee
where $\gamma$ is the Lorentz factor between centre-of-mass and
FRW laboratory frame (cf.\ Section 1). 
The condition for pair creation is obvious,
\be  \label{condgamma1}
\bar \omega^{2} \equiv E \omega  \geq m_{e}^{2} \ .
\ee
If we insert once more the exceptionally large CMB photon energy
that we referred to before in Sections 1 and 2,
$\omega \approx 0.003 ~{\rm eV} \approx 5 \, \langle \omega \rangle \, $,
the threshold for $E$ is almost $ 10 ~{\rm TeV}$.
If we repeat the considerations about the CMB photon density
and the cross-section for this pair creation, in analogy
to Subsection 1.3, also the observed energies $E$ of photons 
coming from far distances (outside our galaxy) 
might be puzzling \cite{GammaTeV}.
The mean free path length $\ell_{\gamma \gamma}$ is given by a 
formula similar to eq.\ (\ref{taueq}),
\be  \label{lgg}
\frac{1}{\ell_{\gamma \gamma}(E)} = \frac{1}{8 E^{2}}
\int_{\omega_{\rm min}(E)}^{\infty} d \omega \, 
\frac{d n_{\gamma} / d \omega}{\omega^{2}} 
\int_{(2 m_{e})^{2}}^{4 E \omega} ds \, s \, \sigma_{\gamma \gamma} (s) \ .
\ee
As in eq.\ (\ref{Deltacon}), $s$ is one of the (LI)
Mandelstam variables.
The differential photon density of the CMB follows
Planck's formula (\ref{planckeq}). The minimal CMB photon
energy is $\omega_{\rm min}(E) = m_{e}^{2}/E \, $, see eq.\
(\ref{condgamma1}). The boundaries for the angular integral 
over $s$ are also obvious, and $\sigma (s)$
is the cross-section. It peaks just above the threshold
$s = (2 m_{e})^{2}$, and at high energy ($ s \gg m_{e}^{2}$)
it decays as \cite{BhaSig}
\be
\sigma_{\gamma \gamma} \propto \frac{1}{s} \ln \frac{s}{2 m_{e}^{2}} \ .
\ee
Ref.\ \cite{Mkn501prob} evaluated this formula and found
for instance for $E = 10 ~{\rm TeV}$ a path length
$\ell_{\gamma \gamma} \approx 10~{\rm Mpc}$, 
so that the radiation from Markarian 421 and 501
appears amazing. Ref.\ \cite{Mkn501prob}
suggested a generalisation of formula (\ref{lgg}) to a LIV form,
inserting {\it ad hoc} a deformed photon dispersion relation,
which extends $\ell_{\gamma \gamma}$. However,
the question if this puzzle persists in the LI form,
after taking all corrections into account, is 
controversial \cite{Mkn501noprob,kappadisp,JLM}.
One issue is our poor knowledge about the radio background radiation, 
which dominates over the CMB at wave lengths $\gsim 1~{\rm m}$
(cf.\ footnote \ref{radio}). Ref.\ \cite{BhaSig} studied the effective
penetration length $\ell_{\rm cascade}$ of the photon cascade under 
different assumptions for the extragalactic magnetic field and the radio
background. Over a broad interval for these two parameters,
Figure 14 in that work implies
\be
\ell_{\rm cascade} \approx \left\{
\begin{array}{ccc}
100 ~ {\rm Mpc} & {\rm at} & E = 30~{\rm TeV} \\
300 ~ {\rm Mpc} & {\rm at} & ~E = 10~{\rm TeV} \ . 
\end{array} \right.
\ee
In light of these numbers the observations of Markarian
421 and 501 appear less puzzling, but the issue is
still under investigation.
As in the case of the GZK cutoff we discuss possible solutions
for such a puzzle --- if it exists --- based on LIV.

Also this {\em multi-TeV $\gamma$-puzzle} could in principle 
be solved by assuming 
different MAVs (cf.\ Sections 2.4, 2.5) for the two particle types
involved \cite{SteGla},
\be
c_{e} = c_{\gamma} + \varepsilon_{e} \ .
\ee
For a head-on collision the condition (\ref{condgamma1}) is modified to
\be  \label{condgamma2}
E \omega  > m_{e}^{2} + \varepsilon_{e} E^{2} + O(\varepsilon_{e}^{2}) \ ,
\ee
in analogy to eq.\ (\ref{epsGZK}).
Therefore a parameter $\varepsilon_{e} >0$ could increase the energy 
threshold and --- if $\varepsilon_{e}$ reaches a critical value 
$\varepsilon_{c}$ --- exclude the electron-positron pair creation.
Then the Universe becomes transparent for photons
of higher energy. At the energies that we considered above,
this happens for
\be
\varepsilon_{e} > \varepsilon_{c} = 2 \, \frac{E \omega - m_{e}^{2}}{E^{2}}
\vert_{E = 20 \, {\rm TeV}, \ \omega = 5 \, \langle \omega \rangle =
0.003 \, {\rm eV}} \simeq
2 \cdot 10^{-15} \ .
\ee
This critical value is much larger than the one for the
proton in eq.\ (\ref{epsc1}). However, what matters is
that $\varepsilon_{c}$ is well below the upper bound on 
$\varepsilon_{e}$ that one obtains from the absence
of vacuum \v Cerenkov radiation, given in eq.\ (\ref{epsele}).
Therefore the MAV scenario could provide a conceivable solution to
the (possible) multi-TEV $\gamma$-puzzle as well.

Ref.\ \cite{JLMth} contains a discussion in the spirit of SME 
and finds that LIV, in addition to modifying the minimal energy threshold,
could also imply an {\em upper} energy bound for this pair production.

\subsection{$\gamma$-Ray-Bursts (GRBs)}

We now consider cosmic $\gamma$-rays of lower energy
and focus on {\em $\gamma$-Ray-Bursts} (GRBs). These bursts are emitted
in powerful eruptions over short periods, usually of a few seconds 
or minutes (in extreme cases of $(0.1 \dots 10^{3})~{\rm s}$).
They typically include photons with  energies 
$E_{\gamma} \approx (10^{4} \dots 10^{8})~{\rm eV}$.
Temporarily they are the brightest spots in the sky.
The size of these sources is rather small.
About the GRB origin we only have speculations,
such as the merger of two neutron stars, or of two black holes,
or of a neutron star and a black hole, or hypernovae
(a theoretically predicted type of supernova, where an exceptionally
massive star collapses). There are, however, sophisticated
models for the GRB generation, dealing in particular
with fireball shock scenarios or a gravitational core collapse
\cite{GBRfireball}.\footnote{It has also been proposed that the 
sources of GRB and UHECR could coincide \cite{Wax,GRBasUHECRsource}.}

GRBs have been observed since 1967. 
Most of them were discovered first by satellites
(such as Hubble); subsequently ground-based telescopes could
focus on them and identify the distance to the source.
Today also the initial detection from Earth is possible, in
particular from the observatory in La Silla (Chile) \cite{LaSilla},
which enables an immediate evaluation of the redshift.
A spectacular GRB was observed in 2005 from the satellite SWIFT:
it came from a distance of $\sim 4 ~{\rm Gpc}$; 
these photons travelled to us directly
from the Universe when it was only $9 \cdot 10^{8}$ years old.

\begin{figure}
\begin{center}
\includegraphics[angle=270,width=.6\linewidth]{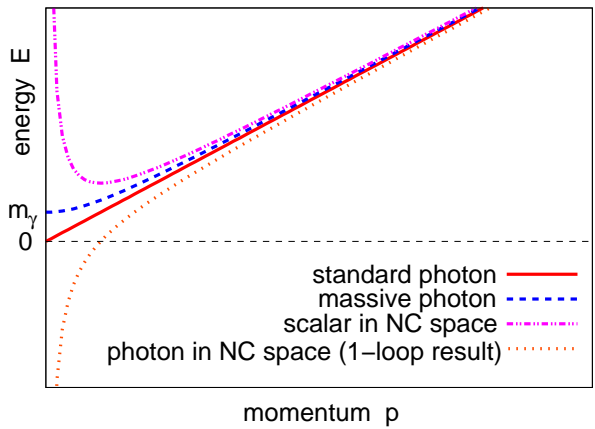}
\end{center}
\caption{\it Qualitative comparison of various dispersion relations
in the IR regime: standard photon (mass zero in commutative space),
massive photon (in commutative space), massless scalar particle
in an NC space and massless photon in an NC space. The latter
corresponds to the 1-loop result, but its
feature changes non-perturbatively, see Subsection 3.3.2.}
\label{photodisp}
\end{figure}

The very long journey of photons with different energies but
similar emission times is suggestive for a test of the {\em photon 
dispersion relation.} This test probes if the group velocity
$v_{\gamma}$ is really independent of the
momentum \cite{Camel}. The simplest approach for a deviation of this
standard assumption is the introduction of a photon mass
$m_{\gamma} > 0$ (which requires a modification
of the Higgs mechanism). This obviously leads to
\be
v_{\gamma} (p) 
= \frac{p}{\sqrt{p^{2} + m_{\gamma}^{2}}} \neq const.
\ee
as sketched in Figure \ref{photodisp} ($c=1$ is still the photon MAV).
From the observation that the photons in a GRB arrive at approximately
the same time one can deduce 
$m_{\gamma} < 10^{-6}~{\rm eV}$ \cite{Camel}. However, the bound 
obtained in the laboratories is superior by many orders of magnitude,
$m_{\gamma} < 10^{-17}~{\rm eV}$ \cite{particledata}.

Therefore the option of introducing a photon mass is not
attractive in this context --- even if it exists below the
experimental bound, its effect would be invisible to us
in GRBs. In the following we stay with $m_{\gamma} =0$. 
The MAV ansatz $E^{2} = p^{2} c_{\gamma}^{2}$ does
not help as a theoretical basis for a possible momentum dependence 
of the speed of photons. 
On the effective level one can still study LIV modified
Maxwell equations and evaluate the dispersion together with the
rotation of the polarisation plane \cite{bireastro,Tina}, see 
Subsection 3.5. However, we now proceed to another
theoretical approach for new physics, including a non-linear
photon dispersion relation, which is highly fashionable.
That approach is basically different from the effective action
concepts that we considered in Section 2.\footnote{A further
alternative are ``varying speed of light theories'', 
which are reviewed in Ref.\ \cite{VSL}.}

\subsection{Non-commutative (NC) field theory}

As a brief introduction, we embed this booming field of 
research into a historic perspective. Galilei's
system of coordinate transformation did not have any constant
of Nature (hence it was in some sense more relativistic than
the relativity theory of the $20^{\rm th}$ century). 
Einstein did introduce the speed of light $c$
as such an invariant; it is a corner stone of both,
Special and General Relativity Theory. Now there are speculations
that there might be yet another constant of Nature in addition
to $c$. Ref.\ \cite{DSR} denotes this line of thought as 
``Doubly Special Relativity'' (according to Ref.\ \cite{VSL} it
emerged as an effective model for ``varying speed of light'' in a 
quantum space-time). This second constant is usually assumed to be a large,
observer-independent energy.

One often refers 
to gravity effects becoming important in field theory at very
large energy. This is supposed to be a generic property,
which is likely to hold even though we do not have an established 
theory of quantum gravity \cite{Shomer}. Therefore it appears natural 
to relate this second constant to the Planck mass (\ref{planckscale})
(though there may be alternatives \cite{Klink}). Geometrically
this means the absence of sharp points, but a space where
points are washed out on the scale of the Planck length
\be
L_{\rm Planck} = M_{\rm Planck}^{-1} \simeq 1.6 \cdot 10^{-35}~{\rm m} \ .
\ee
(J.\ von Neumann dubbed it ``pointless geometry'', though this term 
applies to the phase space in standard quantum mechanics as well).

It can be compared to the lattice regularisations, which
truncates short distances {\it ad hoc}. For numerous quantum field theories 
this is the only regularisation scheme, which works 
beyond perturbation theory:\footnote{The ``fuzzy sphere'' \cite{Madore}
has recently been considered as a possible alternative, but numerical
work in that regularisation \cite{numfuz} is at an early stage.}
it is not only a basis for the numerical measurement of observables,
but it actually {\em defines} these theories on the 
non-perturbative level. Nevertheless
hardly anyone would assume a lattice structure to be physical ---
it could at best mimic the geometry on the
Planck scale. A more subtle way to introduce a geometric UV cutoff
places field theory on a space with a 
{\em non-commutative geometry.} Let us briefly review the
historic construction in the first publication on this subject \cite{Sny}.

We know that quantum mechanics is endowed with a natural
discretisation of angular momenta based on the uncertainty
relation. The idea is now to make use of this mechanism
for space coordinates as well: the coordinate components
should be described by operators with discrete spectra,
as it is familiar for the angular momentum components.
This can be achieved from a higher dimensional perspective.
We start from a 5d Minkowski-type space with the invariant
quantity
\be
S = x_{0}^{2} - x_{1}^{2} - x_{2}^{2} - x_{3}^{2} - x_{4}^{2} \ .
\ee
We specify a surface $S = const. > 0$, which is a 4d de Sitter
space inside the 5d light cone. Now we list the generators 
of transformations that leave $S$ invariant,\footnote{In general
we use natural units, but at this point it is instructive to
make an exception and write $\hbar$ explicitly.}
\bea
L_{3} &=& \frac{\hbar}{\rm i} ( x_{1} \partial_{2} -
x_{2} \partial_{1} ) \ , \quad {\rm invariant~:} \quad
x_{1}^{2} + x_{2}^{2}\, , \, x_{0}\, , \, x_{3}\, ,\, x_{4} \ , \nn \\
M_{1} &=& {\rm i} \hbar ( x_{0} \partial_{1} -
x_{1} \partial_{0} ) \ , \quad {\rm invariant~:} \quad
x_{0}^{2} - x_{1}^{2}\, , \, x_{2}\, , \, x_{3}\, ,\, x_{4} \ , \nn \\
X_{1} &=& \frac{a}{\rm i} ( x_{1} \partial_{4} -
x_{4} \partial_{1} ) \ , \quad {\rm invariant~:} \quad
x_{1}^{2} + x_{4}^{2} \, ,\, x_{0}\, ,\, x_{2}\, ,\, x_{3} \ , \nn \\
T &=& {\rm i} a ( x_{0} \partial_{4} -
x_{4} \partial_{0} ) \ , \quad {\rm invariant~:} \quad
x_{0}^{2} - x_{4}^{2}\, ,\, x_{1}\, ,\, x_{2}\, ,\, x_{3} \ . \nn
\eea
The operators $L_{1}, \, L_{2}$,  $M_{2}, \, M_{3}$
and $X_{2}, \, X_{3}$ are analogous. They all conserve 5d LI.
$L_{i}$ and $M_{i}$ generate the familiar rotations
and Lorentz boosts, so they preserve in addition also 4d LI
(at $x_{4} =0$). 
The latter is broken, however, by the $X_{i}$ and $T$. Here
a parameter $a$ enters in a r\^{o}le similar to $\hbar$
in the 4d transformations.

To show that the spectra of $X_{i}$ are discrete,
we follow the procedure which is well-known from quantum mechanics.
We write the rotated plane in polar coordinates,
$(x_{1}, x_{4}) = r ( \sin \varphi , \cos \varphi )$, so that
$X_{1} = -{\rm i} a \partial_{\varphi} \, $. For an eigenfunction
$\psi$, $X_{1} \psi = \lambda \psi$, we make the ansatz
\begin{displaymath}
\psi \propto \exp ( {\rm i} \varphi \lambda /a ) \ ,
\ \, {\rm and} \quad \psi ( \varphi ) = \psi ( \varphi + 2 \pi ) \
\Rightarrow \ \lambda  = n a \quad ( n \in \Z ) \ .
\end{displaymath}
As another property that is familiar from the angular momentum
algebra, these position operators are {\em non-commutative (NC)}
\be
[ X_{i} , X_{j} ] = 
\frac{{\rm i} a^{2}}{\hbar} \epsilon_{ijk} L_{k} \ , \quad
[ T , X_{i} ] = \frac{{\rm i} a^{2}}{\hbar} M_{i} \ .
\ee
That property obviously implies {\em new uncertainty relations}
for these space-time coordinate operators (without involving momenta !)
\be  \label{spaceuncert}
\Delta X_{i} \ \Delta T \ , \
\Delta X_{i} \ \Delta X_{j} \geq O( a^{2} ) \ , \qquad
(i \neq j) \ .
\ee
In this sense $a$ takes indeed the r\^{o}le of a minimal
length, and thus of a new constant of Nature.
A field theory living on this space is necessarily {\em non-local}
over the characteristic extent $a$. LI is violated in the 4d
subspace, though its 5d version is preserved. The latter property
shows that this way to implement a minimal length is far
more subtle than a crude space-time lattice\footnote{Alternatively, 
a discrete 4d space-time with Poisson distributed random sites
can be compatible with local LI \cite{Rafael}. The LI dispersion
relation can be reproduced exactly, even on regular lattices, if one 
implements a ``perfect lattice action'' based on the renormalisation group
(see {\it e.g.}\ Ref.\ \cite{habi} and refs.\ therein).}: some facet
of relativity persists, and the mechanism is organically
embedded into the general framework of quantum physics.\footnote{Based
on such considerations, it is tempting to speculate that
extra dimensions could in fact have a physical meaning. The literature
on this issue is tremendous; we do not try to review it here, but
we only mention one point: \\
Further arguments for extra dimensions deal with the {\em mass hierarchy 
problem,} {\it i.e.}\ the question why the known particle masses 
are so far below the Planck mass (\ref{planckscale})
(or in QCD: $m_{u,d} \ll \Lambda_{\rm QCD}$).
For instance, fermions naturally undergo
a renormalisation, which puts their mass {\em non-perturbatively}
on the cutoff scale (unless some artificial fine-tuning of the
bare mass is performed). An extra dimension helps to some
extent to implement approximate chirality and therefore a small fermion
mass \cite{braneworld}. If one is willing to assume SUSY in addition, 
then this mechanism is transmitted to the boson masses as well.}

Snyder constructed the 4-momentum as \cite{Sny}
\be  \label{Snymom}
P_{\mu} = \frac{\hbar}{a} \, \frac{x_{\mu}}{x_{4}} 
\qquad (\mu = 0 \dots 3) \ ,
\ee
which is consistent with $L_{k} = \epsilon_{ijk}
( X_{i} P_{j} - X_{j} P_{i}) $,
$M_{i} = X_{i} P_{0} + T P_{i}$. The commutators
\be
[X_{i} ,P_{j}] = {\rm i} \hbar \Big[ \delta_{ij} + 
\Big( \frac{a}{\hbar} \Big)^{2}
P_{i} P_{j} \Big] \ 
\ee
yield Heisenberg uncertainties, which are modified to 
$O(a^{2} P_{i} P_{j} / \hbar)$. For a tiny parameter
$a$ we have --- {\em on tree level} ---
again the situation that the modification
is only visible at huge momenta.

Based on the uncertainty relations (\ref{spaceuncert}), 
the identification of $a$ as the
Planck length now leads to a consistent picture
of the space-time uncertainty range as the event horizon
of a mini black hole \cite{DFR}. In fact, a (hypothetical)
measurement of a length of $O(L_{\rm Planck})$
requires (according to the Heisenberg uncertainty)
an enormous energy density, which gives rise to such
an event horizon --- the notion of detectable events 
then requires inequalities (\ref{spaceuncert})
to hold with $a \sim L_{\rm Planck}$. 
This {\it Gedankenexperiment} suggests that points
should indeed be washed out over a range of 
$O(L_{\rm Planck})$. If several directions are involved,
as in relation (\ref{spaceuncert}), this is
practically equivalent to non-commutativity.

As a variant we may just combine the uncertainty over
the gravitational radius $\Delta x /2 \sim L_{\rm Planck}^{2} \cdot
\Delta p$ with Heisenberg's uncertainty to arrive at a modification
of the latter, even in one dimension \cite{uncert1d},
\be
\Delta x \gsim {\rm max} \Big( \hbar / (2 \Delta p) , \ 2
L_{\rm Planck}^{2} \Delta p / \hbar \Big) \gsim L_{\rm Planck} \ .
\ee
However, this modification can lead to dangerous processes,
setting in at the 1-loop level,
where it is not obvious how to avoid a contradiction to
known phenomenology \cite{uncert1dang}. We now return
to the NC space --- although it is plagued by
dangers of this kind as well --- keeping in mind that basic
progress in physics has been repeatedly related to the realisation 
that specific quantities are ``unspeakable'' \cite{Suda}. \\

Above we followed the historic path to non-commutativity.
The currently popular version (for a review, see {\it e.g.}\
Ref.\ \cite{DougNek}) inserts for the coordinate 
commutators a constant ``tensor field'',\footnote{It is a tensor
under a twisted Poincar\'{e} group, and the action
can be called LI if one also adapts a twisted particle statistics
\cite{Bal}.} 
\be  \label{thetaconst}
[ X_{\mu} , X_{\nu} ] = {\rm i} \Theta_{\mu \nu} \ , \quad
{\rm min} ( \Delta X_{\mu} \ \Delta X_{\nu} ) = 
\frac{1}{2} | \Theta_{\mu \nu} | \ .
\ee
Thus we have the case of an {\em ``active LIV''} (see Subsection 2.3), 
and agreement with the general scope of DSR.
The question of its CPT invariance is 
discussed in Refs.\ \cite{NCQED-CPT,Mozo,CPTinNC}.

The original motivation for introducing NC coordinates
was the hope to tame the notorious UV divergences 
in quantum field theory.  In fact, they are removed in the
non-planar diagrams, but in the planar diagrams they persist
\cite{Filk}. Moreover, the non-planar diagrams give rise to new 
divergences in the IR sector \cite{MRS}, which causes the
dangerous processes that we announced.

A famous example is the one-particle irreducible 2-point function
$\langle \phi (-p) \phi (p) \rangle$ in the 4d NC $\lambda \phi^{4}$ 
model. The planar 1-loop contributions diverges $\propto \Lambda^{2}$
as in the commutative case ($\Lambda$ being a UV cutoff), 
whereas the non-planar diagram 
picks up an additional phase factor,
\be  \label{phasefa}
\langle \phi (-p) \phi (p) \rangle_{\rm 1-loop,~non-planar} =
\frac{\lambda}{6} \int \frac{d^{4}k}{(2 \pi )^{4}} \, 
\frac{e^{ik_{\mu} \Theta^{\mu \nu} p_{\nu}}}{k^{2} + m^{2}} \ .
\ee
The exponential phase factor is generic for each crossing
between lines of external propagators and loops.
At finite $\Theta^{\mu \nu}$ and $p_{\nu}$ the integral
$\int d k_{\mu}$ converges now thanks to the rapid oscillation
at large $| k_{\mu}|$. As a consequence, for this non-planar diagram
the UV cutoff turns into \cite{MRS}
\be
\Lambda_{\rm eff}^{2} = \frac{1}{1/ \Lambda^{2} - 
p_{\mu} (\Theta^{2})^{\mu \nu} p_{\nu}} \ .
\ee
The UV limit $\Lambda \to \infty$ can in fact be taken in this
term, as long as the field momentum $p$ remains finite.
In addition to the IR divergence due to the latter condition, we
also see that the limit $\Vert \Theta \Vert \to 0$ is 
{\em not smooth}. 

So far no systematic solution has been found for the treatment 
of multi-loop diagrams, which involve such singularities
at both ends of the momentum scale. Hence renormalisation does
not become easier but even more complicated due to
non-commutativity. This problem is known as {\em UV/IR mixing}.
Intuitively this is not too surprising
in view of the uncertainty relation (\ref{thetaconst}).

Part of the literature deals with a truncated
expansion in small $\Vert \Theta \Vert$, which allows for instance
for a generalised formulation of the Standard Model \cite{NCSM}.
This generalisation may be of interest by itself, but due to
the UV/IR mixing effects --- and the related discontinuous transition
to standard commutativity, $\Vert \Theta \Vert \to 0$
--- it is {\em fundamentally different} from
complete quantum field theory on a NC space; for instance, locality
is restored in that truncated expansion. Ref.\ \cite{KostelNC} also refers
to such a truncated expansion to embed NC field theory into
the SME (cf.\ Subsection 2.3). However, since this
alters basic properties, the SME --- or other low energy effective
field theory approaches to include LIV terms --- and (full-fledged) 
NC field theory are basically different.

Despite --- or because of --- the amazing UV/IR mixing effects,
NC field theory has attracted a lot of attention;
over the last 10 years, about 2500 works
have been written about it. This boom was triggered by the
observation that it can be mapped onto certain string theories
at low energy \cite{SeiWit}. In that framework one could 
also relate the non-locality range $\sqrt{\Vert \Theta \Vert}$ to the 
string extent; this was the inspiration of Refs.\ \cite{uncert1d}.
A general analysis of the spatial uncertainty in string
theory is given in Ref.\ \cite{stringuncert}, and Ref.\ 
\cite{stringuncertnum} presents numerical results related to 
the emerging geometry in the IIB matrix model (cf.\ footnote \ref{IIB}).

NC field theory can be viewed as the second
main theoretical framework for LIV, in addition to the 
approach described in Subsections 2.3 and 2.4.
In both cases it was string theory which inspired
a new concept in particle physics,\footnote{The history
is reviewed from this perspective in Ref.\ \cite{Iorio}.}
but the way that the formulations are elaborated now they can 
be introduced and studied as modifications of standard field 
theory without any reference to strings. 

\subsubsection{The photon in a NC world}

Due to non-commutativity, gauge transformations and translations
are intertwined. As a consequence, even the pure $U(1)$ gauge field
picks up a Yang-Mills type self-interaction term, {\it i.e.}\ the $U(1)$
gauge field becomes NC as well. 

This can be seen easily if we switch to the alternative
formulation, which uses ordinary coordinates, but all field
multiplications are performed with the {\em star product}
(or Groenewold-Moyal product) \cite{Moyal}
\be  \label{starproduct}
\phi (x) \star \psi (x) := \phi (x) \exp \Big( \frac{\rm i}{2} 
\backderi \ \!\!\! _{\mu} \Theta^{\mu \nu} \! \forderi \ \!\!\! _{\nu} 
\Big) \psi (x) \ .
\ee
Its equivalence to the formulation referred to above can be shown
by a plane wave decomposition, which allows for a transition between
fields and Weyl operators (see {\it e.g.}\ Ref.\ \cite{Szabo}). 
The star-commu\-ta\-tor 
$[x_{\mu}, x_{\nu}]_{\star} 
= {\rm i} \Theta_{\mu \nu}$
makes this link plausible. In case of a pure $U(N)$ gauge action, 
\be
S [A] = - \frac{1}{4} \int d^{d}x \, {\rm Tr} \,
F_{\mu \nu} \star F^{\mu \nu} \ , \quad
F_{\mu \nu} = \partial_{\mu} A_{\nu} - \partial_{\nu} A_{\mu}
+ {\rm i} g [ A_{\mu}, A_{\nu}]_{\star} \,  \nn 
\ee
the commutator term does not vanish, even if the
gauge group is $U(1)$, since it is promoted to a star-commutator,
$[ A_{\mu}, A_{\nu}]_{\star} := 
A_{\mu} \star A_{\nu} - A_{\nu} \star A_{\mu}$.

This formulation applies
to all NC $U(N)$ gauge theories, but for the gauge groups
$SU(N)$ it fails: in commutative spaces the algebra of the
generators $A^{a}$ closes thanks to the identity
$ {\rm Tr} [ A^{a} , A^{b}] \equiv 0$, which does not hold, 
however, for the star-commutator. In view of phenomenology,
the group $U(1)$ is therefore of primary interest.

The Yang-Mills type self-interaction in NC spaces
leads to a distorted photon dispersion relation. 
On tree level the non-commutativity only affects very short 
distances and therefore the UV behaviour of dispersion
relations. But on quantum level it causes in addition IR 
singularities, as we pointed out before. 
Ref.\ \cite{MST} performed 
a 1-loop calculation of the effective potential in NC
$U(1)$ gauge theory and derived a dispersion relation of the form
\bea
E^{2} &=& \vec p^{\, 2} 
+ C \frac{g^{2}}{p \circ p} + O(g^{4}) \ , \nn \\
&& p \circ p  :=  - \Theta^{\mu \nu} p_{\nu} 
\Theta_{\mu}^{~ \sigma} p_{\sigma} > 0 \ , \quad C = const.
\label{dispmod}
\eea
This suggests that a small non-commutativity has a stronger 
impact than a large one, and based on
GRB data one could therefore be tempted to claim a {\em lower} bound
\cite{CamelNC},\footnote{Ref.\ \cite{HellYou} presents a similar 
consideration based on 1-loop effects in the 
potential for NC scalars, which becomes strong for very small
$\Vert \Theta \Vert > 0$.}
which is large compared to the Planck length scale,
\be
\Vert \Theta \Vert > 10^{-44}~{\rm m^{2}} \simeq
(10^{13} \cdot L_{\rm Planck})^{2} \ .
\ee
However, one should better not insist on this apparently spectacular
inequality (as Ref.\ \cite{CamelNC} also noticed) for a number of 
reasons:

\begin{itemize}

\item First we stress that this does not rule out
the standard commutative world, $\Theta = 0$, where no extra
term of this kind occurs. 
Eq.\ (\ref{dispmod}) shows again that the transition
$\Vert \Theta \Vert \to 0$ is not smooth.

Part of the literature deals with an expansion
in $1/\Vert \Theta \Vert$, which does have a smooth limit
to a commutative field theory at $\Vert \Theta \Vert \to \infty$
--- at least in perturbation theory around the free field vacuum
--- because the extra phase factor 
$\exp ({\rm i} k_{\mu} \Theta^{\mu \nu} p_{\nu})$
(for the crossing of propagator lines, see eq.\ (\ref{phasefa}))
cancels the non-planar diagrams at finite external momentum $p$.
However, that commutative limit does not coincide with the standard 
setting $\Theta = 0$. Moreover, in case of SSB the above
perturbative argument collapses, because it does not
refer to an expansion around a vacuum state anymore, and
therefore NC effects may persist \cite{thetainf}.

\item Such a strong non-commutativity is well above the upper
bound established from other observations \cite{CamelNC}.
Does this mean that NC field theory is ruled out already ? \\
Further aspects make this issue more complicated.

\item A careful evaluation \cite{LLT} revealed that the NC term
in eq.\ (\ref{dispmod}) is actually {\em negative,} 
\be  \label{Ruizeq}
C = - \frac{2}{\pi^{2}} \ .
\ee
Therefore the system appears {\em IR unstable,} as
illustrated in Figure \ref{photodisp}. In string theory
there are attempts to interpret such systems
(a key word is ``tachyon condensation'' \cite{Sen}), but in simple
terms of quantum physics a system without
a ground state does not have an obvious meaning.
Hence this observation looks like a disaster for NC field
theory --- it seems to be in straight contradiction to
everyday's experience. 

The behaviour with a {\em positive} IR divergence can be realised 
for instance for NC scalar particles. This feature was indeed 
measured numerically in the 3d $\lambda \phi^{4}$ model with a 
NC spatial plane \cite{NCphi4}, as sketched in Figure 
\ref{photodisp}. A condensation of the finite
mode of minimal energy then gives rise to a ``stripe pattern'' 
formation, {\it i.e.} some non-uniform periodic order
dominates the low energy states, which corresponds to
SSB of Poincar\'{e} invariance. 
In perturbation theory the negative IR energy for the NC photon
emerges from the requirement to fix the gauge.

There is extensive literature about electrodynamics
with star products on tree level, ignoring this problem.
However, the question is in which
regime non-commutativity effects could be manifest while 
quantum corrections are not visible. Here the surprises on 
quantum level are even more striking than in the case of
LIV parameters entering effective field theory
\cite{LIVfinetune} because of the UV/IR mixing.
Most people who were concerned with this severe problem moved on 
to a supersymmetric version, where the negative IR divergence is
cancelled by a positive one due to the fermionic partner
of the gauge field, with an opposite loop correction \cite{LLT,AHH}.
Does this mean that non-commutativity
is only sensible if one puts SUSY as yet 
another hypothetical concept on top of it ? \\
Even then the corresponding ${\cal N}=4$ super-Yang-Mills
theory suffers from a hierarchy problem for the
SUSY breaking scale \cite{Mozo}.

\item From our point of view, the apparent quest for 
multiple hypotheses (NC geometry plus SUSY)
to cancel the IR instability makes 
NC field theory less attractive. However, if the 1-loop
result has such amazing features one may wonder if it
should be trusted to extract qualitative properties
of the system. We therefore reconsidered this question
on the {\em non-perturbative} level \cite{NCQED}.

\end{itemize}

\subsubsection{The NC photon revisited non-perturbatively}

A truly non-perturbative investigation of the NC photon --- at finite 
non-commu\-ta\-ti\-vi\-ty and finite self-coupling ---
has to be carried out numerically. This requires a Euclidean
metrics. If we introduce a NC Euclidean time, reflexion positivity
(as one of the Osterwalder-Schrader axioms \cite{OSax}) is lost. Then
it is not known how to achieve a transition to Minkowski signature,
so we keep the Euclidean time commutative.\footnote{The opposite case with 
$[\hat x_{i}, \hat t ] = \frac{\rm i}{\kappa} \hat x_{i}$,
$[\hat x_{i}, \hat x_{j} ] = 0$ \cite{kappaMink}
is known as {\em $\kappa$-Minkowski space}. Work on its impact on
dispersion relations is summarised in Ref.\ \cite{kappadisp}.}

Moreover we need the (anti-symmetric) tensor $\Theta$ to be invertible 
on the NC subspace,\footnote{If this is the case, two operators 
${\cal O}$ at distinct points can be multiplied as \cite{Szabo} \\
${\cal O} (x) {\cal O} (y) = \pi^{-d}
({\rm det} \Theta )^{-1} \int d^{d}z \, {\cal O} (z) \,
\exp [ -2 {\rm i} (\Theta^{-1})_{\mu \nu} (x-z)_{\mu}(y-z)_{\nu} ]$,
where $d$ is the dimension of the NC subspace, which has to be even.}
hence we only introduce a NC spatial plane
$(x_{1}, x_{2})$ with a single non-commutativity parameter $\theta$
\be  \label{NCplane}
[ \hat x_{i} , \hat x_{j} ] = {\rm i} \, \theta \, \epsilon_{ij}
\qquad (i,j = 1,2) \ ,
\ee
while the plane $(x_{3}, x_{4})$ is kept commutative.
(Once the Euclidean time $x_{4}$ commutates, this form
can always be obtained by means of rotations.)
This is the only conceptually clean non-perturbative NC setting 
available, and it can be viewed as a minimally NC formulation.

Next we need a regularisation to a finite number of degrees of
freedom in order to enable a numerical treatment. As in commutative
field theory the lattice is suitable for this purpose. Here we review
the NC lattice formulation briefly, details are explained in
Ref.\ \cite{Szabo}. Although
we do not have sharp points for the lattice sites at hand, a lattice
structure can be imposed by requiring the operator identity
\be  \label{opid}
\exp \Big( {\rm i} \frac{2 \pi}{a} \hat x_{i} \Big) = \hat \uno
\ .
\ee
The length $a$ is now the (analog of a) lattice spacing, {\it i.e.}\
the spacing in the spectra of the operators $\hat x_{i}$ (which cannot
be diagonalised simultaneously).
Again in the spirit of minimal non-commutativity we let the momentum
components commute (as in Snyder's formulation (\ref{Snymom})). 
Their usual periodicity over the Brillouin zone
has remarkable consequences,
\bea
e^{ {\rm i} k_{i} \hat x_{i}} &=& 
e^{ {\rm i} (k_{i} + 2\pi /a) \hat x_{i}} \qquad \Rightarrow \nn \\
\hat \uno &=& e^{ {\rm i} (k_{i} + 2\pi /a) \hat x_{i}}
e^{ - {\rm i} k_{j} \hat x_{j}} \nn \\
&=& \exp  \Big( {\rm i} (k_{i} + \frac{2\pi}{a} ) \hat x_{i}
- {\rm i} k_{j} \hat x_{j} 
+ \frac{1}{2}(k_{i} + \frac{2\pi}{a} ) k_{j} 
[\hat x_{i}, \hat x_{j}] \Big) \nn \\
&=& \hat \uno \ 
\exp \Big( \frac{{\rm i} \pi}{a} \theta (k_{2} - k_{1}) \Big) 
\quad \
\Rightarrow \quad \ 
\frac{\theta}{2a} k_{i} \in \Z \ . \label{discmom}
\eea
In the exponent on the right-hand-side we had to keep track of
the Baker-Campbell-Hausdorff term, which we evaluated by
inserting relation (\ref{NCplane}).
For fixed parameters $\theta$ and $a$ we infer from this exponential
factor that the momentum components in the NC lattice
are discrete, so that this lattice is automatically {\em periodic.}
Of course, this property is in contrast to lattices in a
commutative space. Here we inevitably obtain an {\em UV and IR 
regularisation} at the same time.

If our lattice is periodic over $N$ sites in each of the
NC directions, the momentum components are integer multiples
of $2 \pi / (aN)$, which --- along with relation (\ref{discmom}) ---
implies
\be  \label{NClat}
\theta = \frac{1}{\pi} N a^{2} \ .
\ee
This shows that the extrapolation of interest is the
{\em Double Scaling Limit} (DSL), which leads simultaneously to the
continuum and to infinite volume at fixed $\theta$,
\be  \label{DSL}
\left. \begin{array}{cc}
a \to 0 & {\rm (continuum~limit)} \\
Na \to \infty & {\rm (infinite~volume)} 
\end{array} \right\}
\quad {\rm at} \quad  Na^{2} = const.
\ee
The requirement to take this simultaneous UV and IR limit
is again specific to NC field theory and related to the
UV/IR mixing. Any deviation from this balanced condition can lead
to a limit, which is not really NC (it may lead to $\theta \to 0$
or $\infty$). This can cause confusion, and in other regularisation 
schemes it is less obvious how to control this equilibrium
between UV and IR effects. 

In view of lattice simulations, the construction
of link variables in the NC plane appears difficult ---
note that a compact link variable should now be star-unitary,
$U \star U^{\dagger} = U^{\dagger} \star U = \uno$.
However, there is an exact mapping \cite{AMNS}
onto a $U(N)$ matrix model in one point, the so-called twisted
Eguchi-Kawai model \cite{GAO}, which is numerically tractable.

The DSL of $U(1)$ lattice gauge theory on a NC plane has been
studied in Refs.\ \cite{2dNCU1}. 
Wilson loops of {\em small} areas are real and they follow the
Gross-Witten area law of commutative 2d $U(N \to \infty )$ gauge theory
\cite{GroWit}.
Wilson loops of {\em large} area $A$ are shape dependent.
For most shapes they have a complex phase $A/ \theta$, which 
corresponds to an Aharonov-Bohm effect, if one formally interprets
$1/ \theta$ as a magnetic field across the plane. 
The same interpretation was suggested
long ago based on simple tree level considerations \cite{Peierls},
and it is also inherent in the famous Seiberg-Witten map \cite{SeiWit}
onto open bosonic string theory. \\

Let us return to the 4d $U(1)$ gauge theory and its formulation
on a regular lattice in the commutative plane, and a lattice as
constructed above 
in the NC plane. 
In each direction there is
(approximately) the same number $N$ of lattice sites. 

In the NC plane the self-interaction and the gauge transformations
are again non-local. As a consequence, there are {\em gauge invariant 
open Wilson lines,} in contrast to gauge theory in commutative spaces.
Due to the separation of their end-points they carry non-vanishing
momenta, so these open Wilson lines are suitable order parameters
to probe if translation and rotation symmetry is spontaneously broken
(we refer to the discrete translation and rotation symmetry on the 
lattice, resp.\ the continuous version in the DSL). In
Mean Field Approximation, active Poincar\'{e} symmetry will always 
appear broken because of the constant tensor $\Theta_{\mu\nu}$.
However, the question whether SSB really sets in 
or not is dynamical. We verified it by measuring the 
expectation values of open Wilson lines.

Numerical simulations at varying gauge coupling and different
lattice sizes $N$ revealed that a number of observables --- Wilson
loops and lines and their correlators, matrix spectra
as extent of a dynamically generated space --- stabilise 
(to a good approximation) for \cite{NCQED}
\be  \label{betasqrtN}
\beta \equiv \frac{1}{a g^{2}} \propto \sqrt{N} \ .
\ee
This defines the scale for the lattice spacing, up
to a proportionality constant, 
which corresponds to the choice of $\theta$.
Now the DSL is well-defined and it can be
explored by increasing $N$ and $\beta$, while keeping
the ratio $N / \beta^{2} = const$. 
Thus we extrapolate to the continuum and to infinite volume
at fixed $\theta$. 

\begin{figure}
\begin{center}
\includegraphics[angle=0,width=.67\linewidth]{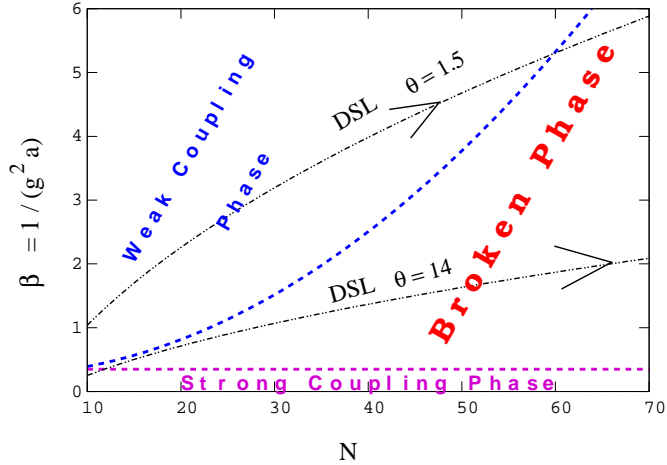}
\end{center}
\vspace*{-6mm}
\caption{\it The phase diagram for NC QED (pure $U(1)$ gauge theory)
based on numerical results \cite{NCQED}. 
At strong coupling ($\beta \lsim 0.35$),
and also at weak coupling, translation and rotation symmetry is intact.
In between, at moderate coupling, there is a phase where it is
spontaneously broken. Perturbation theory perceives the weak coupling
phase. The DSL (\ref{DSL}), however, leads to the
broken phase, which is therefore physically relevant.}
\label{phasedia}
\end{figure}

Our simulation results led to the
phase diagram shown in Figure \ref{phasedia}: at strong
coupling, $\beta \lsim 0.35$ we find for any
$N$ a symmetric phase, and at weak coupling again, but in between
(at moderate coupling) there is a phase of spontaneously 
broken Poincar\'{e} symmetry.\footnote{This observation agrees
qualitatively with later simulations of the 4d twisted Eguchi-Kawai model, 
which were performed with different motivations
\cite{4dTEK}.} Apparently
the strong fluctuations at small $\beta$ avoid SSB on one side,
and at weak coupling the correlation length exceeds the scale
of the mode of minimal energy, so that SSB
only occurs at moderate coupling. The transition line between the
weak and moderate coupling roughly follows the behaviour
$\beta_{c} \propto N^{2}$, which means that the DSL
curve (\ref{betasqrtN}) always ends up
in the {\em broken phase}. So this is the phase that actually
describes the NC continuum limit. Since it stabilises all observables 
that could be measured, the NC photon could be IR stable after all.

\begin{figure}
\begin{center}
\includegraphics[angle=270,width=.5\linewidth]{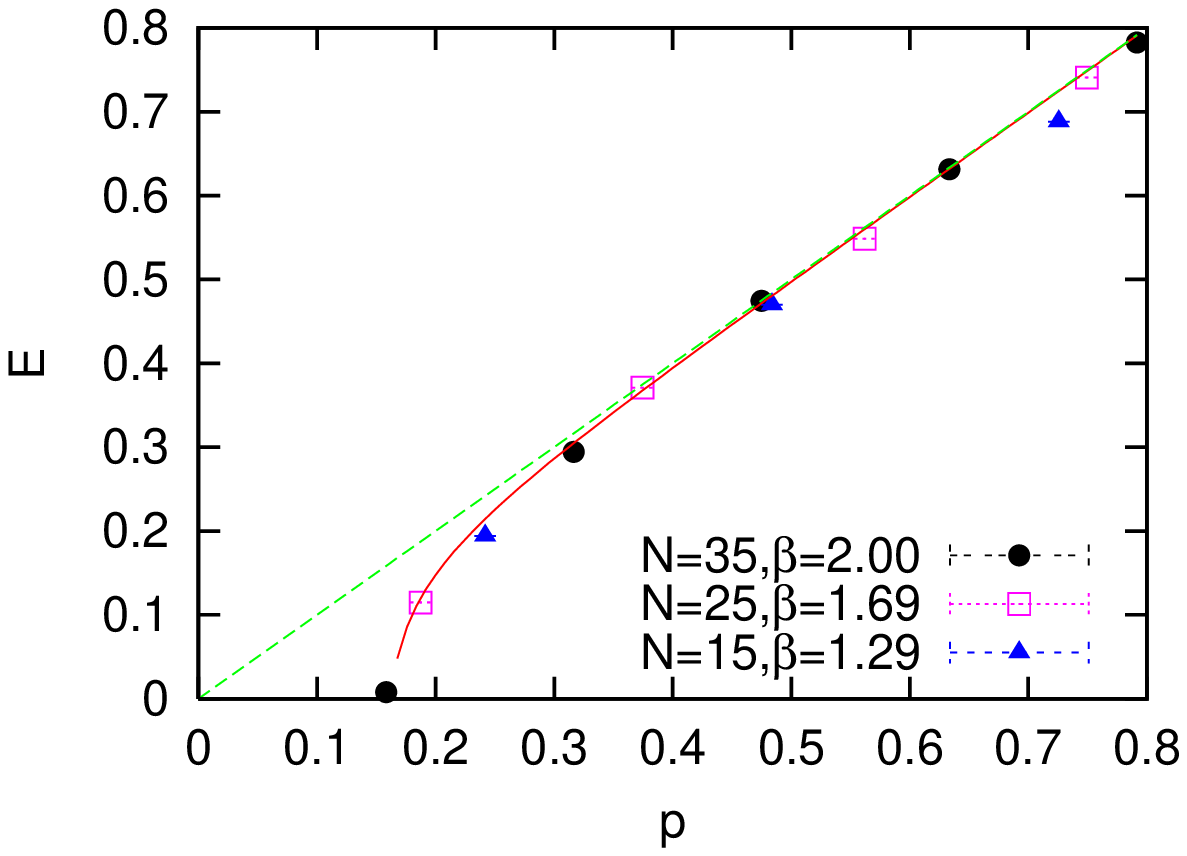}
\hspace*{-3mm}
\includegraphics[angle=270,width=.5\linewidth]{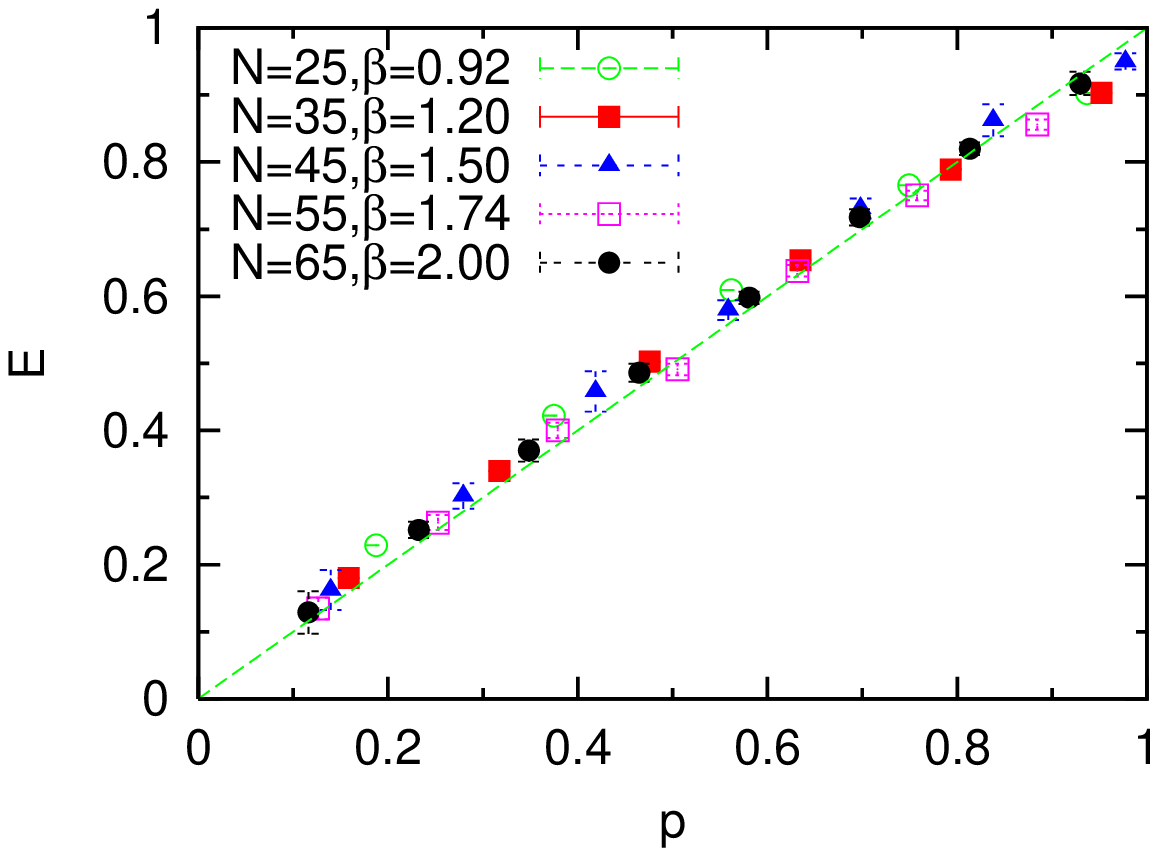}
\end{center}
\vspace*{-5mm}
\caption{\it The dispersion relation of the NC photon in the symmetric
phase (on the left) and in the phase of spontaneously 
broken 
Poincar\'{e} symmetry 
(on the right). The behaviour on the left matches the 1-loop result,
which is IR unstable, and which we also sketched in Figure \ref{photodisp}.
The behaviour on the right was measured in the broken phase, which
describes an IR stable NC continuum and infinite volume 
limit \cite{NCQED}.}
\label{NCphotodisp}
\end{figure}

Figure \ref{NCphotodisp} shows the numerically
measured dispersion relations of the photon in a NC space.
In the symmetric phase at weak coupling it is fully consistent 
with the IR instability found in perturbation theory, cf.\ 
Figure \ref{photodisp}.
Hence we agree with that calculation, but it does not
capture the phase which is really relevant for the NC photon.
The limit to a continuous, infinite NC space (the DSL)
takes it to the broken phase, and there Figure \ref{NCphotodisp} shows
that the linear dispersion relation is restored.

This linear dispersion relation
is consistent with {\em IR stability} and with the
interpretation of the photon as a Nambu-Goldstone boson of
the SSB of Poincar\'{e} symmetry (as it was also suggested in the
SME \cite{ColKos}). This dispersion relation $E(p)$ was
measured in the commutative plane, $p_{1}=p_{2} =0,$ $p = p_{3}$,
where the exponential decay of the correlation function of open Wilson
lines is clearly visible. 

Once IR stability holds, the next step would be to
elaborate also explicit non-perturbative 
results for a NC deformation of the dispersion in the DSL.
Such a study should address 
finite momentum components of the NC directions in the broken
phase dispersion relation. This could then
be confronted with phenomenological data to set bounds on $\theta$.
A NC deformation in the IR sector could also be highly
relevant for cosmic rays in view of the background radiation:
the GZK cutoff and the multi-TeV $\gamma$-puzzle both depend on the 
CMB (and radio background) photons.\footnote{Moreover
the CMB anisotropy (cf.\ footnote \ref{isoCMB} and Ref.\ \cite{Durrer}) 
might be affected by NC effects, if they were sizable in the early 
Universe \cite{NC-CMB}.} 

The results obtained already \cite{NCQED} suggest the important feature
that {\em the photon may survive in a NC world} after all, without 
the necessity of SUSY to cancel a negative IR divergence.
Hence NC QED {\em is} an option for new physics. Of course, the
version with a constant $\Theta$ is probably still just a mimic
of a realistic fuzzy quantum space.

\subsection{Analysis of GRB and blazar flare data}

From the preceding discussion 
we extract the message that
a theoretical basis for a deformed photon dispersion is
conceivable. Similar to Subsection 2.4 we now turn to a
pragmatic point of view and just assume some {\em effective}
deformation. We adapt an obvious parameterisation ansatz \cite{Camel}
for the photon energy $E$ at some momentum $\vec p$,
\be  \label{gammaspeed}
\vec p^{\, 2} = E^{2} \Big( 1 + \frac{E}{M} \Big) 
\quad \to \quad
v_{\gamma} (E) \simeq 1 - \frac{E}{M} \ .
\ee
$M$ is a very large mass, which emerges {\em somehow}, 
for instance from some kind of ``quantum gravity foam''.\footnote{A general 
discussion of possible power series corrections
to dispersion relations and their constraints based on requirements
like positivity, causality and rotation symmetry is given in 
Ref.\ \cite{Lehnert}. Ref.\ \cite{CaMac} studies the impact
on the thermodynamics of a photon gas.}
We keep track of $O(E/M)$, and $v_{\gamma}$ is the photon speed 
(which turns into $c=1$ for $M \to \infty$).

Ref.\ \cite{Ellis} performed a systematic analysis of the GRB data
collected by three satellites named SWIFT, BATSE and HETE\footnote{For
instance HETE distinguished for the GRB photons 4 energy bins and
detected them in time intervals of $6.4 \cdot 10^{-3}~{\rm s}$.}
and confronted the data with ansatz (\ref{gammaspeed}).
In fact there is a distinction in the time of arrival:
photons with higher energy tend to arrive earlier (which would naively 
suggest $M < 0$). However,
the analysis has to verify if this could also be explained without
new physics. We do not know if the photons of different energies in a GRB 
were emitted exactly at the same time 
(we repeat that we have to refer to hypotheses
and models for the origin of GRBs \cite{GBRfireball}). 
We consider photons in energy bins which differ by $\Delta E$.
Without new physics, the delay in the time of arrival can
be described as
\be  \label{delay}
\Delta t = d_{\rm source} \, (1 + z) \qquad (z ~ :~ {\rm redshift}) \ ,
\ee
where $d_{\rm source}$ is the parameter for the possible time-lag at 
the source. This delay is further enhanced on the way to Earth by the
factor $(1+z)$.

\begin{figure}
\begin{center}
\includegraphics[angle=0,width=1.\linewidth]{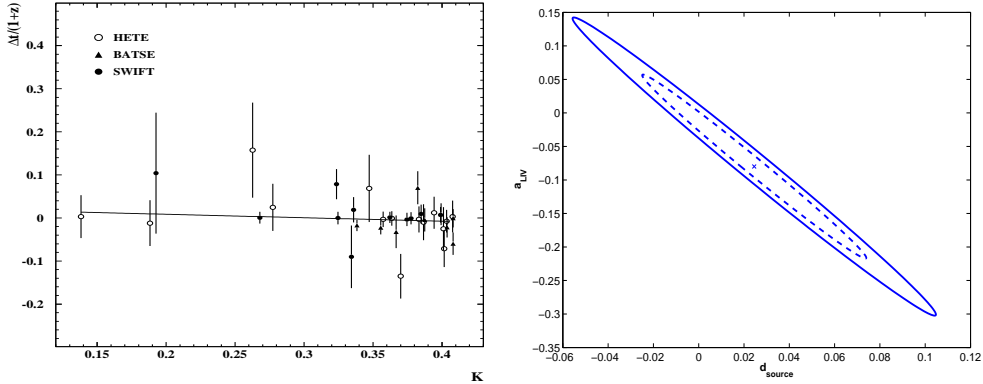}
\end{center}
\caption{\it Illustrations of the GRB data analysis in 
Ref.\ \cite{Ellis}: on the left the left-hand-side of eq.\ 
(\ref{LIVpara}) is plotted against the function $K(z)$.
The question is if the fit requires a non-zero
LIV parameter $a_{\rm LIV}$, i.e.\ if there is a slope with respect
to $K(z)$. On the right the window for the parameters
$d_{\rm source}$ (for a lag at the source) and $a_{\rm LIV}$ is shown
for $68 \, \%$ resp.\ $95 \, \%$ C.L. A strong deviation from
the LI scenario $a_{\rm LIV} =0$ is ruled out, which implies the
bound (\ref{Ellisbound}) (plots adapted from Ref.\ \cite{Ellis}).}
\label{GRBfit}
\end{figure}

Then the question is if the data can be fitted better when we add
a LIV parameter. This leads to the extended parameterisation ansatz
\be  \label{LIVpara}
\frac{\Delta t}{1 + z} = d_{\rm source} + a_{\rm LIV} K(z) \ , \qquad
a_{\rm LIV} := \frac{\Delta E}{M H_{0}} \ ,
\ee 
where $H_{0}$ is the Hubble parameter and $K(z)$ is a correction factor,
which is derived in Ref.\ \cite{Ellis}. Figure \ref{GRBfit} (on the left)
shows the left-hand-side of eq.\ (\ref{LIVpara}) on the vertical
axis, so the question is if the data\footnote{The error bars
were enhanced by the authors of Ref.\ \cite{Ellis} compared to the
satellite data, until they became compatible with a consistent picture.} 
are compatible with a constant, and therefore with LI. 
Although the best linear fit --- shown in the plot --- has a slight 
slope, it is obvious that a constant behaviour is not excluded.
The outcome of a precise evaluation is reproduced in Figure 
\ref{GRBfit} on the right. The inner (outer) depicted area 
captures the parameters $d_{\rm source\,}$ and $a_{\rm LIV}$ with 
$68 \, \%$ ($95 \, \%$) C.L. The crucial point is 
that $a_{\rm LIV}=0$ (absence of LIV) is included in the permitted
region, and $a_{\rm LIV}$ is strongly constrained,
which leads to the cautious conclusion
\be  \label{Ellisbound}
| M | > 1.4 \cdot 10^{25} ~{\rm eV} \approx 0.0012~M_{\rm Planck} \ 
\ee
with $95 \, \%$ C.L.
In some sense this is a negative result: once again the data
do not give evidence for new physics. However, it is impressive
that these observations allow for robust phenomenological
insight about an energy regime, which is not that far below 
the Planck scale.

We add that there are further analyses of GRB data 
along these lines, which do not find evidence for LIV either. 
In some studies the bound for $| M |$ is set even higher, 
{\it e.g.}\ Refs.\ \cite{BWHC,RMP} analysed the data of single
GRBs and concluded $|M| > 0.015 ~M_{\rm Planck}$ resp.\
$|M| > 0.0066 ~M_{\rm Planck}$. \\

Very similar investigation are also carried out based on
the $\gamma$-rays emitted in {\em blazar flares.} Recently the
MAGIC Collaboration used the Imaging Atmospheric \v Cerenkov Telescope 
to analyse a flare of Markarian 501
(cf.\ Subsection 3.1) and arrived even at 
$|M| > 0.022 ~M_{\rm Planck}$ \cite{Magic}.
The Gamma-ray Large Area Space Telescope (GLAST) \cite{GLAST}
is now going to search systematically for photons from GRBs 
and blazar flares in the energy interval of $8~{\rm keV}$ to 
$300~ {\rm GeV}$, which is promising in view of further insight 
even above these bounds \cite{Lamon}.
GLAST has arrived in orbit in June 2008.
 
If we assume some quantum space-time background, then
it is natural that it is subject to fluctuations, which
may affect the speed of photons also stochastically \cite{EFMMN}.
A consequence of this additional effect could be 
the broadening of a flare width, which increases with the
distance. 

At last we add that GRBs are also accompanied
by a burst of neutrinos, which may lead to additional prospects
for LIV tests \cite{LIV-GRB-nu}.

\subsection{Effective Field Theory and vacuum birefringence}

After discussing NC field theory as a LIV approach in Subsection 3.3,
we returned to the Effective Field Theory (EFT) framework in
Subsection 3.4. This is the theoretical framework that we had
already employed in Section 2. It includes the LIV terms, which
are not excluded by symmetries, in an energetic hierarchy.
This concept works successfully in many branches of physics ---
for instance in Chiral Perturbation Theory --- but the example of
NC field theory shows that it is not without alternative (in the 
latter case, the IR divergences are incompatible with EFT).

A systematic EFT analysis has to distinguish the mass dimensions
$n$ of the LIV operators. We summarise this approach briefly,
following the excellent discussion in Ref.\ \cite{JLM}.
We require rotation symmetry and gauge invariance,
and we introduce a time-like unit vector $u$. 
Then an extended QED Lagrangian includes the renormalisable terms
\bea
n=3 &:& b \bar \psi \gamma_{5} u_{\mu} \gamma^{\mu} \psi \\
n=4 &:& K_{\mu \nu \rho \sigma} F^{\mu \nu} F^{\rho \sigma} +
{\rm i} \bar \psi u_{\mu} u_{\nu} \gamma^{\mu} D^{\nu} 
(c + d \gamma_{5}) \psi \ , \quad (b,c,d ~:~ const.) \ . \nn
\eea
We encountered the $n=4$ terms before in Section 2: note that
$K_{\mu \nu \rho \sigma} = u_{[\mu} g_{\nu][\rho} u_{\sigma]}$
has to obey rotation symmetry, which
renders it equivalent to the term (\ref{Bsquare}), and
the last term above is equivalent to (\ref{LIVchiral})
(we proceed to covariant derivatives $D^{\nu}$).

Refs.\ \cite{MyPo,dim5LIV} explored the possibilities to add LIV operators
of mass dimension $n=5$, which are quadratic in the fields (of one kind)
and not reducible to lower dimensions or total derivatives by
applying the equations of motion. They assumed these (CPT odd) terms to be
$O( 1/ M_{\rm Planck})$ suppressed.\footnote{Of course, extending this rule to 
lower dimensions $n$ leads to another manifestation of the unsolved
hierarchy problem, which we commented on before.}
In the pure gauge action, Ref.\ \cite{MyPo} identified only one 
(independent) term,
\be
\frac{\xi}{M_{\rm Planck}} \, u^{\mu} F_{\mu \nu} ( u \cdot \partial )
(u_{\rho} \tilde F^{\rho \nu}) \ , \qquad
( \tilde F^{\rho \nu} := \frac{1}{2}
\epsilon^{\rho \nu \alpha \beta} F_{\alpha \beta} ) \ .
\ee
The LIV parameter $\xi$ affects the photon dispersion relation as
\be  \label{photondispxi}
E_{\pm}^{2} = k^{2} \pm \frac{\xi}{M_{\rm Planck}} k^{3} 
\qquad (k = | \vec k |) \ ,
\ee
where $\pm$ refers to left/right-handed circular 
polarisation.\footnote{Actually radiative corrections amplify the LIV 
term, leading to a fine-tuning problem
even for this dimension 5 term \cite{CriVac}.\\
A perturbative quantisation for the corresponding scalar model --- 
where the Lagrangian involves second order time derivatives --- 
was constructed in Ref.\ \cite{RUV}.}

The LIV modified fermion dispersion is similar, if we assume
$m \ll E \ll M_{\rm Planck}$ (where $m$ is the fermion mass).
Then we can choose the spinor field to be an approximate
eigenvector of $\gamma_{5}$ resp.\ of the chiral projectors,
and we obtain \cite{MyPo} \vspace*{-1mm}
\be  \vspace*{-1mm}
E^{2} = p^{2} + m^{2} + \eta_{\pm} \frac{p^{3}}{M_{\rm Planck}}
\qquad ( p = | \vec p \, |) \ ,
\ee
with $\eta_{\pm}^{\rm fermion} = - \eta_{\mp}^{\rm anti-fermion}$
\cite{JLMS}. Ref.\ \cite{JLM} discusses new QED phenomena, which
may occur due to this modification, including the transitions
\bea  \vspace*{-2mm}
e^{-} \to e^{-} + \gamma && {\mbox{vacuum~\v Cerenkov~radiation}} \nn \\
\gamma \to e^{+} + e^{-} && {\rm photon~decay} \nn \\
e^{\pm} \to e^{\pm} + e^{+} + e^{-} && {\rm pair~emission} \nn \\
\gamma \to 2 \gamma \ , \ 3 \gamma \ \dots && {\rm photon~splitting.}
\label{translist}
\eea
In Subsection 2.5 we already addressed vacuum \v Cerenkov radiation
and the photon decay. In contrast, lepton pair emission and photon splitting 
cannot be arranged for by different MAVs,
but by suitable SME parameters \cite{stabcaus,KosLeh}.
Moreover we have discussed in Subsection 3.1 that the photon absorption
$\gamma + \gamma \to e^{+} + e^{-}$ --- which occurs in standard QED as
well --- may be altered by LIV terms, in particular in view of the
energy threshold.

A careful analysis has to keep track of angular momentum conservation, which
can be achieved {\it e.g.}\ in the photon decay above the energy
threshold, since the momenta do not need to be collinear. Thus all the 
three LIV parameters, $\xi , \ \eta_{+}$ and $\eta_{-}\,$, are involved.
This decay (as well as the pair emission) would occur rapidly above 
threshold, so that the observed high energy photons, in particular from
the Crab Nebula and from blazars (with lower energy but longer time of 
flight, see Subsections 3.1 and 3.2) set combined constraints on these
three parameters \cite{JLM}.

If the GZK cutoff attenuates ultra high energy protons as predicted
by standard field theory, this should also generate high energy photons,
\be
p + \gamma_{\rm IR} \to \Delta^{+} \to p + \pi^{0} \to p + 
2 \gamma_{\rm UV} \ .
\ee
Since it is not observed that UHECRs are accompanied by such UV 
photons \cite{NoTopDown}, 
Ref.\ \cite{UHECRgamma1} concludes that they are also subject to
the expected attenuation, and derived a tiny bound for $| \xi | $,
for the case $\eta_{\pm}=0$. Ref.\ \cite{UHECRgamma2} generalised
this study to obtain combined constraints for $\xi$
and $\eta_{\pm}$, considering in particular the processes
$\gamma \to e^{+} + e^{-}$ and $\gamma \to 3 \gamma$.

Based on the Crab Nebula photons up to $\approx 50~{\rm TeV}$
one obtains extremely stringent constraints, if one relies on the 
standard assumption about the origin of its $\gamma$-radiation (synchrotron 
emission of high energy electrons and positrons, and inverse Compton
scattering of the synchrotron photons) \cite{JLM,MLCK}. \\

However, at this point we would like to be as economic as possible
with our assumptions. Without any hypothesis about the mechanism
causing the radiation, nor any assumption about the interactions in
a LIV modified QED (which is needed for the transitions in the list
({\ref{translist})) we can still address {\em purely kinematic 
effects.}\footnote{Also in this respect there are attempts to
go beyond the EFT approach and motivate the corresponding (parity
breaking) modification of the Maxwell eqs.\ from quantum gravity 
\cite{GaPu}.}

In this regard we have already discussed the energy dependence of the
photon speed in Subsection 3.4. In addition the dispersion relation
(\ref{photondispxi}) also implies a {\em helicity dependence} for 
plane waves of the same momentum $k$. 
Over a distance $\ell$ it leads to a relative delay $\Delta t_{\rm h}$,
\be
v_{\pm} = \frac{\partial E_{\pm}}{d k} 
\simeq 1 \pm \frac{\xi}{2 M_{\rm Planck}} k \quad \to \quad
\Delta t_{\rm h} \simeq \frac{|\xi | \, \ell \, k}{M_{\rm Planck}} \ .
\ee
It is a virtue of this formula that it is not affected by 
the uncertainty about a relative time-lag in the photon emission of
different momenta $k$ (unlike the studies reviewed in Subsection 3.4).
However, one needs of course the (mild) assumption that there is no relative
delay between the helicities at the source.

As we mentioned before,
Markarian 421 and 501 are the two known blazars which emitted
photons that arrived with energies $E \gsim 10~{\rm TeV}$.  The 
quasi-simultaneous arrival of both helicity states implies \cite{JLMS}
\be
| \xi | < 126 \quad \to \quad |M| > 0.008 \, M_{\rm Planck} \ , 
\ee
where $M$ refers to the notation used in Subsection 3.4.
This bound is in the same magnitude as those derived from the
energy dependence. \\

{\bf Birefringence :} The helicity dependent energy 
(\ref{photondispxi}) yields yet
another purely kinematic effect, which leads to a much stronger
bound on the parameter $\xi$ \cite{JLM}.\footnote{Further studies
address possible LIV birefringence effects in gravitational waves 
\cite{Petrov1} and on the electron self-energy \cite{Alfaro2}.}
In addition to the group velocity
$v_{\pm}$ there are also distinct phase velocities. 
As a consequence, the polarisation plane of {\em linearly} polarised 
$\gamma$-radiation of momentum $k$ is rotated after time $t$ by the angle
\be
\alpha (t) = \frac{t}{2} ( E_{+} - E_{-} ) \simeq 
\frac{\xi \, k^{2} t}{M_{\rm Planck}} \ .
\ee
Most useful for our purposes is the effect that a linearly polarised
burst covering momenta $k_{\rm min} \dots k_{\rm max}$ will
be depolarised. The net polarisation is lost completely when
the difference in the polarisation planes arrives at
\be
\alpha_{\max} - \alpha_{\min} = \frac{\xi \, t}{M_{\rm Planck}}
( k_{\rm max}^{2} - k_{\rm min}^{2} ) \approx \frac{\pi}{2} \ .
\ee
We mentioned in Section 2 that this effect has been used to
constrain the $n=3$ Chern-Simons term (\ref{CSterm}) \cite{CFJ},
and it has also been applied to establish bounds on $n=4$ 
LIV terms in the SME \cite{KosMew}. For the $n=5$ term
that we are discussing here, this birefringence property 
was studied by Gleiser and Kozameh \cite{bireastro}.
They referred to $\gamma$-rays originating from galaxies at
a distance of $300 ~ {\rm Mpc}$. For wave lengths ranging from
400~nm to 800~nm a linear polarisation was observed, within
$< 10^{\circ}$ deviation of the polarisation plane.
As a final highlight, let us quote that this yields a
{\em trans-Planckian} bound of
\be
| \xi | \lsim 10^{-4} 
\quad \to \quad |M| > 10^{4} \, M_{\rm Planck} \ .
\ee

Of course this refers to a specific order in the EFT 
framework.\footnote{Returning to NC field theory,
Ref.\ \cite{BireNC} discusses how UV/IR mixing (see Subsection 3.3) 
could affect birefringence. As for the laboratory,
Ref.\ \cite{bireaexp} proposed
a high precision experiment on vacuum birefringence.}
If we restrict the consideration to
CPT even terms --- returning to the concept
outlined in Section 2 --- such operators of dimension $n=5$ are 
excluded. In the EFT scheme we would then naturally proceed to 
$n=6$ operators causing LIV effects of $O(E^{2} / M_{\rm Planck}^{2})$.
For these terms helicity dependent photon velocities are still
possible, but not required anymore \cite{JLM}. 
Here the absence of vacuum \v Cerenkov radiation
for protons of energies up to $\sim 10^{20} ~{\rm eV}$
(see Subsection 2.5.2) becomes most powerful.
As we anticipated in footnote \ref{partonfn}, Ref.\ \cite{GaMo}
discusses a link between SME parameters
and hadronic effects by employing the parton model, though
this connection is not ultimately clarified.

%% file: conclu2.tex
Cosmic rays involve particles at the highest energies in the 
Universe.
Their flux is small at ultra high energies, which
makes measurements difficult, but there is on-going
progress in the techniques and extent of the corresponding observatories.
In addition to the ultra high energy, we can also take advantage 
of their extremely long path to extract
information, which is inaccessible otherwise --- that property
is even more powerful in view of conclusions about scenarios for new 
physics. Therefore cosmic rays provide a unique opportunity to 
access phenomenology at tremendous energies.
We have seen in the last Subsections that this can reach
out to energies only two or three orders of magnitude below 
the Planck scale, and for specific parameters one even
attains trans-Planckian magnitudes. Hence predictions 
in this regime have now arrived at {\em falsifiability,} which is a 
basic criterion for a scientific theory \cite{Popper}.
Although this only leaves a small window for possible LIV parameters,
there is a wide-spread consensus that this scenario deserves
attention --- over the past 10 years a preprint on this subject
appeared on average about once a week, and at least 52 
are top-cited.

We have discussed the challenges to verify
the GZK cutoff for UHECRs, and 
the multi-TeV $\gamma$-puzzle. The question is: 
why is the Universe --- with its Cosmic Background
Radiation (CBR) --- so transparent for protons, or heavier nuclei, 
with energies $E_{p} \gsim 6 \cdot 10^{19}~{\rm eV}$, and for 
photons with $E_{\gamma} \gsim 10 ~ {\rm TeV}$ ?
In both cases it is still controversial if the
puzzle really persists. 

The photopion production of UHECR protons with CMB photons
is actually a process of harmlessly low energy (around $200 ~ {\rm MeV}$)
in the rest frame of the proton. It can be studied in detail in our
laboratories. However, the application of the results
for the cross-section etc.
to UHECRs requires boosts with 
Lorentz factors of the order $\gamma \sim O(10^{11})$.
The applicability of such extreme Lorentz transformations
cannot be tested in the laboratories (which are limited to
$\gamma \lsim O(10^{5})$). Hence the viability of
LI under extreme boosts is a central
issue. If a puzzle persists at last, LIV provides a possible solution.
In Section 2 we discussed a theoretical basis of LIVs
(SME and in particular MAVs),
and its impact on cosmic rays physics --- a tiny LIV
could remove the GZK cutoff.

As an alternative theoretical framework for LIV,
Section 3 addressed field theory in a NC space.
This approach is fundamentally different from the effective
action concepts that we considered in Section 2.
We focused in particular on the dispersion
relation of the photon in a NC world. A possible distortion
of the photon dispersion can be probed best by the
observation of GRBs or blazar flares. We also discussed
the Effective Field Theory approach to LIV terms for cosmic
photons and stringent limits on leading terms,
based on momentum or helicity dependent photon velocities.

Up to now, {\em no experimental evidence for LIV has been
found} anywhere, despite more and more precise tests.
As examples we reviewed attempts with cosmic neutrinos
and photons (other high precision tests are performed in
atomic physics \cite{Matt,LIVboundsKos,Matt2}). 
However, a large number of LIV parameters
are conceivable. For some of them the tested
precision may still be insufficient
to rule out interesting scenarios for new physics,
which could shed completely new light on
the open issues mentioned above. If the GZK cutoff
is indeed confirmed --- as the data by the collaborations
HiRes and Pierre Auger suggest --- then this
could be viewed as {\em indirect evidence in favour of the
validity of LI even at $\gamma \sim O(10^{11})$.} \\

New projects to detect {\em UHECRs} are on the way.
As examples we mention Telescope Array (TA) \cite{TA} 
(another hybrid approach with an air shower array on 
ground and fluorescence telescopes),
the Extreme Universe Space Observatory (EUSO) \cite{EUSO} 
and the Orbiting Wide-angle Light-collectors (OWL) \cite{OWL} 
(both are going to search for air fluorescence light from satellites).
The general prospects for observatories in space are discussed
in Ref.\ \cite{spaceEAS}.

High energy {\em $\gamma$-rays} are detected by satellite-based
telescopes in the AGILE Observatory \cite{AGILE}
(Astro-rivelatore Gamma a Immagini Leggero), and 
in the GLAST Observatory \cite{GLAST}
(systematic observation of cosmic photons over a broad energy).
On ground there are suitable \v Cerenkov detector arrays, such as 
the High Energy Stereoscopic System
H.E.S.S.\ in Namibia \cite{HESS} and MAGIC \cite{Magic}
(we have already referred to GLAST and MAGIC in Section 3).
Ref.\ \cite{VHEGR} gives a recent overview.
Regarding GRBs, the Chinese-French Space-based 
multi-band astronomical Variable Objects Monitor mission (SVOM) 
\cite{SVOM} plans to detect about 80 bursts a year.

As for the upcoming search for high energy cosmic {\em neutrinos} 
(outlined {\it e.g.}\ in Ref.\ \cite{Wax}) we 
mention the IceCube Neutrino Observatory \cite{icecube}, which
detects \v Cerenkov light in the antarctic ice, as well as
ANITA (Antarctic Impulse Transient Antenna, balloon-borne),
ANTARES (Astronomy with a Neutrino Telescope and Abyss 
environmental RESearch, at sites deep in the Mediterranean Sea) and
NESTOR (Neutrino Extended Submarine Telescope with Oceanographic
Research) \cite{ANITAetc}.

Let us also mention a completely new technique, which is currently
under investigation \cite{LOPES}: the idea is to measure the synchrotron 
radiation of air shower leptons in the 
magnetic field of the Earth, in particular radio flashes,
with an array of 30 dipole antennae.

Finally we should emphasise once more
that the Pierre Auger Observatory is still at an early stage;
it is operating successfully since the beginning of the year 2004,
but the final \v Cerenkov detectors have been installed in
2008. Moreover that Collaboration plans another observatory
in the northern hemisphere \cite{PAO}.

This is a very active field of research with exciting open questions.
We may expect enlightening new data in the near future. 
They could lead to new insight in outstanding issues like LIV --- 
or to new puzzles and {\em perhaps} to evidence for new physics.

%% file: AGNhyp2.tex
In November 2007 the Pierre Auger Collaboration published
a spectacular report \cite{AGNHyp}, which is relevant for the 
issues that we discussed in Section 1. They analysed the arrival
directions of the UHECRs that they had detected so far.
We denote their observation as the {\em AGN Hypothesis}: \vspace*{3mm}\\
{\em The UHECRs directions are clustered and correlated with the 
locations of nearby Active Galactic Nuclei (AGN).} \\

This is clearly in contrast to the traditional assumption of an 
isotropic distribution (see Section 1.1).\footnote{For the characteristic
of AGN we refer to footnote \ref{AGNf}, and for an earlier
suggestion that they could be the UHECR source to Ref.\ 
\cite{AGNearly}. The question if they could be endowed with a sufficiently
powerful acceleration mechanism has recently been reconsidered in
Refs.\ \cite{AGNacc}.} 
The possibility that the 
AGN --- or other objects located next to them --- could be the UHECR
sources appears consistent because they refer to
AGN {\em nearby:} then the UHECR may be little deflected by 
magnetic fields, and attenuated only mildly by the CMB.
In this scenario one could assume also AGN far away
to emit UHECRs, but they would arrive with energies below
the GZK cutoff at Earth. Therefore the AGN Hypothesis agrees with
the validity of the GZK cutoff; it disfavours the
possibility that UHECRs indicate new physics.
In particular Ref.\ \cite{ScuSte08} deduces an upper bound
for the MAV difference $c_{\pi} - c_{p} < 5 \cdot 10^{-23}$.
On the other hand, it could be the starting point of
a new epoch of {\em cosmic ray astronomy.}

The Pierre Auger Collaboration started from the UHECR data
collected in the period from January 2004 to May 2006.
Their ana\-ly\-sis involved 3 parameters \\
\hspace*{5mm} $E_{\rm min}$ : threshold energy to be counted as an UHECR \\
\hspace*{5mm} $\psi $ ~~~  : angular interval around an UHECR direction \\
\hspace*{5mm} $R_{\rm max}$ : maximal distance to a ``nearby AGN''.

\noindent
Thanks to the analysis of hybrid events, which were detected
by the extended air shower {\em and} by the fluorescence
telescopes, the energy has a relatively small uncertainty around 
$22 \, \%$ (cf.\ Subsection 1.4), and the angular accuracy is 
about $1^{\circ}$ \cite{PAflux}.
$E_{\rm min}$ and $R_{\rm max}$ were
of course assumed in the order of the GZK cutoff and the
predicted super-GZK path length 
(eqs.\ (\ref{cutoff}) and (\ref{ellmax})).
The distance to the AGN is known from
the redshift of their $\gamma$-radiation.

These 3 parameters were tuned to the values
\be  \label{PApar}
(E_{\rm min}, \ \psi , \ R_{\rm max}) =
(5.6 \cdot 10^{19}~{\rm eV}, \ 3.1^{\circ}, \ 75 ~ {\rm Mpc}) \ ,
\ee
which gave maximal support for the AGN Hypothesis:
it captures 12 out of 15 UHECRs, whereas only $3.2$ would be expected 
for hypothetical isotropic rays
(although in that case one could question if 
one should keep these three parameters fixed).

As a clean test this Collaboration then analysed --- still with fixed 
parameters (\ref{PApar}) --- the UHECRs that they detected later,
in the period May 2006 to August 2007. In that case it captures
8 out of 13 UHECRs. So the ratio went down, but the number 8 
is still large compared to 2.8, which would be the expectation
for isotropic rays.

\begin{figure}
\begin{center}
\includegraphics[angle=0,width=1.\linewidth]{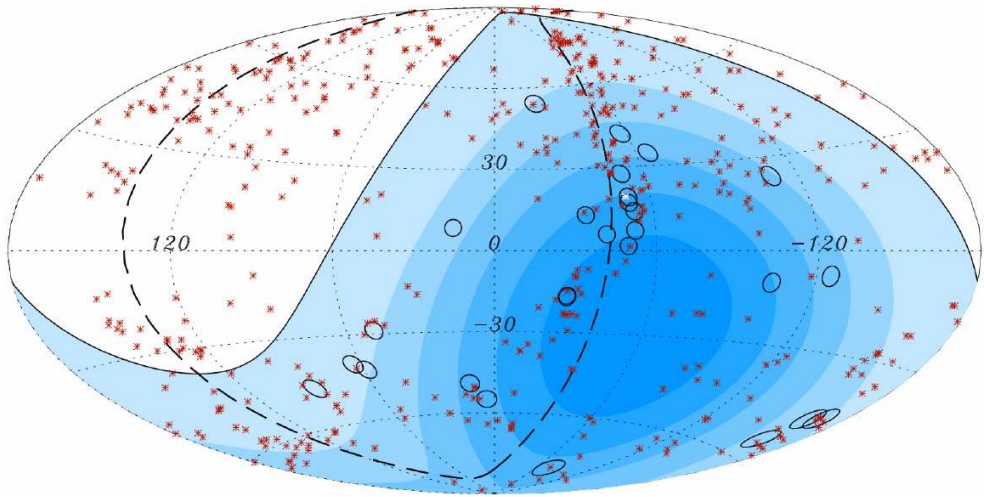}
\end{center}
\caption{\it A projection of the celestial sphere
(plot adapted from Ref.\ \cite{AGNHyp}):
the circles mark the arrival directions of the
27 UHECRs detected by the Pierre Auger Collaboration, and
the asterisks the locations of 472 known nearby
AGN. In particular for Centaurus A a white symbol is used.
The dashed line follows the super-galactic plane.
The darkness of the areas corresponds to the exposure 
(in the white area it vanishes).}
\label{skymap}
\end{figure}

The ensemble of UHECR directions observed by the Pierre Auger
Collaboration in the whole period January 2004 to August 2007
is shown in the Figure \ref{skymap}; it covers their 27 top UHECRs
($E > 5.7 \cdot 10^{19} ~ {\rm eV}$). 
The black circles mark the range within
$3.1^{\circ}$ of their directions of arrival, and the asterisks
correspond to the 472 known nearby AGN (up to the
distance $R_{\rm max}$ in eq.\ (\ref{PApar}), which corresponds
to redshift $z = 0.018$), 
according to the V\'eron-Cetty \& V\'eron Catalogue \cite{VCV}.
The alternative use of the Swift BAT Catalogue leads 
qualitatively to the same results \cite{AGNSwift}.
In further investigations
Ref.\ \cite{spiral} (Ref.\ \cite{Chile}) reports a correlation 
between these 27 UHECR directions and the locations of
spiral galaxies (extended nearby radiogalaxies)
by using the HIPASS Catalogue (several catalogues).

\subsection{Comments on the AGN Hypothesis}

The AGN Hypothesis consists actually of two parts,
and we consider it worthwhile discussing them separately: \vspace*{1mm}\\
\ (a) Clustering of the UHECR directions. \\
\ (b) Correlation of the clustered directions with 
locations of nearby 
AGN.

\vspace*{1mm}
\noindent
Part (b) depends on (a), but
the validity of (a) only could be an option. \\

Ref.\ \cite{AGNHyp} reports $99 \, \%$ C.L. for part (a)
and here only two tuned parameters
enter, $E_{\rm min}$ and $\psi$. Still this new UHECR statistics
is in the same magnitude as the total previous world data (and
the same holds for the total exposure so far, {\it i.e.}\ 
[detection time] $\times$ [area];
a comparative Table is given in Ref.\ \cite{Kampert}).
Hence one may wonder why this clustering did not become obvious earlier. 
Of course, previous data 
were also analysed with this respect, and some indication
for clustering was found in the world data available in
1995 (from Volcano Ranch, Haverah Park, Yakutsk and
AGASA at an early stage) \cite{Stanevaniso}. 
This inspired the AGASA Collaboration
to perform a careful analysis of the directions where they found
UHECRs \cite{AGASAaniso}. They could see a slight
clustering near the super-galactic plane,\footnote{The super-galactic
plane is a sheet-like structure that contains the local super-clusters.
It corresponds to the dash line in Figure
\ref{skymap}.} but it was weaker than in the previous
world data; otherwise they found consistence with isotropy. 
Not even their slight anisotropy
signal was confirmed by HiRes, which reported full 
isotropy \cite{anticlust}. 
A new analysis referring to an updated
definition of the super-galactic plane confirms a
clustering of the 27 top UHECR directions observed by
the Pierre Auger Collaboration, but for a similar number
of previous world data events (in the same energy magnitude)
clustering is still not evident \cite{Stanev08}.\\

With respect to part (b) of the AGN Hypothesis the statements
about the C.L.\ are less striking in Ref.\ \cite{AGNHyp}.
The analysis in Ref.\ \cite{PAO2} is more complete, for instance
regarding the variation of the parameters (\ref{PApar})
(see also Ref.\ \cite{LuLin}).
First support for this Hypothesis beyond the Pierre Auger
Observatory was provided by a new analysis of the
Yakutsk data \cite{AGNHyp2}, but {\em not} by the HiRes data
\cite{HiRes08}: in that case, only 2 out of 13 events fulfil
the correlation criterion for the parameter set (\ref{PApar}).
Ref.\ \cite{Dermer08} comments that the effect could reflect the 
property that the known AGN, as well as the detected nearby UHECR sources, 
are both preferentially located near the super-galactic plane.

Critics was also expressed in Ref.\ \cite{GTTT}, which starts from
the statement that the arriving flux of a source at 
distance $R$ should be $\propto 1/R^{2}$.
Hence according to the AGN Hypothesis
the {\em nearest} AGN should contribute most. 
Therefore Ref.\ \cite{GTTT} expects Centaurus A and Virgo --- 
which contain the two nearest AGN --- to contribute
each about 6 out of the 27 UHECR in the sky-map of Figure \ref{skymap},
if the AGN Hypothesis holds.
The direction of the galaxy Centaurus A does contribute, and (with
some tolerance about the angular window) it is compatible
with this expectation,\footnote{See Ref.\ \cite{KOT,AABMOPSY} for 
discussions of the high energy radiation mechanism of Centaurus A.}
but {\em nothing} is found in the direction of the Virgo 
cluster of galaxies (in Figure \ref{skymap}
it is located in the upper part near the dashed line, which
follows the super-galactic plane). That deficit
is the main point in the analysis of Ref.\ \cite{GTTT},
which claims that the data {\em disfavour} the AGN
Hypothesis with $99 \, \%$ C.L.
One might remark that this reasoning seems to refer to a scenario
where all AGN are UHECR sources with continuous luminosity. If 
AGN are episodic sources, 
then this argument might be circumvented.\footnote{Ref.\ \cite{Sigl}
compares the scenarios of continuous vs.\ episodic UHECR sources,
and arrives at different predictions for the energy dependence
of the $\gamma$ and $\nu$ flux.}

For sure the matter distribution in the GZK-sphere is inhomogeneous.
Therefore the AGN hypothesis could indeed lead to a new branch of
astronomy based on cosmic rays, as we mentioned in the beginning
of this appendix.
A possible correlation of the nearby matter distribution
with the magnetic field strength is another controversial issue 
\cite{Kron}. The way it could affect the UHECR propagation is 
discussed --- for various scenarios of the
magnetic field structure --- in Ref.\ \cite{GMT}.

Finally we add another interpretation put forward
in Ref.\ \cite{Fargi}, which concludes from the UHECR 
path limitation --- according to the analysis in Ref.\ \cite{AGNHyp}
--- that the primary particles at source could be mostly nuclei
heavier than protons. In fact the distance $R_{\rm max}$
to the nearby AGN is a bit shorter than required for
the maximal super-GZK path length of protons (eq. (\ref{ellmax})),
hence better suited for other nuclei.
A new calculation of the effective energy loss of UHECRs
in Ref.\ \cite{Dermer08} supports the conjecture that these
rays are mainly emitted as light nuclei, up to an atomic mass number
$A \approx 24$. 
This interpretation caused a surge of interest in the
CBR attenuation of nuclei, which has also been revisited 
in Refs.\ \cite{cosnuc,ABDOP}.\footnote{One expects UHECRs to be 
accompanied by a neutrino flux generated in CMB interactions 
(see {\it e.g.}\ Ref.\ \cite{Wax}), which also depends on the 
atomic mass $A$ \cite{AABMOPSY}.}
Ref.\ \cite{LGM} added that this could affect the magnitude to which
LIV parameters are ruled out if the AGN Hypothesis holds, since
possible LIV effects in the dispersion of $N$ bound particles
are amplified $\propto N^{2}$. On the other hand, for a given
energy the Lorentz factor between the FRW laboratory frame and
the ray rest frame shrinks $\propto 1/A$.

After a long journey, Ref.\ \cite{ABDOP} expects either protons or Fe 
nuclei (or perhaps sub-Fe nuclei) to survive as UHECR primary particle. 
In this regard, a statistical analysis of the penetration depth 
into the atmosphere (cf.\ eq.\ (\ref{Dmax})) suggests a mixed 
composition \cite{Unger}. However, also that observation appears 
to be consistent only in part with the HiRes data \cite{Fedorova}
(a comparison is shown in Figure 4 of Ref.\ \cite{Petrera}).

In any case, more statistics will be needed to identify the nature
of the UHECR primary particles directly. Above $10^{19}~{\rm eV}$
not much this is known in this respect. 
We may hope for new insight based on upcoming data, which
the Pierre Auger Collaboration intends to release in 2009.

%% file: short2.tex
For convenience we add an alphabetic list of the short-hand 
notations used in the text, with some hints where the corresponding
terms are introduced and used:


\begin{tabbing}
..................\= \kill
{\bf AGN}  \> Active Galactic Nuclei, see footnote \ref{AGNf}. \\
   \>  They play a central r\^{o}le in Appendix A.\\
{\bf C.L.}  \> Confidence Level (for statistical results).\\
{\bf CPT}  \> Charge conjugation combined with 
parity and time reflexion. \\ 
   \> Subsection 2.1 comments on its relation to LIV.\\
{\bf CMB}  \> Cosmic Microwave Background, see Subsection 1.2.\\
{\bf DSL}  \> Double Scaling Limit, see Subsection 3.3.2.\\
{\bf DSR}  \> Doubly Special Relativity, see Subsection 3.3.\\
{\bf EFT}  \> Effective Field Theory, see Subsection 3.5. \\
{\bf FRW}  \> Friedmann-Robertson-Walker \\
  \> (or Friedmann-Lema\^{\i}tre-Robertson-Walker) \\
  \> A line element in the FRW metrics is given by \\
  \> $ds^{2} = dt^{2} - a(t)^{2} (dr^{2} + \bar r^{2} d \Omega )$ 
  \ \ ($d\Omega$: spherical element)\\
  \> $a(t)$: scale factor of the Universe, $\bar r$ : covariant distance \\
  \> $\bar r = r$ (no curvature), $\bar r = R \sin (r/R)$ 
  ($\bar r = R \sinh (r/R)$) \\
  \> for positive (negative) curvature with radius $R$.\\ 
{\bf GRB} \> Gamma Ray Burst, see Subsection 3.2.\\
{\bf GRT} \> General Relativity Theory \\
{\bf GZK} \> Greisen-Zatsepin-Kuz'min \\
   \> The GZK cutoff for the energy of cosmic rays, \\ 
   \>  $E_{\rm GZK} \simeq 6 \cdot 10^{19} {\rm eV}$, 
       is discussed in Subsection 1.3.\\
{\bf LI} \> Lorentz Invariance (or Lorentz Invariant) \\
{\bf LIV} \> Lorentz Invariance Violation, addressed in Sections 2 and 3.\\
{\bf MAV} \> Maximal Attainable Velocity, introduced in Subsection 2.4.\\
{\bf NC} \> Non-Commutative, see Subsection 3.3. \\
{\bf SSB} \> Spontaneous Symmetry Breaking \\
{\bf SME} \> Standard Model Extension, see Subsection 2.3.\\
{\bf SRT} \> Special Relativity Theory \\
{\bf SUSY} \> Supersymmetry \\ 
{\bf UHECR} \> Ultra High Energy Cosmic Ray, a cosmic ray with energy \\
   \> near (or above) the magnitude of $E_{\rm GZK}$ 
\end{tabbing}

\vspace{-2mm}

For completeness we add that the literature often expresses
the huge energies involved in cosmic rays in units of {\bf PeV}, 
{\bf EeV}  and {\bf ZeV}, where the initial capital letters mean
Peta : $10^{15}$, Exa : $10^{18}$, Zetta : $10^{21}$.


%% file: acknow2.tex
\vspace*{1cm}

\noindent
{\small
{\bf Acknowledgements :} \
This review is based on talks presented at
Humboldt-Universit\"{a}t zu Berlin, Universidad Central
de Caracas, Universit\"{a}t Bern, IHEP (Beijing)
and Beijing University. I thank Marco Panero for his careful
reading of the draft version and many helpful comments. 
I am indebted to Maria Teresa Dova, Pedro Facal, 
Anamaria Font, Heinrich Rebel, Markus Risse, Mar\-kus Roth, 
Michael Unger and Luis Villase\~{n}or for instructive information,
and to Wolfram Schroers for technical support.
Jorge Alfaro, Aiyalam Balachandran, Robert Bluhm, Anosh Joseph,
Nikolaos Mavromatos, Richard Obousy, Albert Petrov and 
Alexander Sakharov attracted my attention
to relevant references. I also benefitted from inspiring remarks 
by Brett Altschul, Dorothea Bahns, Yu Jia, Hermann Kolanowski, 
Bo-Qiang Ma, Uwe-Jens Wiese, Ulli Wolff and Zhi Xiao. 
Finally I thank my collaborators for their contributions
to the joint works quoted here, and the authors of
Refs.\ \cite{hires,PAflux,HMR,PAO,Niteroi,DMBO06,MACRO,Ellis,AGNHyp}
for the kind permission to reproduce some of their plots.
}

%% file: cosmo2.bbl
\begin{thebibliography}{100}


\bibitem{Bergwitz} K.\ Bergwitz,
Habilitation Thesis, Braunschweig (1910).

\bibitem{Gockel} A.\ Gockel,
{\em Physik.\ Zeitschr.} {\bf 11} (1910) 280;
{\em ibid.} {\bf 12} (1911) 595.

\bibitem{Wulf} Th.\ Wulf,
{\em Physik.\ Zeitschr.} {\bf 11} (1910) 811.

\bibitem{Hess} V.F.\ Hess,
{\em Physik.\ Zeitschr.} {\bf 13} (1912) 1084.

\bibitem{Kol} W.\ Kolh\"{o}rster,
{\em Physik.\ Zeitschr.} {\bf 14} (1913) 1153. 

\bibitem{Hess2} V.F.\ Hess,
{\em Physik.\ Zeitschr.} {\bf 14} (1913) 610.

\bibitem{Auger1} P.\ Auger, R.\ Maze and T.\ Grivet-Mayer,
{\em Compt.\ Rend.\ Acad.\ Sci.} {\bf 206} (1938) 1721.

\bibitem{KolMatWeb} W.\ Kolh\"{o}rster, I.\ Mathes and E.\ Weber,
{\em Naturwissenschaften} {\bf 26} (1938) 576.

\bibitem{Auger2} P.\ Auger, P.\ Ehrenfest, R.\ Maze, 
J.\ Daudin and A.\ Fr\'eon Robley,
{\em Rev.\ Mod.\ Phys.} {\bf 11} (1939) 288. 

\bibitem{hires} http://www.cosmic-ray.org/

\bibitem{PAflux} Pierre Auger Collaboration (J.\ Abraham {\it et al.}),
{\em Phys.\ Rev.\ Lett.} {\bf 101} (2008) 061101. 

\bibitem{GMT} G.A.\ Medina-Tanco,
{\tt astro-ph/0607543.}

\bibitem{KASCADE} T.\ Antoni {\it et al.} (KASCADE Collaboration),
{\em Astropart.\ Phys.} {\bf 24} (2005) 1. 

\bibitem{Fermi} E.\ Fermi, {\em Phys.\ Rev.} {\bf 75} (1949) 1169.

\bibitem{Fermi2} W.I.\ Axelford, E.\ Leer and G.\ Skadron, 
{\em Proc.\ 15$^{\, th}$ Int.\ Cosmic Ray Conference} {\bf 11} (1977) 132.
G.F.\ Krymsky, {\em Dok.\ Acad.\ Nauk.\ USSR} {\bf 234} (1977) 1306.
Supernovae as cosmic ray sources were suggested already by
W.\ Baade and F.\ Zwicky, {\em Proc.\ Nat.\ Acad.\ Sci.} {\bf 20}
(1934) 259. Shock wave acceleration in supernova remnants is now the 
most popular scenario for {\em galactic} cosmic rays, see {\it e.g.}\ 
S.\ Gabici, {\tt arXiv:0811.0836 [astro-ph]}, which do, however, 
not include UHECRs.

\bibitem{Rev2002} A.V. Olinto,
{\em Phys.\ Rept.} {\bf 333} (2000) 329. 
L.\ Anchordoqui, T.\ Paul, S.\ Reucroft and J.\ Swain,
{\em Int.\ J.\ Mod.\ Phys.} {\bf A 18} (2003) 2229. 

\bibitem{pulsars} W.\ Bednarek and M.\ Bartosik,
{\em Astron.\ Astrophys.} {\bf 423} (2004) 405. 

\bibitem{FarBier} G.R.\ Farrar and P.L.\ Biermann,
{\em Phys.\ Rev.\ Lett.} {\bf 81} (1998) 3579. 

\bibitem{UHECRsources} A.\ Meli, J.K.\ Becker and J.J.\ Quenby, 
{\tt arXiv:0709.3031 [astro-ph].}

\bibitem{DarRu} A.\ Dar and A.\ De R\'{u}jula,
{\tt hep-ph/0606199.}

\bibitem{BhaSig} P.\ Bhattacharjee and G.\ Sigl,
{\em Phys.\ Rept.} {\bf 327} (2000) 109. 

\bibitem{topdown} V.S.\ Berezinsky and A.\ Vilenkin,
{\em Phys.\ Rev.} {\bf D 62} (2000) 083512. 
V.A.\ Kuz'min and I.I.\ Tkachev,
{\em Phys.\ Rept.} {\bf 320} (1999) 199. 
S.\ Sarkar and R.\ Toldra,
{\em Nucl.\ Phys.} {\bf B 621} (2002) 495. 

\bibitem{monopole} C.T.\ Hill,
{\em Nucl.\ Phys.} {\bf B 224} (1983) 469.
P.\ Bhattacharjee and G.\ Sigl,
{\em Phys.\ Rev.} {\bf D 51} (1995) 4079. 
J.J.\ Blanco-Pillado and K.D.\ Olum, 
{\em Phys.\ Rev.} {\bf D 60} (1999) 083001. 
E.\ Huguet and P.\ Peter,
{\em Astropart.\ Phys.} {\bf 12} (2000) 277. 

\bibitem{wimpzilla} D.J.H.\ Chung, E.W.\ Kolb, A.\ Riotto and I.I.\ Tkachev, 
{\em Phys.\ Rev.} {\bf D 62} (2000) 043508. 
H.\ Ziaeepour,  
{\em Astropart.\ Phys.} {\bf 16} (2001) 101. 
E.W.\ Kolb and A.A.\ Starobinsky,
{\em JCAP} {\bf 0707} (2007) 005. 

\bibitem{NoTopDown} Pierre Auger Collaboration 
(J.\ Abraham {\it et al.}), 
{\em Astropart.\ Phys.} {\bf 29} (2008) 243. 

\bibitem{NoTopDown2} M.\ Risse and P.\ Homola,
{\em Mod.\ Phys.\ Lett.} {\bf A 22} (2007) 749. 
M.\ Kachelrie\ss,
{\tt arXiv:0810.3017 [astro-ph].}

\bibitem{Hoer} J.R.\ H\"{o}randel,
{\tt arXiv:0803.3040 [astro-ph].}

\bibitem{Rub} G.I.\ Rubtsov {\it et al.},
{\em Phys.\ Rev.} {\bf D 73} (2006) 063009. 

\bibitem{PenWil} A.A.\ Penzias and R.W.\ Wilson,
{\em Astrophys.\ J.} {\bf 142} (1965) 419.

\bibitem{CMBrev} M.\ Kamionkowski and A.\ Kosowsky,
{\em Ann.\ Rev.\ Nucl.\ Part.\ Sci.} {\bf 49} (1999) 77. 
D.\ Samtleben, S.\ Staggs and B.\ Winstein,
{\em Annu.\ Rev.\ Nucl.\ Part.\ Sci.} {\bf 57} (2007) 245. 
W.\ Hu,
{\tt arXiv:0802.3688 [astro-ph].}

\bibitem{COBE} D.J.\ Fixsen, E.S.\ Cheng, J.M.\ Gales, J.C.\ Mather
and R.A.\ Shafer,
{\em Astrophys.\ J.} {\bf 473} (1996) 576. 

\bibitem{particledata} W.-M. Yao {\it et al.} (Particle Data Group),
{\em J.\ Phys.} {\bf G 33} (2006) 1.

\bibitem{Durrer} R.\ Durrer, ``The cosmic microwave background'',
Cambridge University Press, Cambridge U.K.\ (2008).

\bibitem{Grei} K.\ Greisen, {\em Phys.\ Rev.\ Lett.} {\bf 16}
(1966) 748.

\bibitem{ZK} G.T.\ Zatsepin and V.A. Kuz'min, 
{\em Sov.\ Phys. JETP Lett.} {\bf 4} (1966) 78.

\bibitem{Stecker} F.W.\ Stecker, {\em Phys.\ Rev.\ Lett.} {\bf 21}
(1968) 1016.

\bibitem{KKW} R.R.\ Wilson, {\em Phys.\ Rev.} {\bf 110} (1958) 1212.

\bibitem{sigmaK} J.P.\ Rachen and P.L.\ Biermann,
{\em Astron.\ Astrophys.} {\bf 272} (1993) 161. 

\bibitem{photodesint} F.W.\ Stecker and M.H.\ Salamon,
{\em Astrophys.\ J.} {\bf 512} (1999) 521. 

\bibitem{RevGZK}  V.\ Berezinsky, A.Z.\ Gazizov and S.I.\ Grigorieva,
{\em Phys.\ Rev.} {\bf D 74} (2006) 043005. 

\bibitem{HMR} D.\ Harari, S.\ Mollerach and E.\ Roulet,
{\em JCAP} {\bf 0611} (2006) 012. 

\bibitem{PAO} http://www.auger.org/

\bibitem{ee-pairdip} V.\ Berezinsky, A.Z.\ Gazizov and S.I.\ Grigorieva,
{\em Phys.\ Lett.} {\bf B 612} (2005) 147. 
R.\ Aloisio, V.\ Berezinsky, P.\ Blasi, 
A.\ Gazizov and S.\ Grigorieva,
{\em Astropart.\ Phys.} {\bf 27} (2007) 76. 

\bibitem{Linsley} J.\ Linsley, {\em Phys.\ Rev.\ Lett.} {\bf 10}
(1963) 146.

\bibitem{Suga} K.\ Suga, H.\ Sakuyama, S.\ Kawaguchi and T.\ Hara,
{\em Phys.\ Rev.\ Lett.} {\bf 27} (1971) 1604.

\bibitem{Sato} H.\ Sato,
{\em Prog.\ Theor.\ (Phys.\ Suppl.)} {\bf 163} (2006) 163.

\bibitem{Bird93} D.J.\ Bird {\it et al.}, 
{\em Phys.\ Rev.\ Lett.} {\bf 71} (1993) 3401.

\bibitem{Yakutsk} S.P.\ Knurenko {\it et al.} (Yakutsk Collaboration),
{\em Int.\ J.\ Mod.\ Phys.} {\bf A 20} (2005) 6878. 

\bibitem{Haverah} M.A.\ Lawrence, R.J.O.\ Reid and A.A.\ Watson,
{\em J.\ Phys.} {\bf G 17} (1991) 733.

\bibitem{Niteroi} C.E.\ Navia, C.R.A.\ Augusto and K.H.\ Tsui,
{\tt arXiv:0707.1896 [astro-ph].}

\bibitem{HiResPap} R.\ Abbasi {\it et al.} (HiRes Collaboration),
{\em Phys.\ Rev.\ Lett.} {\bf 100} (2008) 101101. 

\bibitem{Kampert} K.-H.\ Kampert,
{\em J.\ Phys.\ Conf.\ Ser.} {\bf 120} (2008) 062002. 

\bibitem{DMBO06} D.\ De Marco, P.\ Blasi and A.V.\ Olinto,
{\em JCAP} {\bf 0601} (2006) 002. 

\bibitem{MBO03} D.\ De Marco, P.\ Blasi and A.V.\ Olinto,
{\em Astropart.\ Phys.} {\bf 20} (2003) 53. 

\bibitem{Matthews} J.\ Matthews,
{\em Astropart.\ Phys.} {\bf 22} (2005) 387.

\bibitem{LaPlata} L.\ Anchordoqui, M.T.\ Dova, A.\ Mariazzi, 
T.\ McCauley, T.\ Paul, S.\ Reucroft and J.\ Swain,
{\em Annals Phys.} {\bf 314} (2004) 145. 
M.T.\ Dova,
{\tt astro-ph/0505583.}

\bibitem{Roth} M.\ Roth (for the Pierre Auger Collaboration),
{\tt arXiv:0706.2096 [astro-ph].}

\bibitem{Yamamoto} T.\ Yamamoto (for the Pierre Auger Collaboration),\\
{\tt arXiv:0707.2638 [astro-ph].}

\bibitem{FeldCou} G.J.\ Feldman and R.D.\ Cousins,
{\em Phys.\ Rev.} {\bf D 57} (1998) 3873. 


\bibitem{Ramond} P.\ Ramond, ``Field Theory: A Modern Primer''
(second edition), Frontiers in Physics V.74, Addison-Wesley (1980).

\bibitem{Shomer} A.\ Shomer,
{\tt arXiv:0709.3555 [hep-th].}

\bibitem{SMEgrav} V.A.\ Kosteleck\'{y},
{\em Phys.\ Rev.} {\bf D 69} (2004) 105009. 
Q.G.\ Bailey and V.A.\ Kosteleck\'{y},
{\em Phys.\ Rev.} {\bf D 74} (2006) 045001. 

\bibitem{Alfaro2} J.\ Alfaro,
{\em Phys.\ Rev.\ Lett.} {\bf 94} (2005) 221302. 

\bibitem{SSBinSME} R.\ Bluhm and V.A.\ Kosteleck\'{y},
{\em Phys.\ Rev.} {\bf D 71} (2005) 065008. 
R.\ Bluhm, S.-H.\ Fung and V.A.\ Kosteleck\'{y},
{\em Phys.\ Rev.} {\bf D 77} (2008) 065020. 

\bibitem{Alfaro1} J.\ Alfaro,
{\em Phys.\ Rev.} {\bf D 72} (2005) 024027. 

\bibitem{LIVprehisto} P.A.M.\ Dirac, 
{\em Nature} {\bf 168} (1951) 906.
J.D.\ Bjorken,
{\em Annals Phys.} {\bf 24} (1963) 174.
D.I.\ Blokhintsev and G.I.\ Kolerov, 
{\em Nuovo Cimento} {\bf 34} (1964) 163.

\bibitem{GLud} W.\ Pauli, {\em Nuovo Cimento} {\bf 6} (1957) 204.
G.\ L\"{u}ders, {\em Ann.\ Phys.} {\bf 2} (1957) 1.

\bibitem{Jost} R.\ Jost, {\em Helv.\ Phys.\ Acta} {\bf 30} (1957) 409.

\bibitem{SpinCPT} J.\ Fr\"{o}hlich, 
{\tt arXiv:0801.2724 [math-ph].} 

\bibitem{Greenberg} O.W.\ Greenberg,
{\em Phys.\ Rev.\ Lett.} {\bf 89} (2002) 231602.

\bibitem{ColGla} S.R.\ Coleman and S.L.\ Glashow,
{\em Phys.\ Rev.} {\bf D 59} (1999) 116008. 

\bibitem{Matt} D.\ Mattingly,
{\em Living Rev.\ Rel.} {\bf 8} (2005) 5. 

\bibitem{CPTtest} M.\ Antonelli and
G.\ D'Ambrosio, \\ 
{\tt http://pdg.lbl.gov/2008/reviews/cpt\_s011254.pdf}

\bibitem{magmo} R.\ Bluhm, V.A.\ Kosteleck\'y and N.\ Russell,
{\em Phys.\ Rev.\ Lett.} {\bf 79} (1997) 1432. 
H.\ Belich, L.P.\ Colatto, T.\ Costa-Soares, 
J.A.\ Helay\"{e}l-Neto and M.T.D.\ Orlando, 
{\tt arXiv:0806.1253 [hep-th].}

\bibitem{Fischbach} E.\ Fischbach, M.P.\ Haugan, 
D.\ Tadic and H.-Y.\ Cheng,
{\em Phys.\ Rev.} {\bf D 32} (1985) 154.

\bibitem{JLM} T.\ Jacobson, S.\ Liberati and D.\ Mattingly,
{\em Annals Phys.} {\bf 321} (2006) 150. 

\bibitem{LIVboundsKos} V.A.\ Kosteleck\'{y} and N.\ Russell,
{\tt arXiv:0801.0287 [hep-ph].}

\bibitem{Matt2} D.\ Mattingly,
{\tt arXiv:0802.1561 [gr-qc].}

\bibitem{IKKT} N.\ Ishibashi, H.\ Kawai, Y.\ Kitazawa and A.\ Tsuchiya,
{\em Nucl.\ Phys.} {\bf B 498} (1997) 467. 

\bibitem{10dIIB} J.\ Nishimura and G.\ Vernizzi,
{\em JHEP} {\bf 0004} (2000) 015; 
{\em Phys.\ Rev.\ Lett.} {\bf 85} (2000) 4664. 
J.\ Ambj\o rn, K.N.\ Anagnostopoulos, W.\ Bietenholz,
T. Hotta and J. Nishimura, 
{\em JHEP} {\bf 0007} (2000) 011. 

\bibitem{4dIIB} P.\ Bialas, Z.\ Burda, B.\ Petersson and J.\ Tabaczek,
{\em Nucl.\ Phys.} {\bf B 592} (2001) 391. 
J.\ Ambj\o rn, K.N.\ Anagnostopoulos, W.\ Bietenholz and F.\ Hofheinz,
{\em Phys.\ Rev.} {\bf D 65} (2002) 086001. 

\bibitem{Obousy} R.K.\ Obousy and G.\ Cleaver,
{\tt arXiv:0805.0019 [gr-qc].}

\bibitem{AlfPal} J.\ Alfaro and G.\ Palma,
{\em Phys.\ Rev.} {\bf D 65} (2002) 103516; 
{\em Phys.\ Rev.} {\bf D 67} (2003) 083003. 

\bibitem{UNAM} D.\ Sudarsky, L.\ Urrutia and H.\ Vucetich,
{\em Phys.\ Rev.\ Lett.} {\bf 89} (2002) 231301; 
{\em Phys.\ Rev.} {\bf D 68} (2003) 024010. 
G.\ Amelino-Camelia, C.\ L\"{a}mmerzahl, A.\ Macias and H.\ M\"{u}ller,
{\em AIP Conf.\ Proc.} {\bf 758} (2005) 30. 

\bibitem{Lust} D.\ L\"{u}st, S.\ Stieberger and T.R.\ Taylor,
{\tt arXiv:0807.3333 [hep-th].}

\bibitem{LHCsafty} J.R.\ Ellis, G.\ Giudice, M.L.\ Mangano, 
I.\ Tkachev and U.\ Wiedemann,
{\em J.\ Phys.} {\bf G 35} (2008) 115004. 

\bibitem{ColKos} D.\ Colladay and V.A.\ Kosteleck\'y,
{\em Phys.\ Rev.} {\bf D 58} (1998) 116002. 

\bibitem{Bluhm} R.\ Bluhm,
{\em Lect.\ Notes Phys.} {\bf 702} (2006) 191 {\tt [hep-ph/0506054].}

\bibitem{KosSam} V.A.\ Kosteleck\'y and S.\ Samuel,
{\em Phys.\ Rev.} {\bf D 39} (1989) 683.

\bibitem{Kosphilo} V.A.\ Kosteleck\'y,
{\tt hep-ph/0104227.}

\bibitem{CFJ} S.M.\ Carroll, G.B.\ Field and R.\ Jackiw,
{\em Phys.\ Rev.} {\bf D 41} (1990) 1231.

\bibitem{Petrov2} M.\ Gomes, J.R.\ Nascimento, E.\ Passos, 
A.Yu.\ Petrov and A.J.\ da Silva,
{\em Phys.\ Rev.} {\bf D 76} (2007) 047701. 
T.\ Mariz, J.R.\ Nascimento, A.Yu.\ Petrov, L.Y.\ Santos and 
A.J.\ da Silva,
{\em Phys.\ Lett.} {\bf B 661} (2008) 312. 

\bibitem{IHEP} J.\ Gamboa, J.\ Lopez-Sarrion and A.P.\ Polychronakos,
{\em Phys.\ Lett.} {\bf B 634} (2006) 471. 
J.-Q.\ Xia, H.\ Li, X.-L.\ Wang and X-M.\ Zhang,
{\em Astron.\ Astrophys.} {\bf 483} (2008) 715. 

\bibitem{ColKosCPT} D.\ Colladay and V.A.\ Kosteleck\'y,
{\em Phys.\ Rev.} {\bf D 55} (1997) 6760. 

\bibitem{stabcaus} V.A.\ Kosteleck\'{y} and R.\ Lehnert,
{\em Phys.\ Rev.} {\bf D 63} (2001) 065008. 

\bibitem{KosPot} V.A.\ Kosteleck\'y and R.\ Potting,
{\em Phys.\ Rev.} {\bf D 51} (1995) 3923. 

\bibitem{LIVfinetune} J.\ Collins, A.\ Perez, D.\ Sudarsky, 
L.\ Urrutia and H.\ Vucetich,
{\em Phys.\ Rev.\ Lett.} {\bf 93} (2004) 191301. 
J.\ Collins, A.\ Perez and D.\ Sudarsky,
{\tt hep-th/0603002.}

\bibitem{MAVtestinEAS} E.E.\ Antonov, L.G.\ Dedenko, A.A.\ Kirillov, 
T.M.\ Roganova, G.F.\ Fedorova and E.Yu.\ Fedunin,
{\em JETP Lett.} {\bf 73} (2001) 446. 

\bibitem{Altschulbeta} B.\ Altschul,
{\tt arXiv:0805.0781 [hep-ph].}

\bibitem{ColGlaPLB} S.R.\ Coleman and S.L.\ Glashow,
{\em Phys.\ Lett.} {\bf B 405} (1997) 249. 

\bibitem{SatoTati} H.\ Sato and T.\ Tati,
{\em Prog.\ Theor.\ Phys.} {\bf 47} (1972) 1788.
D.A.\ Kirzhnits and V.A.\ Chechin,
{\em Yad.\ Fiz.} {\bf 15} (1972) 1051.

\bibitem{Mofa} O.\ Bertolami and C.S.\ Carvalho,
{\em Phys.\ Rev.} {\bf D 61} (2000) 103002. 
J.W.\ Moffat,
{\em Int.\ J.\ Mod.\ Phys.} {\bf D 12} (2003) 1279. 

\bibitem{GaMo} O.\ Gagnon and G.D.\ Moore,
{\em Phys.\ Rev.} {\bf D 70} (2004) 065002. 

\bibitem{AltschulMEVpi} B.\ Altschul, 
{\em Phys.\ Rev.} {\bf D 77} (2008) 105018. 

\bibitem{DubTin} S.L.\ Dubovsky and P.G.\ Tinyakov,
{\em Astropart.\ Phys.} {\bf 18} (2002) 89. 

\bibitem{neutralprimary} A.M.\ Atoyan and C.D.\ Dermer,
{\em Astrophys.\ J.} {\bf 586} (2003) 79. 
P.G.\ Tinyakov and I.I.\ Tkachev,
{\em J.\ Exp.\ Theor.\ Phys.} {\bf 106} (2008) 481. 


\bibitem{neutralprimarygamma}
G.\ Gelmini, O.\ Kalashev and D.V.\ Semikoz,
{\em Astropart.\ Phys.} {\bf 28} (2007) 390; 
{\em JCAP} {\bf 0711} (2007) 002. 

\bibitem{neutralprimarynu} J.\ Bordes, H.-M.\ Chan, J.\ Faridani, 
J.\ Pfaudler and S.T.\ Tsou,
{\em Astropart.\ Phys.} {\bf 8} (1998) 135. 
P.\ Jain, D.W.\ McKay, S.\ Panda and J.P.\ Ralston,
{\em Phys.\ Lett.} {\bf B 484} (2000) 267. 

\bibitem{neutralprimarySUSY} D.J.H.\ Chung, G.R.\ Farrar and 
E.W.\ Kolb,
{\em Phys.\ Rev.} {\bf D 57} (1998) 4606. 

\bibitem{KosMew04} V.A.\ Kosteleck\'{y} and M.\ Mewes,  
{\em Phys.\ Rev.} {\bf D 70} (2004) 076002. 

\bibitem{numixnopureMAV} P.\ Lipari and M.\ Lusignoli,
{\em Phys.\ Rev.} {\bf D 60} (1999) 013003. 

\bibitem{K2Ka} G.L.\ Fogli, E.\ Lisi, A.\ Marrone and G.\ Scioscia,
{\em Phys.\ Rev.} {\bf D 60} (1999) 053006. 

\bibitem{MACRO} G.\ Battistoni {\it et al.},
{\em Phys.\ Lett.} {\bf B 615} (2005) 14. 

\bibitem{K2Kb}M.C.\ Gonzalez-Garcia and M.\ Maltoni,
{\em Phys.\ Rev.} {\bf D 70} (2004) 033010. 

\bibitem{Auerbach} L.B.\ Auerbach {\it et al.} (LSND Collaboration),
{\em Phys.\ Rev.} {\bf D 72} (2005) 076004. 

\bibitem{XiaoMa} Z.\ Xiao and B.-Q.\ Ma, 
{\tt arXiv:0805.2012 [hep-ph].}


\bibitem{crab} T.~Tanimori~{\it et~al.}~(CANGAROO Collaboration),\\
{\tt astro-ph/9710272.}
S.V.\ Godambe {\it et al.},
{\tt arXiv:0804.1473 [astro-ph].}

\bibitem{HEGRA} F.A.\ Aharonian {\it et al.} (HEGRA Collaboration),
{\em Astron.\ Astrophys.} {\bf 349} (1999) 11.

\bibitem{Mkn421} F.\ Krennrich {\it et al.},
{\tt astro-ph/9808333.}

\bibitem{Camel} G.\ Amelino-Camelia, J.R.\ Ellis, N.E.\ Mavromatos, 
D.V.\ Nanopoulos and S.\ Sarkar,
{\em Nature} {\bf 393} (1998) 763. 

\bibitem{GammaTeV} G.\ Amelino-Camelia and T.\ Piran,
{\em Phys.\ Lett.} {\bf B 497} (2001) 265; 
{\em Phys.\ Rev.} {\bf D 64} (2001) 036005. 

\bibitem{Mkn501prob} R.J.\ Protheroe and H.\ Meyer,
{\em Phys.\ Lett.} {\bf B 493} (2000) 1. 

\bibitem{Mkn501noprob} A.K.\ Konopelko, A.\ Mastichiadis, 
J.G.\ Kirk, O.C.\ de Jager and F.W.\ Stecker,
{\em Astrophys.\ J.} {\bf 597} (2003) 851. 
F.W.\ Stecker,
{\em Astropart.\ Phys.} {\bf 20} (2003) 85. 
A.\ Franceschini, G.\ Rodighiero and M.\ Vaccari,
{\tt arXiv:0805.1841 [astro-ph].}

\bibitem{kappadisp} G.\ Amelino-Camelia,
{\em New J.\ Phys.} {\bf 6} (2004) 188. 

\bibitem{SteGla} F.W.\ Stecker and S.L.\ Glashow,
{\em Astropart.\ Phys.} {\bf 16} (2001) 97. 

\bibitem{JLMth} T.\ Jacobson, S.\ Liberati and D.\ Mattingly,
{\em Phys.\ Rev.} {\bf D 67} (2003) 124011. 

\bibitem{GBRfireball} P.\ M\'{e}sz\'{a}ros,
{\em Annual Rev.\ of Astronomy and Astrophysics}
{\bf 40} (2002) 137.
T.\ Piran,
{\em Rev.\ Mod.\ Phys.} {\bf 76} (2004) 1143. 
Y.-Z.\ Fan and T.\ Piran, 
{\em Front.\ Phys.\ China} {\bf 3} (2008) 306. 
V.V.\ Sokolov,
{\tt arXiv:0805.3262 [astro-ph].}

\bibitem{Wax} E.\ Waxman,
{\em Phil.\ Trans.\ Roy.\ Soc.\ Lond.} {\bf A 365} (2007) 1323. 

\bibitem{GRBasUHECRsource} C.D.\ Dermer,
{\tt arXiv:0711.2804 [astro-ph].}

\bibitem{LaSilla} http://www.eso.org

\bibitem{bireastro} R.J.\ Gleiser and C.N.\ Kozameh,
{\em Phys.\ Rev.} {\bf D 64} (2001) 083007. 

\bibitem{Tina} C.\ L\"{a}mmerzahl, A.\ Macias and H.\ M\"{u}ller,
{\em Phys.\ Rev.} {\bf D 71} (2005) 025007. 
T.\ Kahniashvili, G.\ Gogoberidze and B.\ Ratra,
{\em Phys.\ Lett.} {\bf B 643} (2006) 81. 
F.R.\ Klinkhamer and M.\ Risse,
{\em Phys.\ Rev.} {\bf D 77} (2008) 016002. 

\bibitem{VSL} J.\ Magueijo,
{\em Rept.\ Prog.\ Phys.} {\bf 66} (2003) 2025. 

\bibitem{DSR} G.\ Amelino-Camelia,
{\em Nature} {\bf 418} (2002) 34. 

\bibitem{Klink} F.R.\ Klinkhamer,
{\em JETP Lett.} {\bf 86} (2007) 73. 

\bibitem{Madore} J.\ Madore, 
{\em Class.\ and Quant.\ Grav.} {\bf 9} (1992) 69.

\bibitem{numfuz} T.\ Azuma, S.\ Bal, K.\ Nagao and J.\ Nishimura,
{\em JHEP} {\bf 0405} (2004) 005. 
X.\ Martin, 
{\em JHEP} {\bf 0404} (2004) 077. 
M.\ Panero,
{\em JHEP} {\bf 0705} (2007) 082. 
D.\ O'Connor and B.\ Ydri,
{\em JHEP} {\bf 0611} (2006) 016. 
J.\ Medina, W.\ Bietenholz and D.\ O'Connor,
{\em JHEP} {\bf 04} (2008) 041. 
W.\ Bietenholz,
{\tt arXiv:0808.2387 [hep-th].}

\bibitem{Sny} H.S.\ Snyder,
{\em Phys.\ Rev.} {\bf 71} (1971) 38.

\bibitem{Rafael} F.\ Dowker, J.\ Henson and R.D.\ Sorkin, 
{\em Mod.\ Phys.\ Lett.} {\bf A 19} (2004) 1829. 

\bibitem{habi} W.\ Bietenholz, 
{\em Fortsch.\ Phys.} {\bf 56} (2008) 107. 

\bibitem{braneworld} D.B.\ Kaplan,
{\em Phys.\ Lett.} {\bf B 288} (1992) 342. 
W.\ Bietenholz, A.\ Gfeller and U.-J.\ Wiese,
{\em JHEP} {\bf 0310} (2003) 018. 

\bibitem{DFR} S.\ Doplicher, K.\ Fredenhagen and J.E.\ Roberts,
{\em Phys.\ Lett.} {\bf B 331} (1994) 39;
{\em Commun.\ Math.\ Phys.} {\bf 172} (1995) 187. 

\bibitem{uncert1d} D.\ Amati, M.\ Ciafaloni and G.\ Veneziano,
{\em Phys.\ Lett.} {\bf B 216} (1989) 41.
K.\ Konishi, G.\ Paffuti and P.\ Provero,
{\em Phys.\ Lett.} {\bf B 234} (1990) 276.
M.\ Maggiore,
{\em Phys.\ Lett.} {\bf B 304} (1993) 65. 

\bibitem{uncert1dang} C.\ Bambi and K.\ Freese,
{\em Class.\ Quant.\ Grav.} {\bf 25} (2008) 195013.

\bibitem{Suda} D.\ Sudarsky,
{\em Int.\ J.\ Mod.\ Phys.} {\bf D 17} (2008) 425. 

\bibitem{DougNek} M.R.\ Douglas and N.A.\ Nekrasov,
{\em Rev.\ Mod.\ Phys.} {\bf 73} (2001) 977. 

\bibitem{Bal} A.P.\ Balachandran, T.R.\ Govindarajan, G.\ Mangano, 
A.\ Pinzul, B.A.\ Qureshi and S.\ Vaidya,
{\em Phys.\ Rev.} {\bf D 75} (2007) 045009. 
A.P.\ Balachandran, A.\ Pinzul, B.A. Qureshi and S.\ Vaidya,
{\em Phys.\ Rev.} {\bf D 77} (2008) 025020. 

\bibitem{NCQED-CPT} M.M.\ Sheikh-Jabbari,
{\em Phys.\ Rev.\ Lett.} {\bf 84} (2000) 5265. 

\bibitem{Mozo} L.\ \'{A}lvarez-Gaum\'{e} and M.A.\ V\'{a}zquez-Mozo,
{\em Nucl.\ Phys.} {\bf B 668} (2003) 293. 

\bibitem{CPTinNC} E.\ Akofor, A.P.\ Balachandran, S.G.\ Jo and A.\ Joseph,
{\em JHEP} {\bf 0708} (2007) 045. 

\bibitem{Filk} T.\ Filk,
{\em Phys.\ Lett.} {\bf B 376} (1996) 53.

\bibitem{MRS} S.\ Minwalla, M.\ Van Raamsdonk and N.\ Seiberg,
{\em JHEP} {\bf 0002} (2000) 020. 

\bibitem{NCSM} X.\ Calmet, B.\ Jurco, P.\ Schupp and J.\ Wess,
{\em Eur.\ Phys.\ J.} {\bf C 23} (2002) 363. 

\bibitem{KostelNC} S.M.\ Carroll, 
J.A.\ Harvey, V.A.\ Kosteleck\'{y}, C.D.\ Lane and T.\ Okamoto,
{\em Phys.\ Rev.\ Lett.} {\bf 87} (2001) 141601. 

\bibitem{SeiWit} N.\ Seiberg and E.\ Witten,
{\em JHEP} {\bf 9909} (1999) 032. 
For earlier work along these lines, see {\it e.g.}
A.\ Abouelsaood, C.G.\ Callan, C.R.\ Nappi 
and S.A.\ Yost,
{\em Nucl.\ Phys.} {\bf B 280} (1987) 599.

\bibitem{stringuncert} M.\ Li and T.\ Yoneya
{\em Phys.\ Rev.\ Lett.} {\bf 78} (1997) 1219. 
T.\ Yoneya,
{\em Prog.\ Theor.\ Phys.} {\bf 103} (2000) 1081. 

\bibitem{stringuncertnum} J.\ Ambj\o rn, K.N.\ Anagnostopoulos, 
W.\ Bietenholz, T. Hotta and J. Nishimura,
{\em JHEP} {\bf 0007} (2000) 013. 

\bibitem{Iorio} A.\ Iorio,
{\em J.\ Phys.\ (Conf.\ Ser.)} {\bf 67} (2007) 012008. 

\bibitem{Moyal} H.J.\ Groenewold,
{\em Physica} {\bf 12} (1946) 405.
J.E.\ Moyal,
{\em Proc.\ Cambridge Phil.\ Soc.} {\bf 45} (1949) 99.

\bibitem{Szabo} R.J.\ Szabo,
{\em Phys.\ Rept.} {\bf 378} (2003) 207. 

\bibitem{MST} A.\ Matusis, L.\ Susskind and N.\ Toumbas, 
{\em JHEP} {\bf 0012} (2000) 002. 

\bibitem{CamelNC} G.\ Amelino-Camelia, L.\ Doplicher, S.-K.\ Nam
and Y.-S.\ Seo,
{\em Phys.\ Rev.} {\bf D 67} (2003) 085008. 

\bibitem{HellYou} R.C.\ Helling and J.\ You,
{\tt arXiv:0707.1885 [hep-th].}

\bibitem{thetainf} W.\ Bietenholz, F.\ Hofheinz and J.\ Nishimura,
{\em JHEP} {\bf 0405} (2004) 047. 

\bibitem{LLT} F.\ Ruiz Ruiz, 
{\em Phys.\ Lett.} {\bf B 502} (2001) 274. 
K.\ Landsteiner, E.\ Lopez and M.H.G.\ Tytgat,
{\em JHEP} {\bf 0106} (2001) 055. 
Z.\ Guralnik, R.C.\ Helling, K.\ Landsteiner and E.\ Lopez,
{\em JHEP} {\bf 0205} (2002) 025. 
M.\ Van Raamsdonk,
{\em JHEP} {\bf 0111} (2001) 006. 

\bibitem{Sen} A.\ Sen, {\em JHEP} {\bf 9808} (1998) 012. 

\bibitem{NCphi4} W.\ Bietenholz, F.\ Hofheinz and J.\ Nishimura,
{\em JHEP} {\bf 0406} (2004) 042. 

\bibitem{AHH} T.\ Azeyanagi, M.\ Hanada and T.\ Hirata,
{\tt arXiv:0806.3252 [hep-th].}

\bibitem{NCQED} W.\ Bietenholz, J.\ Nishimura, Y.\ Susaki and J.\ Volkholz,
{\em JHEP} {\bf 0610} (2006) 042. 
W.\ Bietenholz, A.\ Bigarini, J.\ Nishimura, Y.\ Susaki and
A.\ Torrielli,
{\em PoS(LATTICE2007)049}. 

\bibitem{OSax} K.\ Osterwalder and R.\ Schrader,
{\em Commun.\ Math.\ Phys.} {\bf 31} (1973) 83; 
{\em Commun.\ Math.\ Phys.} {\bf 42} (1975) 281.

\bibitem{kappaMink} S.\ Majid and H.\ Ruegg,
{\em Phys.\ Lett.} {\bf B 334} (1994) 348. 

\bibitem{AMNS} J.~Ambj{\o}rn, Y.~Makeenko, J.~Nishimura and R.J.~Szabo, 
{\it JHEP} \textbf{9911} (1999) 29; 
{\it Phys.\ Lett.} \textbf{B 480} (2000) 399; 
{\it JHEP} \textbf{05} (2000) 023. 

\bibitem{GAO} A.\ Gonzalez-Arroyo and M.\ Okawa,
{\em Phys.\ Rev.} {\bf D 27} (1983) 2397.
A.\ Gonzalez-Arroyo and C.P.\ Korthals Altes,
{\em Phys.\ Lett.} {\bf B 131} (1983) 396.

\bibitem{2dNCU1} W.\ Bietenholz, F.\ Hofheinz and J.\ Nishimura,
{\em JHEP} {\bf 0209} (2002) 009. 
W.\ Bietenholz, A.\ Bigarini and A.\ Torrielli,
{\em JHEP} {\bf 0708} (2007) 041. 

\bibitem{GroWit} D.J.\ Gross and E.\ Witten,
{\em Phys.\ Rev.} {\bf D 21} (1980) 446.

\bibitem{Peierls} R.\ Peierls, {\em Z.\ Phys.} {\bf 80} (1933) 763.

\bibitem{4dTEK} M.\ Teper and H.\ Vairinhos,
{\em Phys.\ Lett.} {\bf B 652} (2007) 359. 
T.\ Azeyanagi, M.\ Hanada, T.\ Hirata and T.\ Ishikawa,
{\em JHEP} {\bf 0801} (2008) 025. 

\bibitem{NC-CMB} Q.-G.\ Huang and M.\ Li,
{\em JHEP} {\bf 0306} (2003) 014; 
{\em JCAP} {\bf 0311} (2003) 001. 
S.\ Tsujikawa, R.\ Maartens and R.\ Brandenberger,
{\em Phys.\ Lett.} {\bf B 574} (2003) 141. 
A.H.\ Fatollahi and M.\ Hajirahimi,
{\em Europhys.\ Lett.} {\bf 75} (2006) 542; 
{\em Phys.\ Lett.} {\bf B 641} (2006) 381. 
A.P.\ Balachandran, A.R.\ Queiroz, A.M.\ Marques and P.\ Teotonio-Sobrinho,
{\em Phys.\ Rev.} {\bf D 77} (2008) 105032. 
E.\ Akofor, A.P.\ Balachandran, S.G.\ Jo, A.\ Joseph and B.A.\ Qureshi,
{\em JHEP} {\bf 0805} (2008) 092. 
L.\ Barosi, F.A.\ Brito and A.R.\ Queiroz,
{\em JCAP} {\bf 0804} (2008) 005. 
E.\ Akofor, A.P.\ Balachandran, A.\ Joseph, L.\ Pekowsky and B.A.\ Qureshi,
{\tt arXiv:0806.2458 [astro-ph].}

\bibitem{Lehnert} R.\ Lehnert,
{\em Phys.\ Rev.} {\bf D 68} (2003) 085003. 

\bibitem{CaMac} A.\ Camacho and A.\ Macias,
{\em Gen.\ Rel.\ Grav.} {\bf 39} (2007) 1175. 

\bibitem{Ellis} J.R.\ Ellis, N.E.\ Mavromatos, D.V.\ Nanopoulos, 
A.S.\ Sakharov and E.K.G.\ Sarkisyan,
{\em Astropart.\ Phys.} {\bf 25} (2006) 402; 
Erratum {\tt arXiv:0712.2781 [astro-ph].}

\bibitem{BWHC} S.E.\ Boggs, C.B.\ Wunderer, K.\ Hurley and 
W.\ Coburn,
{\em Astrophys.\ J.} {\bf 611} (2004) L77. 

\bibitem{RMP} M.\ Rodriguez Martinez and T.\ Piran,
{\em JCAP} {\bf 0605} (2006) 017. 

\bibitem{Magic} J.\ Albert {\it et al.} (MAGIC Collaboration),
and J.R.\ Ellis, N.E.\ Mavromatos, D.V.\ Nanopoulos, 
A.S.\ Sakharov and E.K.G.\ Sarkisyan,
{\em Phys.\ Lett.} {\bf B 668} (2008) 253. 

\bibitem{GLAST} http://glast.gsfc.nasa.gov/science/

\bibitem{Lamon} R.\ Lamon,
{\em JCAP} {\bf 0808} (2008) 022. 

\bibitem{EFMMN} J.R.\ Ellis, K.\ Farakos, N.E.\ Mavromatos, 
V.A.\ Mitsou, D.V.\ Nanopoulos,
{\em Astrophys.\ J.} {\bf 535} (2000) 139. 

\bibitem{LIV-GRB-nu} J.R.\ Ellis, N.E.\ Mavromatos, D.V.\ Nanopoulos 
and G.\ Volkov,
{\em Gen.\ Rel.\ Grav.} {\bf 32} (2000) 1777. 
V.\ Ammosov and G.\ Volkov,
{\tt hep-ph/0008032.}
U.\ Jacob and T.\ Piran,
{\em Nature Phys.} {\bf 3} (2007) 87. 
M.\ Biesiada and A.\ Pi\'{o}rkowska,
{\em JCAP} {\bf 05} (2007) 011. 
J.R.\ Ellis, N.\ Harries, A.\ Meregaglia, A.\ Rubbia and A.\ Sakharov,
{\tt arXiv:0805.0253 [hep-ph].}

\bibitem{MyPo} R.C.\ Myers and M.\ Pospelov,
{\em Phys.\ Rev.\ Lett.} {\bf 90} (2003) 211601. 

\bibitem{dim5LIV} P.A.\ Bolokhov and M.\ Pospelov,
{\em Phys.\ Rev.} {\bf D 77} (2008) 025022. 

\bibitem{CriVac} P.M.\ Crichigno and H.\ Vucetich,
{\em Phys.\ Lett.} {\bf B 651} (2007) 313. 

\bibitem{RUV} C.M.\ Reyes, L.\ Urrutia and J.D.\ Vergara,
{\tt arXiv:0806.2166 [hep-ph].}

\bibitem{JLMS} T.A.\ Jacobson, S.\ Liberati, D.\ Mattingly and
F.W. Stecker,
{\em Phys.\ Rev.\ Lett.} {\bf 93} (2004) 021101. 

\bibitem{KosLeh} V.A.\ Kosteleck\'{y} and A.G.M.\ Pickering,
{\em Phys.\ Rev.\ Lett.} {\bf 91} (2003) 031801. 

\bibitem{UHECRgamma1} M.\ Galaverni and G.\ Sigl,
{\em Phys.\ Rev.\ Lett.} {\bf 100} (2008) 021102.

\bibitem{UHECRgamma2} L.\ Maccione and S.\ Liberati, 
{\tt arXiv:0805.2548 [astro-ph].}

\bibitem{MLCK} L.\ Maccione, S.\ Liberati, A.\ Celotti and J.G.\ Kirk,
{\em JCAP} {\bf 0710} (2007) 013. 

\bibitem{GaPu} R.\ Gambini and J.\ Pullin,
{\em Phys.\ Rev.} {\bf D 59} (1999) 124021. 

\bibitem{Petrov1} A.F.\ Ferrari, M.\ Gomes, J.R.\ Nascimento, 
E.\ Passos, A.Yu.\ Petrov and A.J.\ da Silva,
{\em Phys.\ Lett.} {\bf B 652} (2007) 174. 

\bibitem{KosMew} V.A.\ Kosteleck\'{y} and M.\ Mewes,
{\em Phys.\ Rev.\ Lett.} {\bf 87} (2001) 251304. 

\bibitem{BireNC} S.A.\ Abel, J.\ Jaeckel, V.V.\ Khoze and A.\ Ringwald,
{\em JHEP} {\bf 0609} (2006) 074. 

\bibitem{bireaexp} T.\ Heinzl, B.\ Liesfeld, K.-U.\ Amthor, 
H.\ Schwoerer, R.\ Sauerbrey and A.\ Wipf,
{\em Opt.\ Commun.} {\bf 267} (2006) 318. 


\bibitem{Popper}  K.R.\ Popper, 
``Conjectures and Refutations: The Growth of Scientific Knowledge'',
Routledge, U.K. (1963).

\bibitem{TA} http://www.telescopearray.org/

\bibitem{EUSO} http://euso.riken.go.jp/

\bibitem{OWL} http://owl.gsfc.nasa.gov/

\bibitem{spaceEAS} M.\ Pallavicini, R.\ Pesce, A.\ Petrolini and A.\ Thea,
{\tt arXiv:0810.5711 [astro-ph].}

\bibitem{AGILE} http://agile.asdc.asi.it/

\bibitem{HESS} http://www.mpi-hd.mpg.de/hfm/HESS/

\bibitem{VHEGR} A.\ De Angelis, O.\ Mansutti and M.\ Persic,
{\em Riv.\ Nuovo Cim.} {\bf 31} (2008) 187. 

\bibitem{SVOM} S.\ Basa, J.\ Wei, J.\ Paul and S.N.\ Zhang
(for the SVOM Collaboration),
{\tt arXiv:0811.1154 [astro-ph].}

\bibitem{icecube} http://www.icecube.wisc.edu/ \\
F.\ Halzen, A.\ Kappes and A.\ \'{O} Murchadha,
{\em Phys.\ Rev.} {\bf D 78} (2008) 063004.

\bibitem{ANITAetc} http://amanda.uci.edu/$\sim$anita/ \\
http://antares.in2p3.fr/ \\
http://www.nestor.org.gr/

\bibitem{LOPES} H.\ Falcke {\it et al.} (LOPES Collaboration),
{\em Nature} {\bf 435} (2005) 313. 


\bibitem{AGNHyp} Pierre Auger Collaboration (J.\ Abraham {\it et al.}),
{\em Science} {\bf 318} (2007) 939.

\bibitem{AGNearly} V.\ Berezinsky, A.Z.\ Gazizov and S.I.\ Grigorieva,\\
{\tt astro-ph/0210095.}

\bibitem{AGNacc} S.\ Collin, 
{\tt arXiv:0811.1731 [astro-ph].}
P.L.\ Biermann {\it et al.},
{\tt arXiv:0811.1848 [astro-ph].}

\bibitem{ScuSte08} S.T.\ Scully and F.W.\ Stecker,
{\tt arXiv:0811.2230 [astro-ph].}

\bibitem{VCV} M.-P.\ V\'{e}ron-Cetty and P. V\'{e}ron, 
{\em Astron.\ and Astrophys.} {\bf 455} (2006) 773.

\bibitem{AGNSwift} M.R.\ George, A.C.\ Fabian, W.H.\ Baumgartner, 
R.F.\ Mushotzky and J.\ Tueller,
{\tt arXiv:0805.2053 [astro-ph].}

\bibitem{spiral} G.\ Ghisellini, G.\ Ghirlanda, F.\ Tavecchio, 
F.\ Fraternali and G. Pareschi,
{\tt arXiv:0806.2393 [astro-ph].}

\bibitem{Chile} N.M.\ Nagar and J.\ Matulich,
{\tt arXiv:0806.3220 [astro-ph].}

\bibitem{Stanevaniso} T.\ Stanev, P.L.\ Biermann, J.\ Lloyd-Evans, 
J.P.\ Rachen and A.\ Watson,
{\em Phys.\ Rev.\ Lett.} {\bf 75} (1995) 3056. 

\bibitem{AGASAaniso} N.\ Hayashida {\it et al.}
(AGASA Collaboration),
{\em Phys.\ Rev.\ Lett.} {\bf 77} (1996) 1000.
 
\bibitem{anticlust} R.U.\ Abbasi {\it et al.}
(High Resolution Fly's Eye Collaboration),
{\em Astrophys.\ J.} {\bf 623} (2005) 164. 

\bibitem{Stanev08} T.\ Stanev,
{\tt arXiv:0805.1746 [astro-ph].}

\bibitem{PAO2} Pierre Auger Collaboration (J.\ Abraham {\it et al.}),
{\em Astropart.\ Phys.} {\bf 29} (2008) 188. 

\bibitem{LuLin} C.-C.\ Lu and G.-L.\ Lin,
{\tt arXiv:0804.3122 [astro-ph].}

\bibitem{AGNHyp2} A.A.\ Ivanov (for the Yakutsk Array Group),
{\em Pis'ma v ZhETF} {\bf 87} (2008) 215. 

\bibitem{HiRes08} R.U.\ Abbasi {\it et al.} (HiRes Collaboration),
{\tt arXiv:0804.0382 [astro-ph].}

\bibitem{Dermer08} C.D.\ Dermer,
{\tt arXiv:0804.2466 [astro-ph].}

\bibitem{GTTT} D.\ Gorbunov, P.\ Tinyakov, I.\ Tkachev
and S.\ Troitsky,\\
{\em JETP Letters} {\bf 87} (2008) 461; 
{\tt arXiv:0804.1088 [astro-ph].}

\bibitem{KOT} M.\ Kachelrie\ss, S.\ Ostapchenko and R.\ Tom\`{a}s,
{\tt arXiv:0805.2608 [astro-ph].}

\bibitem{AABMOPSY} D.\ Allard, M.\ Ave, N.\ Busca, M.A.\ Malkan, 
A.V.\ Olinto, E.\ Parizot, F.W.\ Stecker and T.\ Yamamoto,
{\em JCAP} {\bf 0609} (2006) 005. 

\bibitem{Sigl} G.\ Sigl,
{\tt arXiv:0803.3800 [astro-ph].}

\bibitem{Kron} P.P.\ Kronberg,
{\em Space Sci.\ Rev.} {\bf 75} (1996) 387.

\bibitem{Fargi} D.\ Fargion,
{\em Phys.\ Scripta} {\bf 78} (2008) 045901. 

\bibitem{cosnuc} R.\ Aloisio, V.\ Berezinsky and S.\ Grigorieva,
{\tt arXiv:0802.4452 [astro-ph].}
R.\ Aloisio, V.\ Berezinsky and A.\ Gazizov,
{\tt arXiv:0803.2494 [astro-ph].}

\bibitem{LGM} L.\ Gonzalez-Mestres, 
{\tt arXiv:0802.2536 [hep-ph].}

\bibitem{ABDOP} D.\ Allard, N.G.\ Busca, G.\ Decerprit, 
A.V.\ Olinto and E.\ Parizot,
{\tt arXiv:0805.4779 [astro-ph].}

\bibitem{Unger} M.\ Unger (for the Pierre Auger Collaboration),
{\tt arXiv:0706.1495 [astro-ph].}

\bibitem{Fedorova} Y.\ Fedorova {\it et al.}\
(High Resolution Fly's Eye Collaboration),
{\em Proc.\ 30$^{\, th}$ ICRC} (2007) 1236.

\bibitem{Petrera} S.\ Petrera,
{\tt arXiv:0810.4710 [astro-ph].}


\end{thebibliography}
